\documentclass[unpublished]{JHEP3} 

\JHEPspecialurl{http://jhep.sissa.it/JOURNAL/JHEP3.tar.gz}

\usepackage{epsfig,multicol,bbm}

\newcommand\fverb{\setbox\fverbbox=\hbox\bgroup\verb}
\newcommand\fverbdo{\egroup\medskip\noindent%
                        \fbox{\unhbox\fverbbox}\ }
\newcommand\fverbit{\egroup\item[\fbox{\unhbox\fverbbox}]}
\newbox\fverbbox

\usepackage{graphicx}
\usepackage{amsmath}
\usepackage{amsfonts}
\usepackage{amssymb}
\def\msusy{M_{SUSY}}
\def\cvi{CV_1}
\def\cvii{CV_2}
\def\meff{m_{eff}}

\def\mupmum{\mu^+\mu^-}
\def\mmumu{M_{\mupmum}}
\def\mups{M_{\Upsilon(1S)}}
\def\upsi{\Upsilon_{1S}}

\def\upsii{\Upsilon_{2S}}

\def\upsiii{\Upsilon_{3S}}
\def\mupsiii{M_{\upsiii}}
\def\mupsiii{M_{\Upsilon(3S)}}

\def\pbi{~\mbox{pb}^{-1}}
\def\rbt{R_{b/t}}

\def\cabb{C_{ab\anti b}}
\def\catt{C_{at\anti t}}
\def\camumu{C_{a\mu^-\mu^+}}
\def\catautau{C_{a\tau^-\tau^+}}
\def\caibb{C_{\ai b\anti b}}
\def\caitt{C_{\ai t\anti t}}
\def\caimumu{C_{\ai \mu^-\mu^+}}
\def\caitautau{C_{\ai \tau^-\tau^+}}
\def\cmax{\cabb^{\rm max}}

\def\cta{\cos\theta_A}
\def\ctamax{\cta^{\rm max}}
\def\sta{\sin\theta_A}

\def\ma{m_a}
\def\mh{m_h}

\def\hi{h_1}
\def\hii{h_2}
\def\hiii{h_3}
\def\ai{a_1}
\def\aii{a_2}
\def\mhi{m_{h_1}}
\def\mhii{m_{h_2}}

\def\mai{m_{a_1}}

\def\mtau{m_\tau}

\def\mups{M_{\Upsilon(1S)}}
\def\beq{\begin{equation}}
\def\eeq{\end{equation}}
\def\bea{\begin{eqnarray}}
\def\eea{\end{eqnarray}}
\def\ie{{\it i.e.}}

\def\lsim{\mathrel{\raise.3ex\hbox{$<$\kern-.75em\lower1ex\hbox{$\sim$}}}}
\def\gsim{\mathrel{\raise.3ex\hbox{$>$\kern-.75em\lower1ex\hbox{$\sim$}}}}
\def\ifmath#1{\relax\ifmmode #1\else $#1$\fi}

\def\fbi{~{\mbox{fb}^{-1}}}

\def\br{BR}
\def\gev{~{\mbox{GeV}}}

\def\to{\rightarrow}

\def\hsm{h_{SM}}

\def\eps{\epsilon}

\def\anti{\overline}

    \def\fillboxx#1#2{\hbox to #1{\vbox to #2{\vfil}\hfil}   }

\def\tauptaum{\tau^+\tau^-}

\def\gev{~{\rm GeV}}
\def\gam{\gamma}
\def\tanb{\tan\beta}
\def\cotb{\cot\beta}

\def\anti{\overline}

\def\alam{A_{\lambda}}
\def\akap{A_{\kappa}}

\def\epem{e^+e^-}

\newcommand{ \slashchar }[1]{\setbox0=\hbox{$#1$}   
   \dimen0=\wd0                                     
   \setbox1=\hbox{/} \dimen1=\wd1                   
   \ifdim\dimen0>\dimen1                            
      \rlap{\hbox to \dimen0{\hfil/\hfil}}          
      #1                                            
   \else                                            
      \rlap{\hbox to \dimen1{\hfil$#1$\hfil}}       
      /                                             
   \fi}     

\title{\boldmath New constraints on a light CP-odd Higgs
  boson and related NMSSM Ideal Higgs Scenarios.
}

\author{Radovan Dermisek
   \\ Department of Physics, Indiana University, Bloomington, IN 47405} 
\author{John
  F. Gunion \\ Department of Physics, University of California, Davis,
  CA 95616, USA\\ and \\Theory Group, CERN, CH-1211, Geneva 23, Switzerland}

\abstract{Recent BaBar limits on $\br(\Upsilon(3S)\to \gam a\to \gam
  \tauptaum)$ and $\br(\Upsilon(3S)\to \gam a\to \gam \mupmum)$
  provide increased constraints on the $a b\anti b$ coupling of a
  CP-odd Higgs boson, $a$, with $\ma<\mupsiii$. We extract these
  limits from the BaBar data and compare to the limits previously
  obtained using other data sets, especially the CLEO-III
  $\br(\Upsilon(1S)\to \gam\to\tauptaum)$ limits.  Comparisons are
  made to predictions in the context of ``ideal''-Higgs NMSSM
  scenarios, in which the lightest CP-even Higgs boson, $\hi$, can
  have mass below $105\gev$ (as preferred by precision electroweak
  data) and yet can escape old LEP limits by virtue of decays to a
  pair of the lightest CP-odd Higgs bosons, $\hi\to\ai\ai$, with
  $\mai<2m_B$.  Most such scenarios with $\mai<2\mtau$ are eliminated,
  but the bulk of the $\mai>7.5\gev$ scenarios, which are
  theoretically the most favored, survive. We also outline the impact
  of the new ALEPH LEP results in the $\epem\to Z+4\tau$ channel.  For
  $\tanb\geq 3$, only NMSSM ideal Higgs scenarios with $\mhi\gsim
  98\gev$ and $\mai$ close to $2m_B $ satisfy the ALEPH limits.  For
  $\tanb\lsim 2$, the ALEPH limits are easily satisfied for the most
  theoretically preferred NMSSM scenarios, which are those with $\mai$
  close to $2m_B$ and $\mhi\sim 90\gev-100\gev$.  } \received{}
\accepted{} \preprint{} \textheight=9in
\preprint{CERN-PH-TH/2010-031\\IUHET-541} 
\begin{document}

\section{Introduction}

Many motivations for the existence of a light CP-odd $a$ Higgs boson
have emerged in a variety of contexts in recent years.  Of particular
interest is the $\ma<2m_B$ region, for which a light Higgs, $h$, with
SM-like $WW$, $ZZ$ and fermionic couplings can have mass $\mh\sim
100\gev$ while still being consistent with published LEP data by virtue of $h\to
aa\to 4\tau$ or $4~jet$ decays being
dominant~\cite{Dermisek:2005ar,Dermisek:2005gg,Dermisek:2006wr,Dermisek:2007yt}
(see also \cite{Chang:2005ht,Chang:2008cw}).  Such a light Higgs
provides perfect agreement with the rather compelling precision
electroweak constraints, and for $\br(h\to aa)\gsim 0.75$ also
provides an explanation for the $\sim 2.3\sigma$ excess observed at
LEP in $\epem \to Z b\anti b$ in the region $M_{b\anti b}\sim
100\gev$.  This is sometimes referred to as the ``ideal'' Higgs
scenario.  More generally, superstring modeling suggests the
possibility of many light $a$'s.  In this note, we update the analysis
of \cite{Gunion:2008dg} (see also \cite{Domingo:2008rr}), quantifying the increased constraints on a
general CP-odd $a$ arising from recent BaBar limits on the branching
ratio for $\Upsilon(3S)\to \gam a\to \gam\tauptaum$ decays
\cite{Aubert:2009cka} and $\Upsilon(3S)\to \gam a\to \gam\mupmum$
decays \cite{Aubert:2009cp}.  We also quantify the impact of these
constraints, as well as the impact of the new  ALEPH LEP results
in the $\epem\to Z+4\tau$ final state~\cite{cranmer}, on the
Next-to-Minimal Supersymmetric Model (NMSSM) ideal Higgs scenarios.

The possibilities for discovery of an $a$ and limits on the $a$ are
phrased in terms of the $a\mu^-\mu^+$, $a\tau^-\tau^+$, $ab\anti b$
and $at\anti t$ couplings defined via
\beq
{\cal L}_{af\anti f}\equiv i C_{af\anti f}{ig_2m_f\over2m_W}\anti f \gamma_5 f
a\,.
\label{cabbdef}
\eeq 
(Note: when discussing a generic CP-even (CP-odd) Higgs boson, we
will use the notation $h$ ($a$).  When specializing to the NMSSM
context, we will use $\hi,\hii,\hiii$ ($\ai,\aii$) for the mass
ordered Higgs states.) In this paper, we assume a Higgs model in which
$\camumu=\catautau=\cabb$, as typified by a two-Higgs-doublet model
(2HDM) of either type-I or type-II, or more generally if the lepton
and down-type quark masses are generated by the same combination of Higgs
fields. However, one should keep in mind that there are models in
which $r=(\camumu=\catautau) /\cabb\gg 1$ --- such models include
those in which the muon and tau masses are generated by different
Higgs fields than the $b$ mass.  In a 2HDM of type-II and in the MSSM,
$\camumu=\catautau=\cabb=\tanb$ (where $\tanb=h_u/h_d$ is the ratio of
the vacuum expectation values for the doublets giving mass to up-type
quarks vs.  down-type quarks) and $\catt=\cotb$.  These results are
modified in the NMSSM (see, e.g. \cite{Ellis:1988er} and \cite{hhg}).
\footnote{A convenient program for exploring the NMSSM Higgs sector
is NMHDECAY~\cite{Ellwanger:2004xm,Ellwanger:2005dv}.} In the NMSSM, both
$\caitt$ and $\caibb=\caimumu=\caitautau$ are multiplied by a factor
$\cta$, where $\cta$ is defined by
\beq
\ai=\cta a_{MSSM}+\sta a_S,
\eeq
where $\ai$ is the lightest of the 2 CP-odd scalars in the model.
Above, $a_{MSSM}$ is the CP-odd (doublet) scalar in the MSSM sector of
the NMSSM and $a_S$ is the additional CP-odd singlet scalar of the
NMSSM.  In terms of $\cta$, $\caimumu=\caitautau=\caibb=\cta\tanb$ and
$\caitt=\cta\cotb$.  Quite small values of $\cta$ are natural when
$\mai$ is small as a result of being close to the $U(1)_R$ limit of
the model.  In the most general Higgs model, $\camumu$, $\catautau$,
$\cabb$ and $\catt$ will be more complicated functions of the vevs of
the Higgs fields and the structure of the Yukawa couplings. In this
paper, we assume $\camumu=\catautau=\cabb$ and
$\cabb/\catt=\tan^2\beta$. 

For the analysis presented in this paper, we neglect the possible
presence of large corrections at large $\tanb$ to $\cabb$ from SUSY
loops~\cite{Hall:1993gn,Carena:1994bv,Pierce:1996zz}. These are
typically characterized by the quantity $\Delta_b$ which is crudely of
order ${\mu\tanb\over 16\pi^2M_{SUSY}}$. The correction to the
coupling takes the form of $1/(1+\Delta_b)$. Since $\mu$ can have
either sign, $\cabb$ can be either enhanced or suppressed relative to
equality with $\catautau $ (the corrections to which are much smaller)
and $\camumu$ (the corrections to which are negligible). This same
correction factor would apply to $\caibb$ in the NMSSM case.

\begin{figure}
\begin{center}
\includegraphics[width=0.55\textwidth,angle=90]{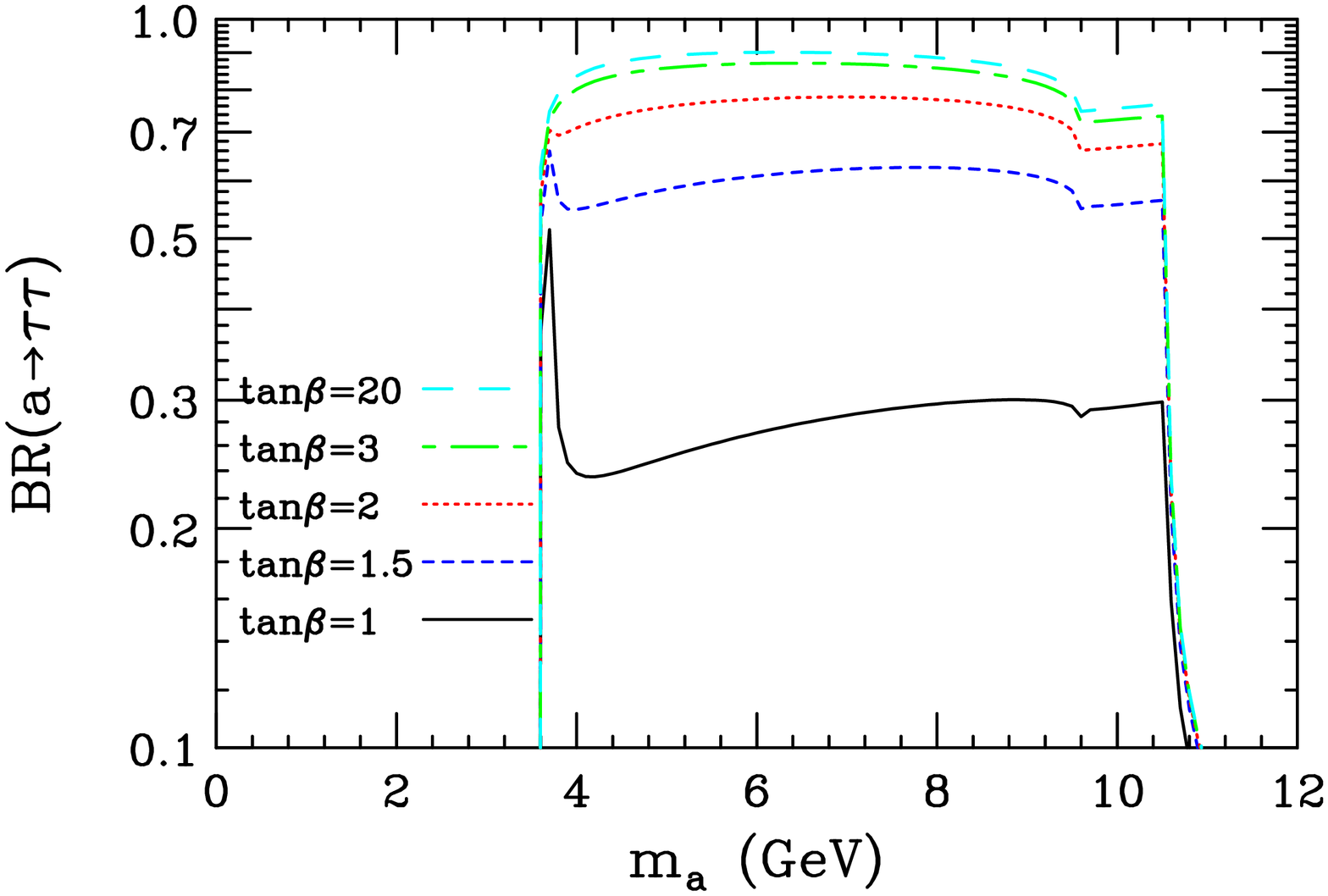}
\end{center}
\vspace*{-.3in}
\caption{$\br(a\to \tauptaum)$ is plotted as a function of $\ma$ for a
  variety of $\tanb$ values. $\br(a\to\tauptaum)$ is independent of $\cta$.
}
\label{bratautau}
\end{figure}

\begin{figure}
\begin{center}
\includegraphics[width=0.55\textwidth,angle=90]{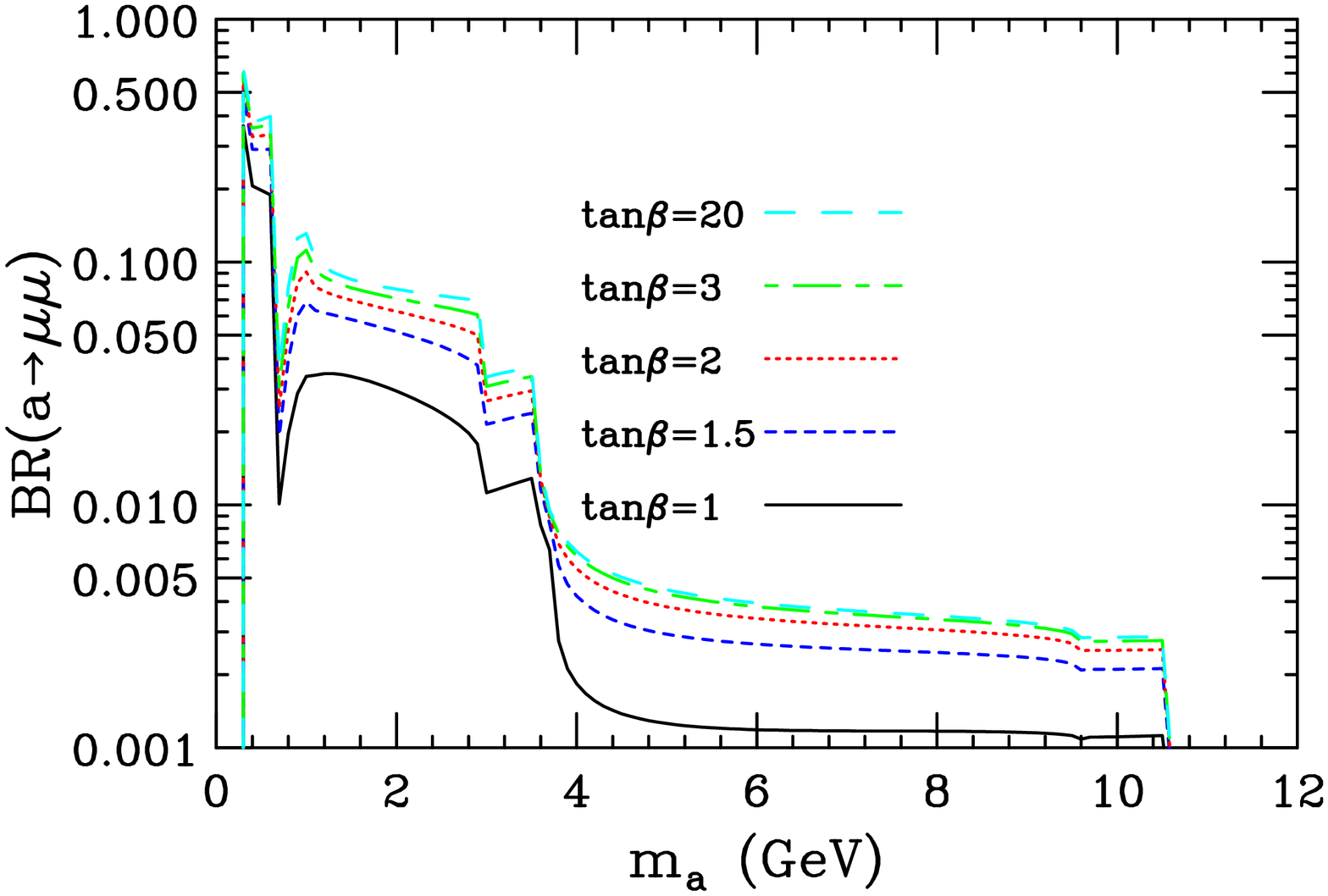}
\end{center}
\vspace*{-.3in}
\caption{$\br(a\to \mupmum)$ is plotted as a function of $\ma$ for a
  variety of $\tanb$ values. $\br(a\to\mupmum)$ is independent of $\cta$.
}
\label{bramumu}
\end{figure}

\begin{figure}
\begin{center}
\includegraphics[width=0.55\textwidth,angle=90]{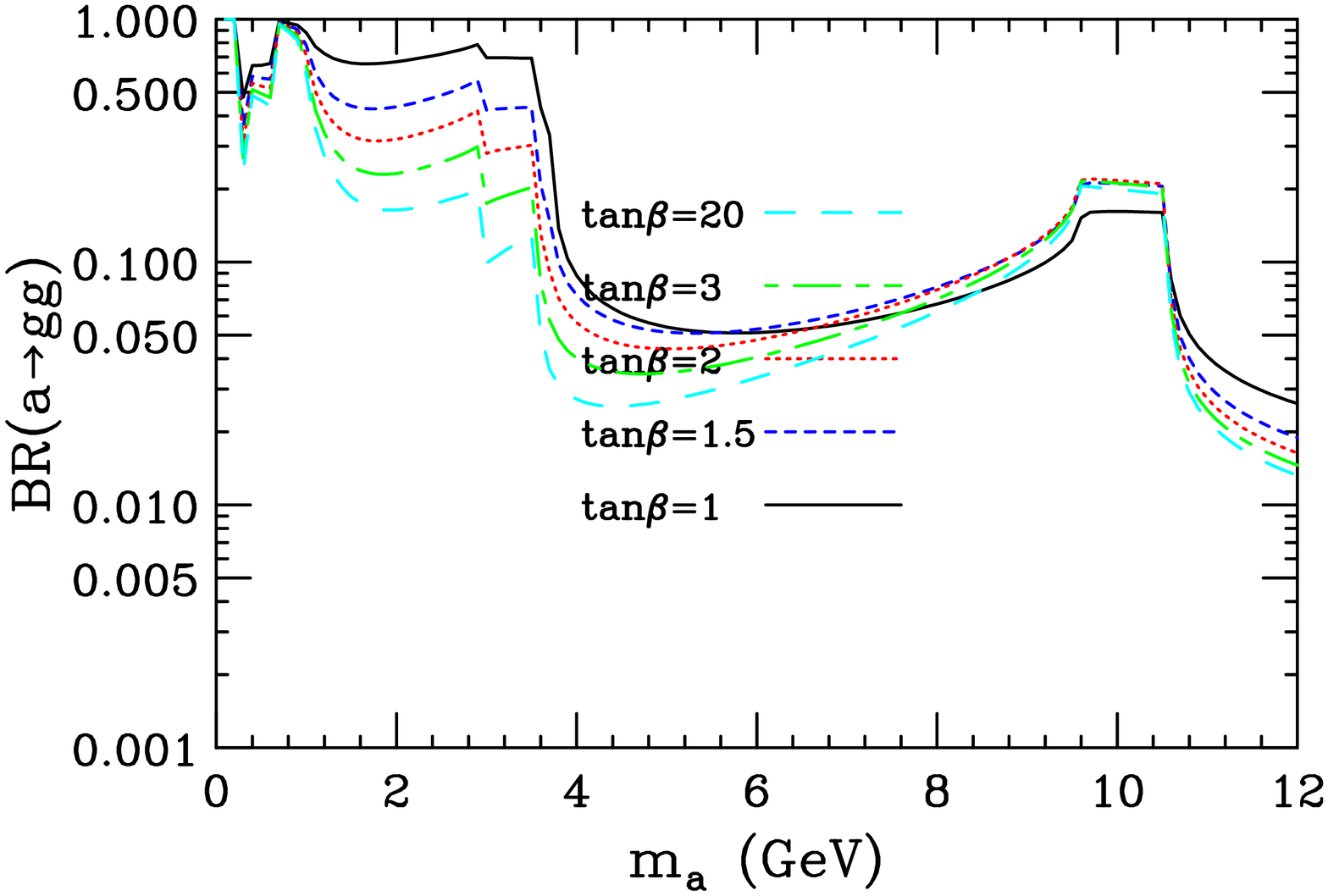}
\end{center}
\vspace*{-.3in}
\caption{$\br(a\to gg)$ is plotted as a function of $\ma$ for a
  variety of $\tanb$ values. 
}
\label{bragg}
\end{figure}

Key ingredients in understanding current limits are the branching
ratios for $a\to \tauptaum$ and $a\to \mupmum$ decays. These branching
ratios are plotted in Figs.~\ref{bratautau} and \ref{bramumu}. (It is
important to note that at tree-level the $a$ branching ratios apply
equally to the $\ai$, independent of $\cta$, due to the absence of
tree-level $a,\ai \to VV$ couplings and similar.)  Note that
$\br(a\to\tauptaum)$ and $\br(a\to\mupmum)$ change very little with
increasing $\tanb$ at any given $\ma$ once $\tanb\gsim 2$. We note
that in the region $\ma<2\mtau$, $\br(a\to\mupmum)$ has some
significant structures that arise from the fact that $\br(a\to gg)$ is
substantial and varies rapidly in that region. The rapid variation in
$\br(a\to gg)$ occurs when $\ma$ crosses the internal quark loop
thresholds. At higher $\ma$, $\br(a\to gg)$ becomes significant for
$\ma$ near $2m_b$. We plot $\br(a\to gg)$ in Fig.~\ref{bragg}. Note
that in the calculation of $\br(a\to gg)$ we have chosen to keep the
loop quark masses equal to the current quark masses in our
calculations, whereas we employ thresholds of $2m_K$ and $2m_D$ for
the strange quark and charm quark final states, respectively.  Some
changes in the structures present, especially in $\br(a\to\mupmum)$,
take place if, instead, the loop quark masses are set equal to the
true physical threshold masses.

\begin{figure}
\begin{center}
\includegraphics[width=0.55\textwidth,angle=90]{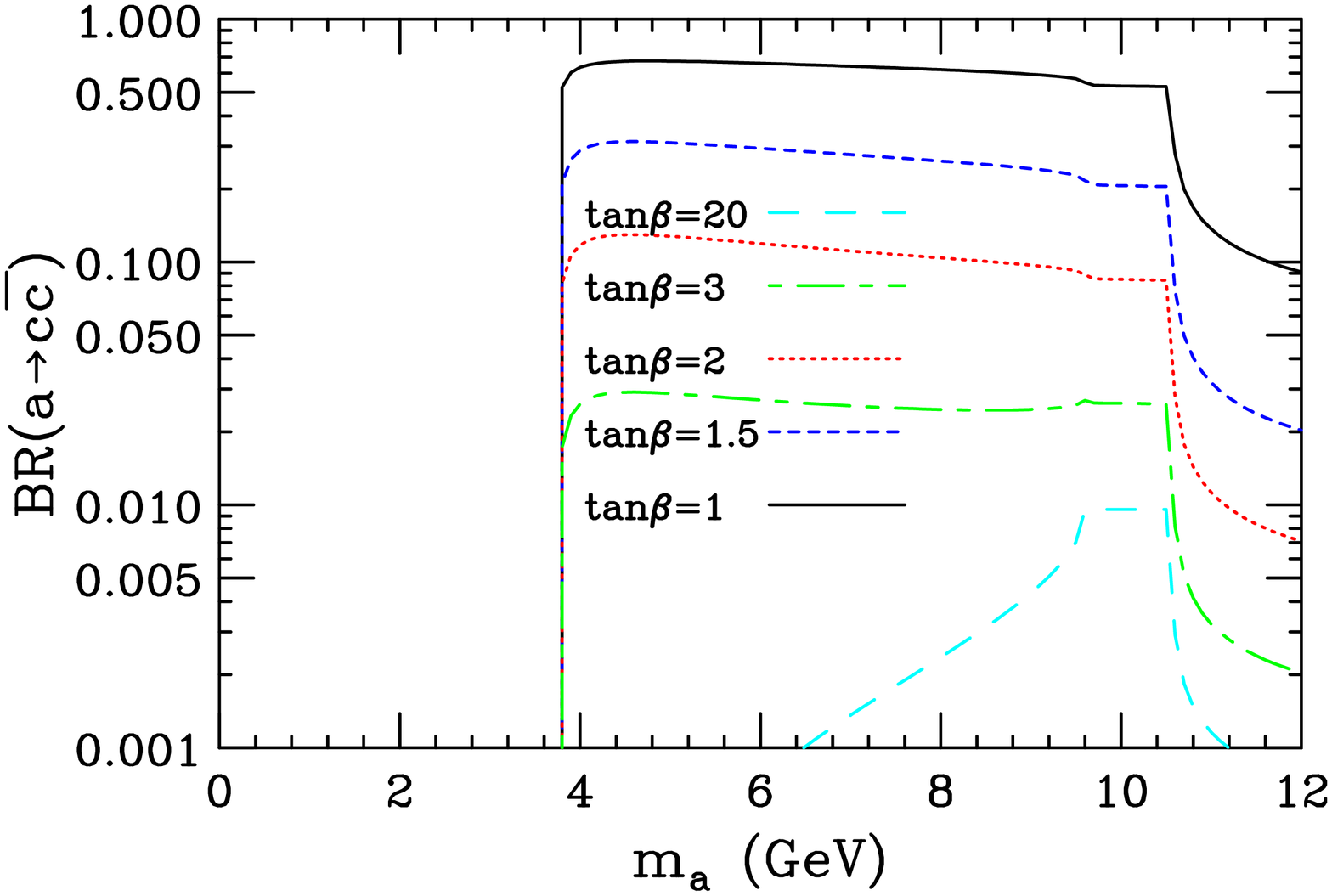}
\end{center}
\vspace*{-.3in}
\caption{$\br(a\to c\anti c)$ is plotted as a function of $\ma$ for a
  variety of $\tanb$ values. 
}
\label{bracc}
\end{figure}

\begin{figure}
\begin{center}
\includegraphics[width=0.55\textwidth,angle=90]{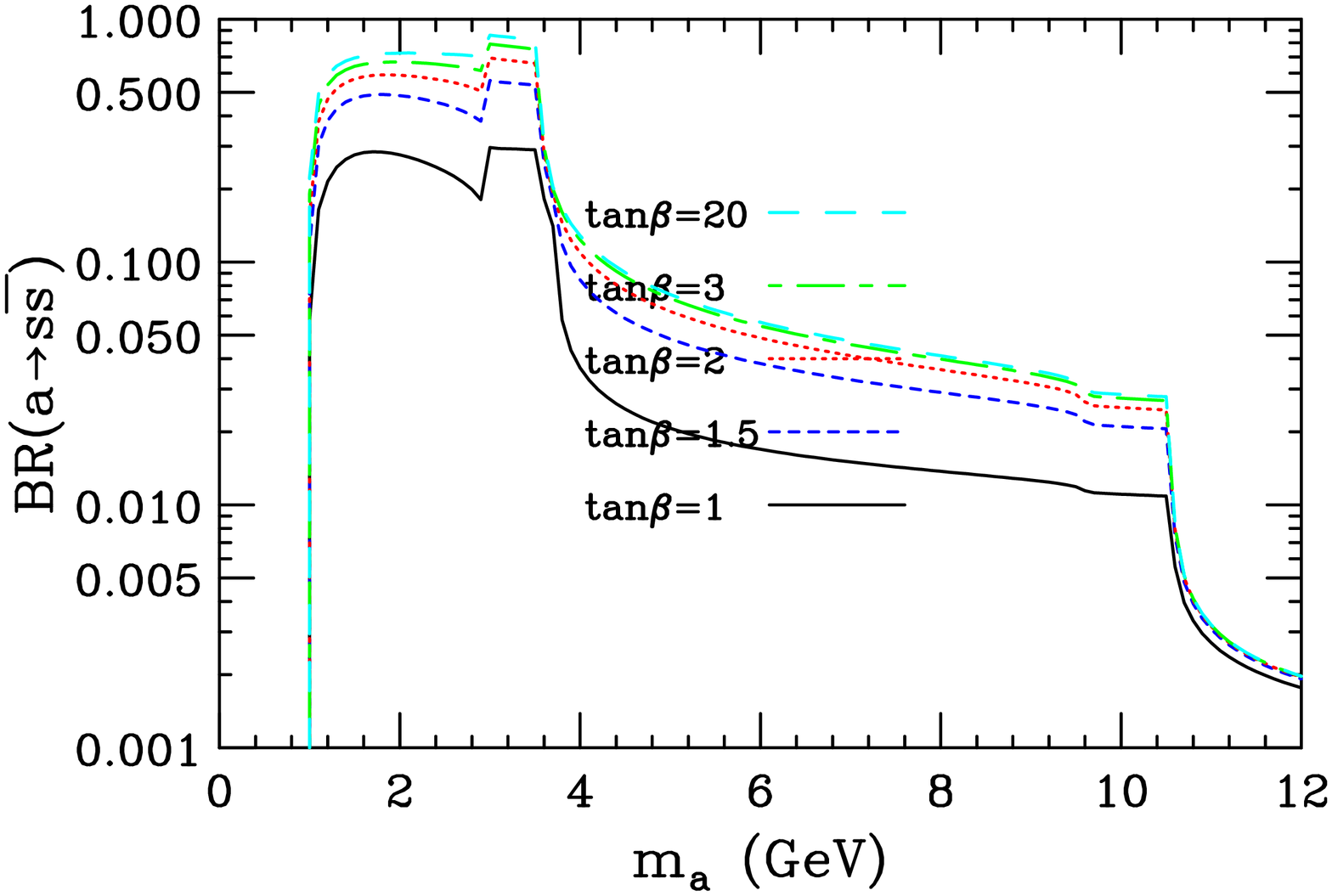}
\end{center}
\vspace*{-.3in}
\caption{$\br(a\to s\anti s)$ is plotted as a function of $\ma$ for a
  variety of $\tanb$ values. 
}
\label{brass}
\end{figure}

Of course, the above branching ratios are impacted by the $a\to c\anti
c$ and $a\to s\anti s$ channels, the latter being a rather important
competitor for smaller $\tanb$ and $\ma>2m_K$.  Plots of these
branching ratios appear in Figs.~\ref{bracc} and \ref{brass},
respectively.

It is relevant to note that both $\br(a\to \mupmum)$ and $\br(a\to
\tauptaum)$ tend to decline slowly as $\ma$ is increased, with a
significant dip in the latter for $\ma$ close to $2m_b$ where the
$b$-loop contribution to the $gga$ coupling is close to the point at
which the internal $b$'s can go on-shell .  This has important
implications for using these channels to probe the $9\gev\lsim\ma\lsim
2m_B$ region in which many parameter choices lead to absence of
light-$\ai$ finetuning in the NMSSM. ``Light-$\ai$'' finetuning is
characterized numerically by a quantity we call $G$, defined in
\cite{Dermisek:2006wr}, that gives the degree of precision with which
the $\alam$ and $\akap$ soft-SUSY-breaking NMSSM parameters must be
chosen in order that $\mai<2m_B$ {\it and} $\br(\hi\to \ai\ai)>0.75$
as required to allow $\mhi\lsim 105\gev$ to be consistent with
published LEP constraints when the $\hi$ has SM-like $\hi ZZ$
coupling. Absence of light-$\ai$ finetuning is equivalent to $G\lsim
20$. Typically, this condition is satisfied only when the light $\ai$
of the NMSSM is mainly singlet.  For example, at
$\tanb=10$, $0.6\lsim |\caibb|\lsim 1.2$ ($0.06\lsim |\cta|\lsim 0.12$)
is required if $G<20$ is imposed as well as requiring $\mai<2m_B$ and
$\br(\hi\to \ai\ai)>0.75$, with $G<10$ achieved only for $\cta\in
[-0.08,-0.1]$, corresponding to $|\caibb|\in [0.8,1]$. The $G<10$ range
for $\tanb=3$ is broader, $\cta\in[-0.28,-0.08]$, while that for
$\tanb=50$ is narrow, $\cta\in[-0.04,-0.06]$, yielding
$|\caibb|\in[0.24,0.84]$ and $|\caibb|\in[2,3]$, respectively. Thus,
lower $\tanb$ values will be harder to probe using direct limits on
the $\ai$.

We emphasize that, given the importance of the exact $a$ or $\ai$ branching
ratios in the analyses that follow, additional attention to the
most precise predictions possible is warranted.  Our $a,\ai$ decay results
employ a branching ratio program that is taken from
HDECAY~\cite{Djouadi:1997yw}.  We note that the $\ai$ branching ratios
obtained using this program are somewhat different than those that one
obtains using the $\ai$ decay formulae in the current version of
NMHDECAY. In particular, the former often predicts smaller $\br(\ai\to
\tauptaum)$ than does the latter.

\section{Upsilon decay limits compared to NMSSM predictions}
\begin{figure}[h!!]
\begin{center}
\includegraphics[width=0.27\textheight,angle=90]{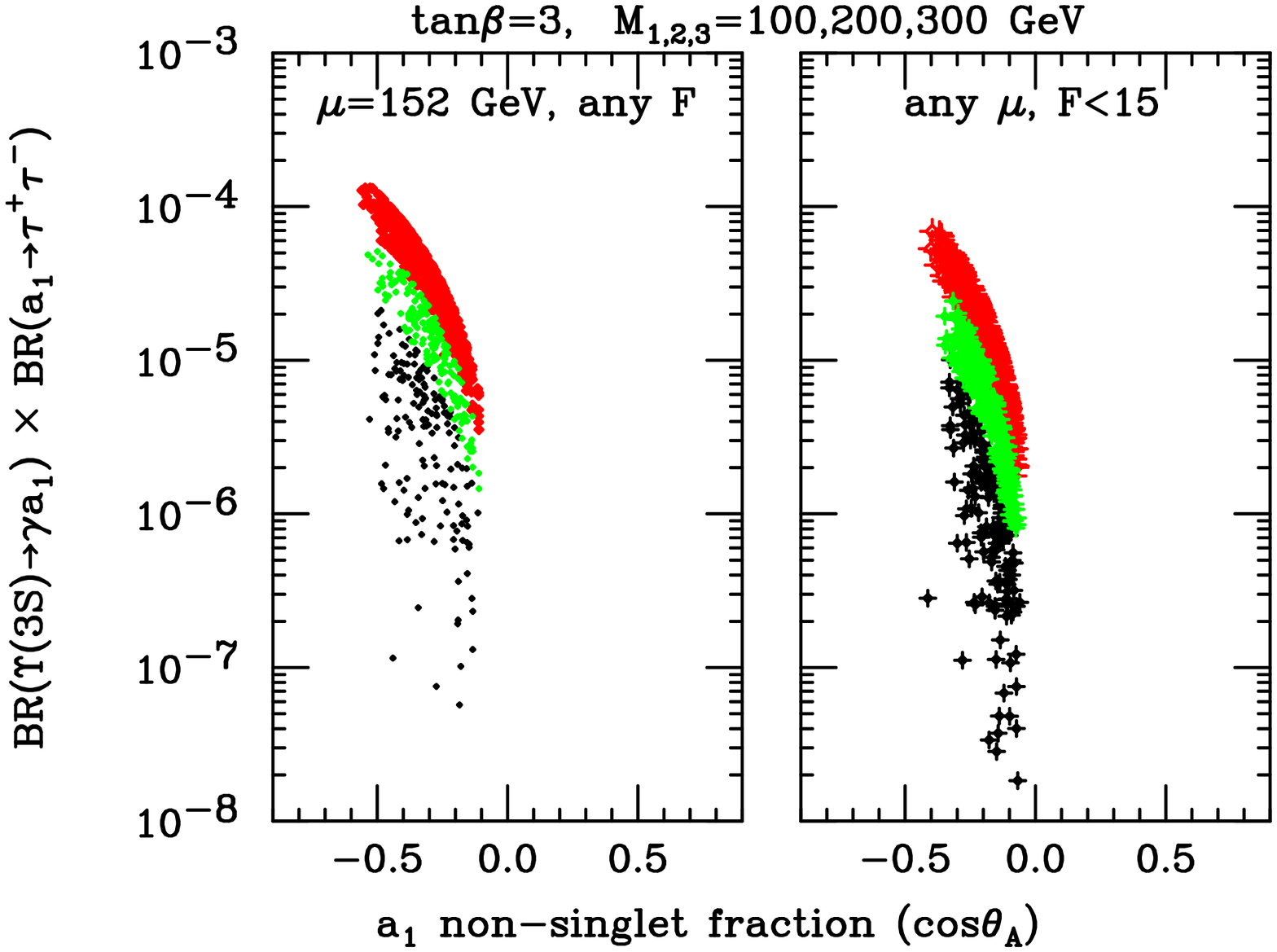}
\includegraphics[width=0.27\textheight,angle=90]{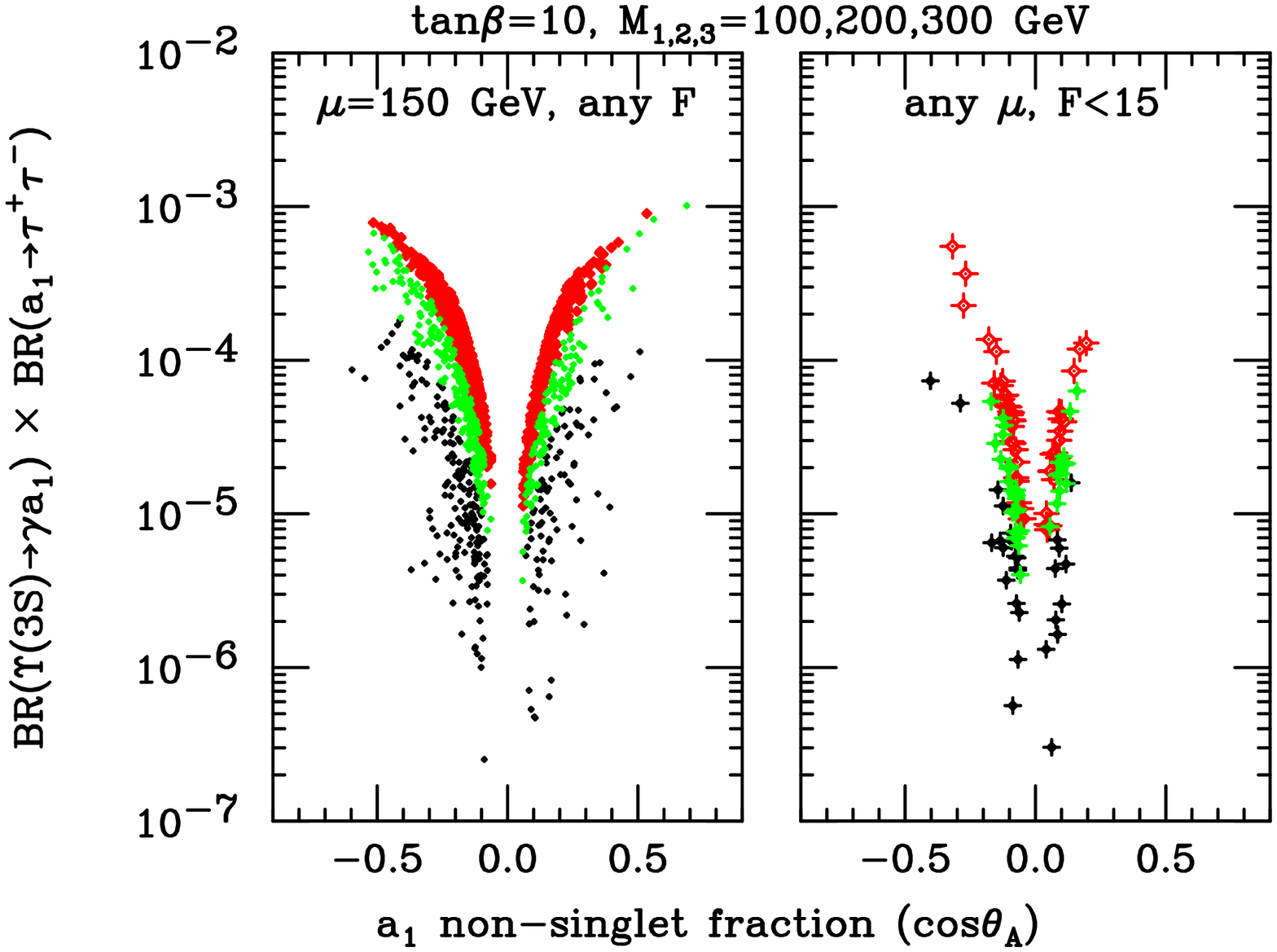}
\includegraphics[width=0.27\textheight,angle=90]{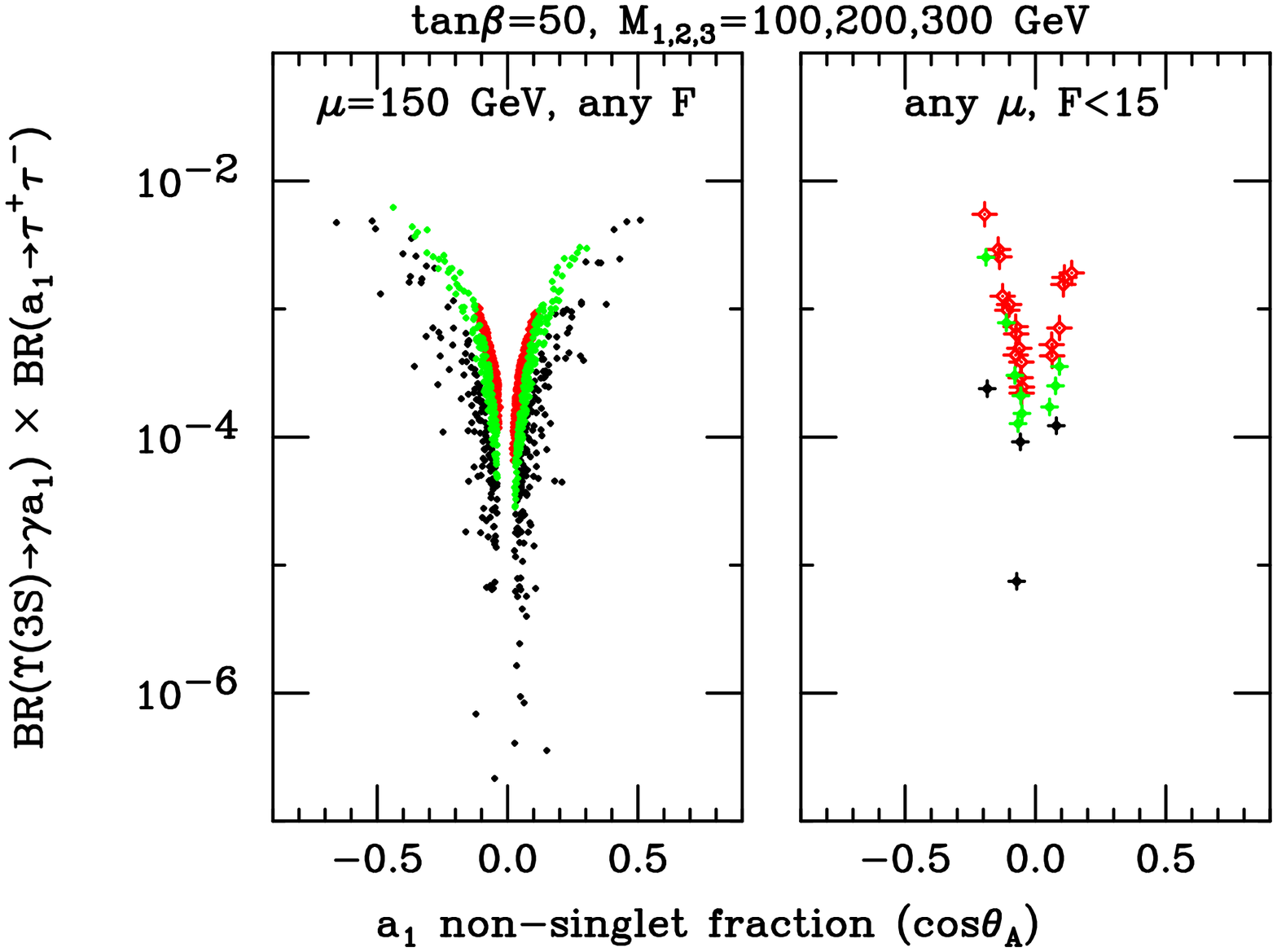}
\end{center}
\vspace*{-.2in}
\caption{$\br(\Upsilon(3S)\to\gam\ai)\times \br(\ai\to \tauptaum)$ for
  NMSSM scenarios with various ranges for $\mai$: medium grey (red) =
  $2\mtau<\mai<7.5\gev$; light grey (green) = $7.5\gev<\mai<8.8\gev$;
  and black = $8.8\gev<\mai<2m_B\gev$.  The plots are for
  $\tanb=3,10,50$, respectively. The left-hand window in each plot
  shows results for a ``fixed-$\mu$-scan'' as defined in the text (and
  in Ref.~\cite{Dermisek:2006py}) The right-hand window shows results
  for $F<15$ points found using a ``full scan'' as defined in the
  text.  }
\label{brups3sbratautau}
\end{figure}

\begin{figure}[h!!!]
\begin{center}
\includegraphics[width=0.27\textheight,angle=90]{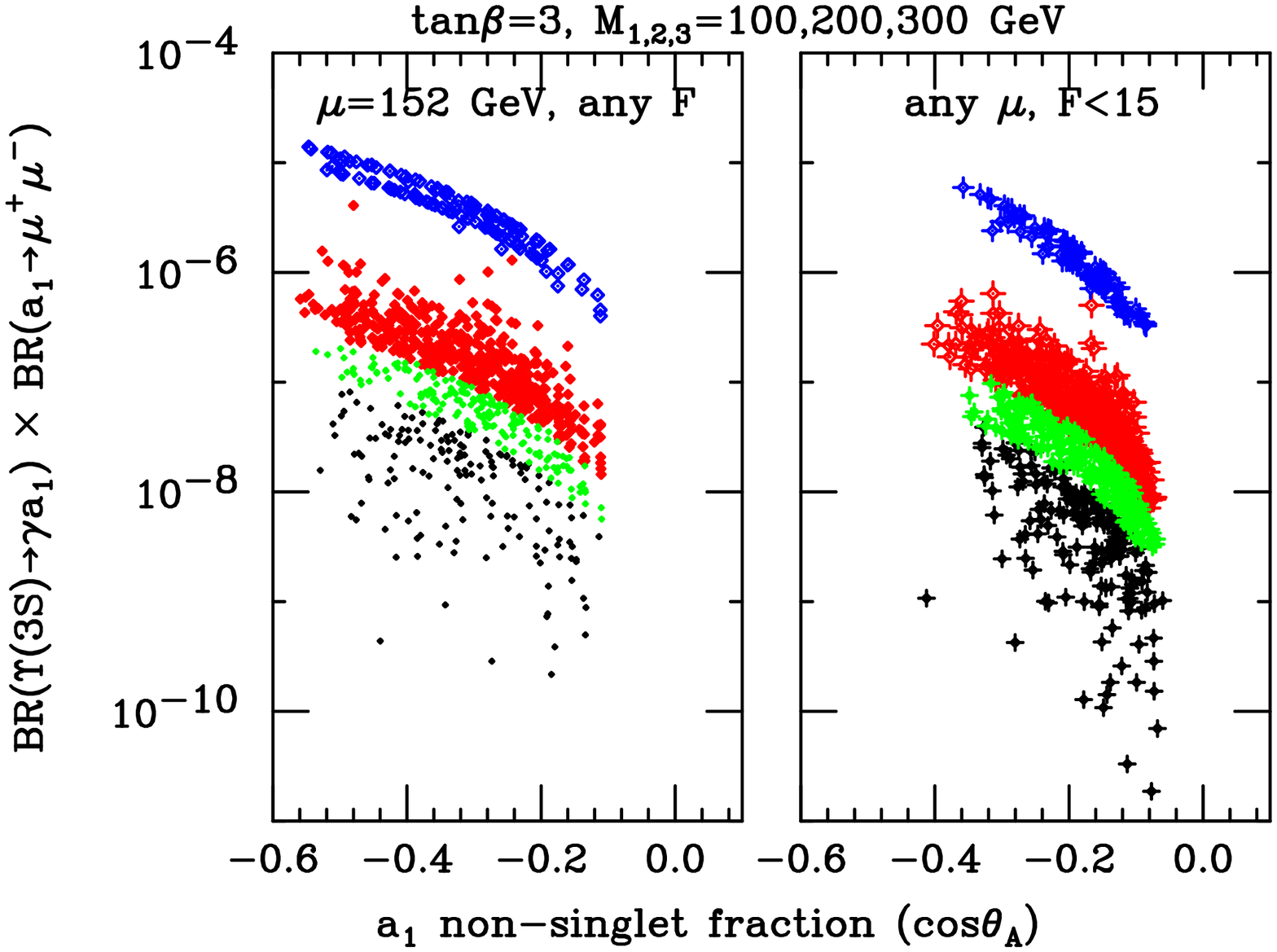}
\includegraphics[width=0.27\textheight,angle=90]{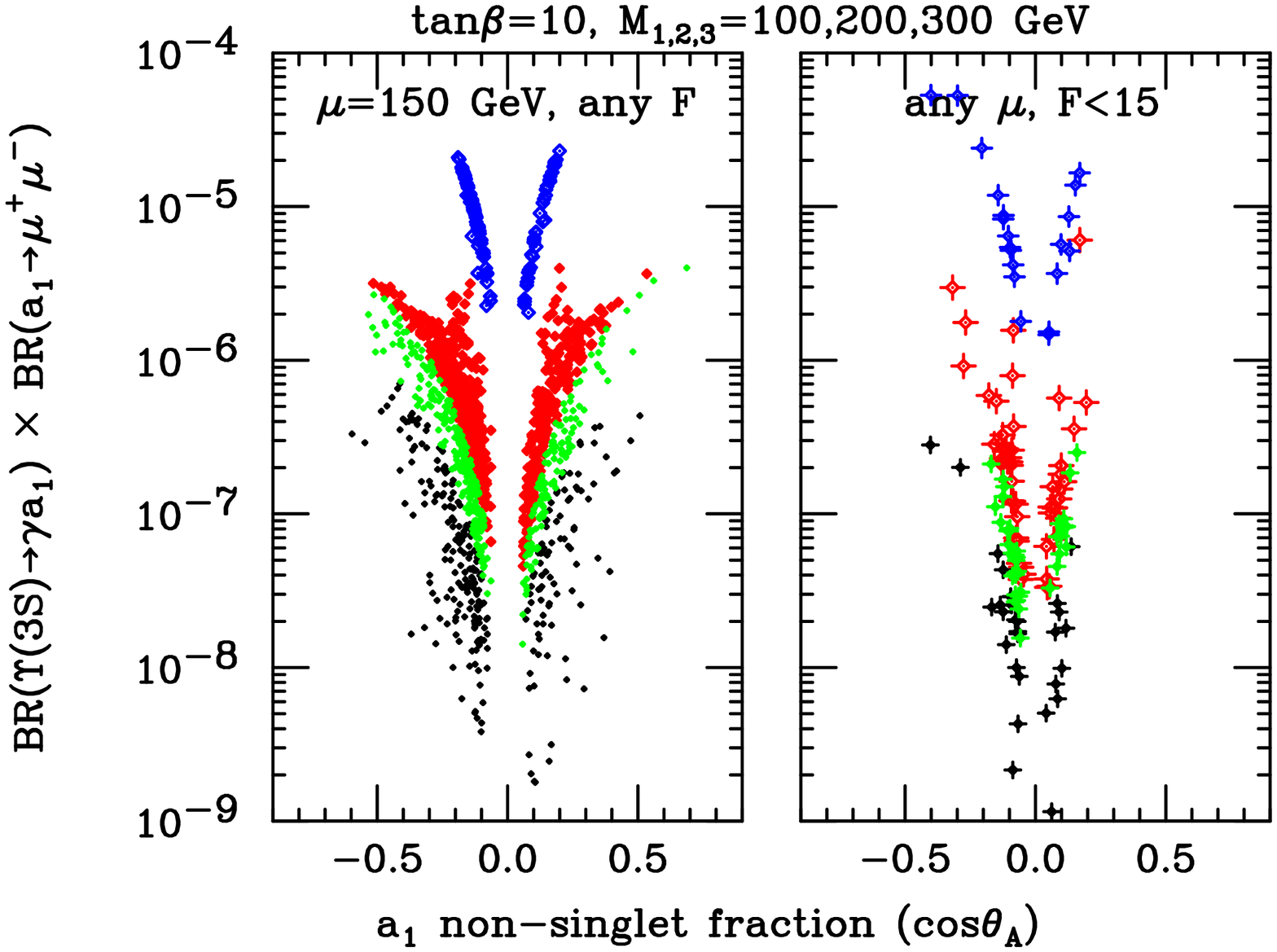}
\includegraphics[width=0.27\textheight,angle=90]{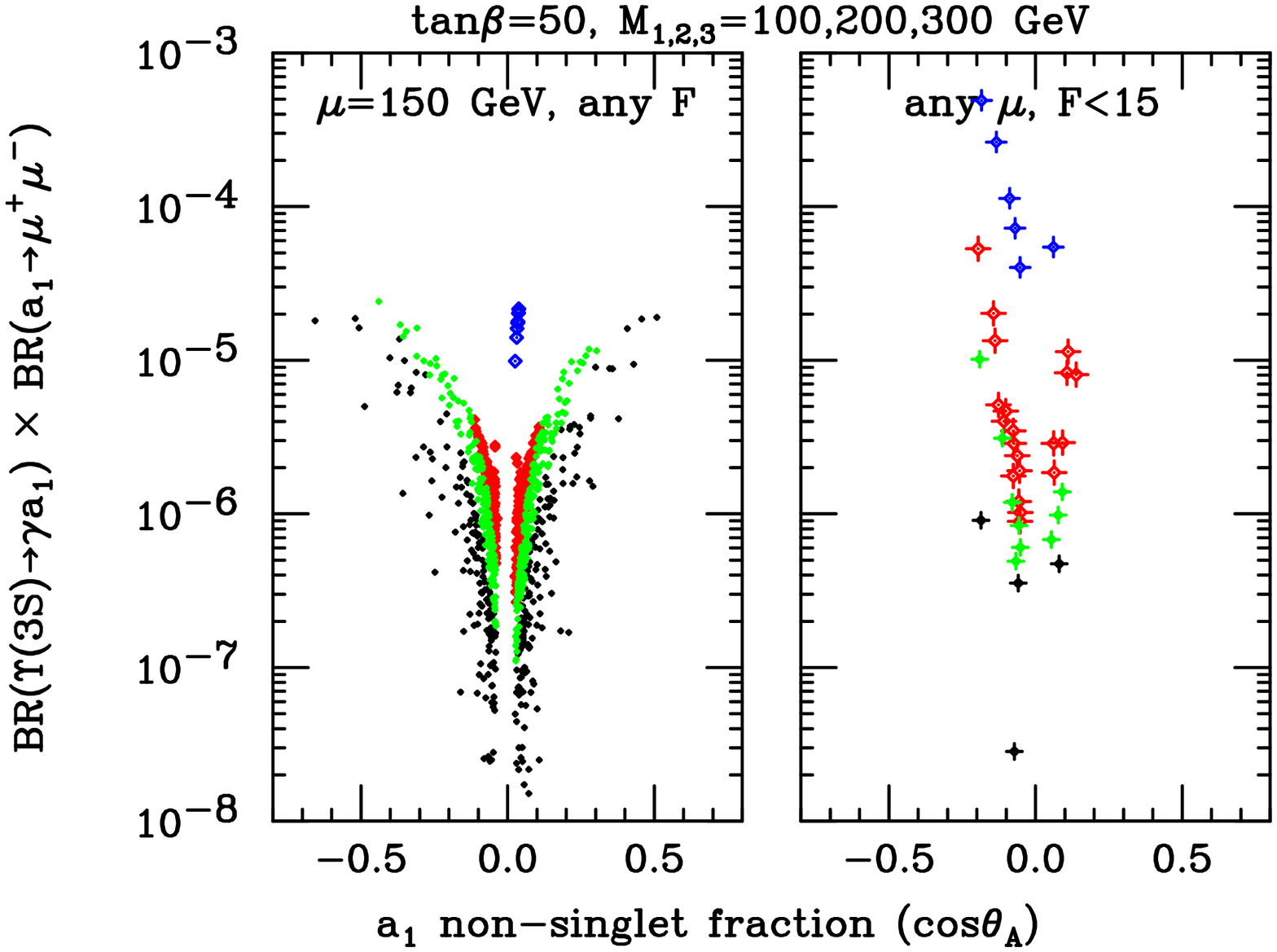}
\end{center}
\vspace*{-.2in}
\caption{We plot $\br(\Upsilon(3S)\to\gam\ai)\times \br(\ai\to
  \mupmum)$ using the same notation and scanning procedures as described
  in the caption of Fig.~\protect\ref{brups3sbratautau}.}
\label{brups3sbramumu}
\end{figure}

Before continuing with the general analysis, it is useful to compare
the limits of \cite{Aubert:2009cka} and \cite{Aubert:2009cp} with the
predictions of the NMSSM.  This comparison is done for the same two
types of scans as in the earlier paper \cite{Dermisek:2006py}, except
that here we focus on the $3S$ state rather than the $1S$ state. In
both scans, we hold the gaugino soft-SUSY-breaking parameters of the
NMSSM fixed at $M_{1,2,3}(m_Z)=100,200,300\gev$ and fix $\tanb$.  In
the first type of scan, called a ``fixed-$\mu$ scan'', we scan over
the NMSSM soft-SUSY-breaking Higgs potential parameters $A_\lambda$
and $A_\kappa$ keeping the effective $\mu$ parameter of the model
fixed at the representative value of $\mu=150\gev$ (at $\tanb=10$ and $50$) or
$\mu=152\gev$ (at $\tanb=3$ for which we must take $\mu=152\gev$
in order to get physically allowable scenarios). In addition, in the
fixed-$\mu$ scans we have kept the scalar soft-SUSY-breaking masses
fixed at common value of $\msusy =300\gev$ and the $A$
soft-SUSY-breaking parameters fixed to a common value of $-300\gev$.
In the second type of scan, termed a ``full scan'', we have allowed
$\mu$ to vary and have also allowed the soft-SUSY-breaking scalar
masses and $A$ parameters to vary (independently of one another).  In
the full scan results presented we have kept only scenarios with very
low electroweak finetuning, as characterized by the parameter $F$ (see
\cite{Dermisek:2005ar} for more details) being smaller than $15$, where
$F<15$ corresponds to absence of electroweak finetuning.  $F<15$
scenarios only arise for $\mhi\lsim 105\gev$ and are thus
automatically ``ideal'' in the precision electroweak sense.  As part
of the fixed-$\mu$ scans and the full scans, we have required that the
CP-even $\hi$ escape published LEP limits by virtue of dominant
$\hi\to \ai\ai\to 4\tau$ or $4~jet$ decays. In the forthcoming plots,
the left-hand windows correspond to fixed-$\mu$ scan results and the
right-hand windows give the results of a full scan for the same
$\tanb$ value.

Our results for the $\tauptaum$ final state are shown in
Fig.~\ref{brups3sbratautau} and those for the $\mupmum$ final state
are shown in Fig.~\ref{brups3sbramumu}.  Let us focus first on the
$\tauptaum$ final state.  The 90\% CL $\br(\Upsilon(3S)\to\gam
a)\times \br(a\to \tauptaum)$ limits from BaBar range from $\sim
10^{-5}$ at $\ma$ just above $2\mtau$ with a long plateau at the
$3-7\times 10^{-5}$ until $\ma$ passes above $10\gev$ where the limit
is of order $10^{-4}$. In Fig.~\ref{brups3sbratautau}, the black
points have high $\mai$ ($8.8\gev< \mai\leq 2m_B$), the light grey
(green) points have $7.5\gev< \mai\leq 8.8\gev$ and the medium grey
(red) points have $2\mtau<\mai\leq 7.5\gev$.  Let us first discuss
$\tanb=10$ results, since these can be compared to those for
$\Upsilon(1S)\to\gam \ai \to \gam \tauptaum$ presented in
Ref.~\cite{Dermisek:2006py}.  From comparing the BaBar limits
summarized above with the relevant plot of
Fig.~\ref{brups3sbratautau}, we see that most of the $\mai<7.5\gev$
points are excluded, about half of the $7.5\gev< \mai\leq 8.8\gev$ are
excluded, but that many fewer of the $\mai>8.8\gev$ points are
excluded. Still, exclusions of this higher $\mai$ region are much
superior to those from the CLEO-III $\Upsilon(1S)$ data
\cite{cleoiii}, which excluded none of the black points, a small
fraction of the green points and about half of the red points. This
ability to probe to higher $\mai$ using the $\Upsilon(3S)$ is
particularly relevant in the NMSSM context since the GUT-scale tunings
of $A_\lambda$ and $A_\kappa$ needed to obtain $\mai<2m_B$ while at
the same time having $\br(\hi\to\ai\ai)\gsim 0.7$, as required in the
ideal Higgs scenario, is minimal for $\mai$ values close to $2m_B$.
For $\tanb=50$, one finds that almost all the $2\mtau<\mai<8.8\gev$
scenarios are excluded, but that lots of $\mai>8.8\gev$ points
survive.  In contrast, for $\tanb=3$ the BaBar results only
significantly constrain the region $2\mtau<\mai\leq 7.5\gev$.

We now turn to the $\mupmum$ final state.  The 90\% CL
$\br(\Upsilon(3S)\to\gam a)\times \br(a\to \mupmum)$ limits from BaBar
are $\sim 1-3.5 \times 10^{-6}$ for $\ma\lsim 1\gev$, $\sim 1-2 \times
10^{-6}$ for $1\lsim \ma<2\mtau$, $\sim 1-3\times 10^{-6}$ for
$2\mtau\lsim \ma\lsim 7.5\gev$, and $\sim 1-5\times 10^{-6}$ for
$7.5\gev\lsim \mai\lsim 9.2\gev$. In Fig.~\ref{brups3sbramumu} the
black points have high $\mai$ ($8.8\gev< \mai\leq 2m_B$), the light
grey (green) points have $7.5\gev< \mai\leq 8.8\gev$, the medium grey
(red) points have $2\mtau<\mai\leq 7.5\gev$ and the darker grey (blue)
points have $\mai<2\mtau$.  At $\tanb=3$, the $\mupmum$ final state
data eliminates more than 4/5 of the NMSSM model points in the
$\mai<2\mtau$ mass range, but only a small number of the NMSSM points
for $2\mtau<\mai<7.5\gev$ and none of the points with
$7.5\gev\lsim\mai$.  At $\tanb=10$, all $\mai<2\mtau$ NMSSM points are
eliminated by the $\mupmum$ data as well as a small fraction of the
$2\mtau<\mai<7.5\gev$ and $7.5\gev<\mai<8.8\gev$ points, but none of
the $8.8\gev<\mai$ points.  At $\tanb=50$, all $\mai<2\mtau$ NMSSM
points are again eliminated, perhaps half of the $2\mtau<\mai<7.5\gev$
points are eliminated, a still significant fraction of the
$7.5\gev<\mai<8.8\gev$ points are eliminated, and even a significant
number of the $8.8\gev<\mai$ points are eliminated.

To summarize, only the $\mupmum$ channel provides constraints for
$\mai<2\mtau$ and almost all the ideal-Higgs-like NMSSM scenarios with
$\tanb\geq 3$ are eliminated.  For $2\mtau<\mai$, the $\tauptaum$
channel provides the most eliminations for all $\tanb$. Certainly,
the BaBar $\Upsilon(3S)$ results are a big stride relative to the
CLEO-III $\Upsilon(1S)$ results, especially at $\mai<2\mtau$ and at
high $\mai$.  Of course, it is important to note that the NMSSM
scenarios most favored in order to minimize light-$\ai$ finetuning
have $\mai$ very near $2m_B$ and thus cannot be limited by Upsilon
decays.

\section{General limits on the $ab\anti b$ coupling}

\begin{figure}[h!]
\begin{center}
\includegraphics[width=0.65\textwidth,angle=90]{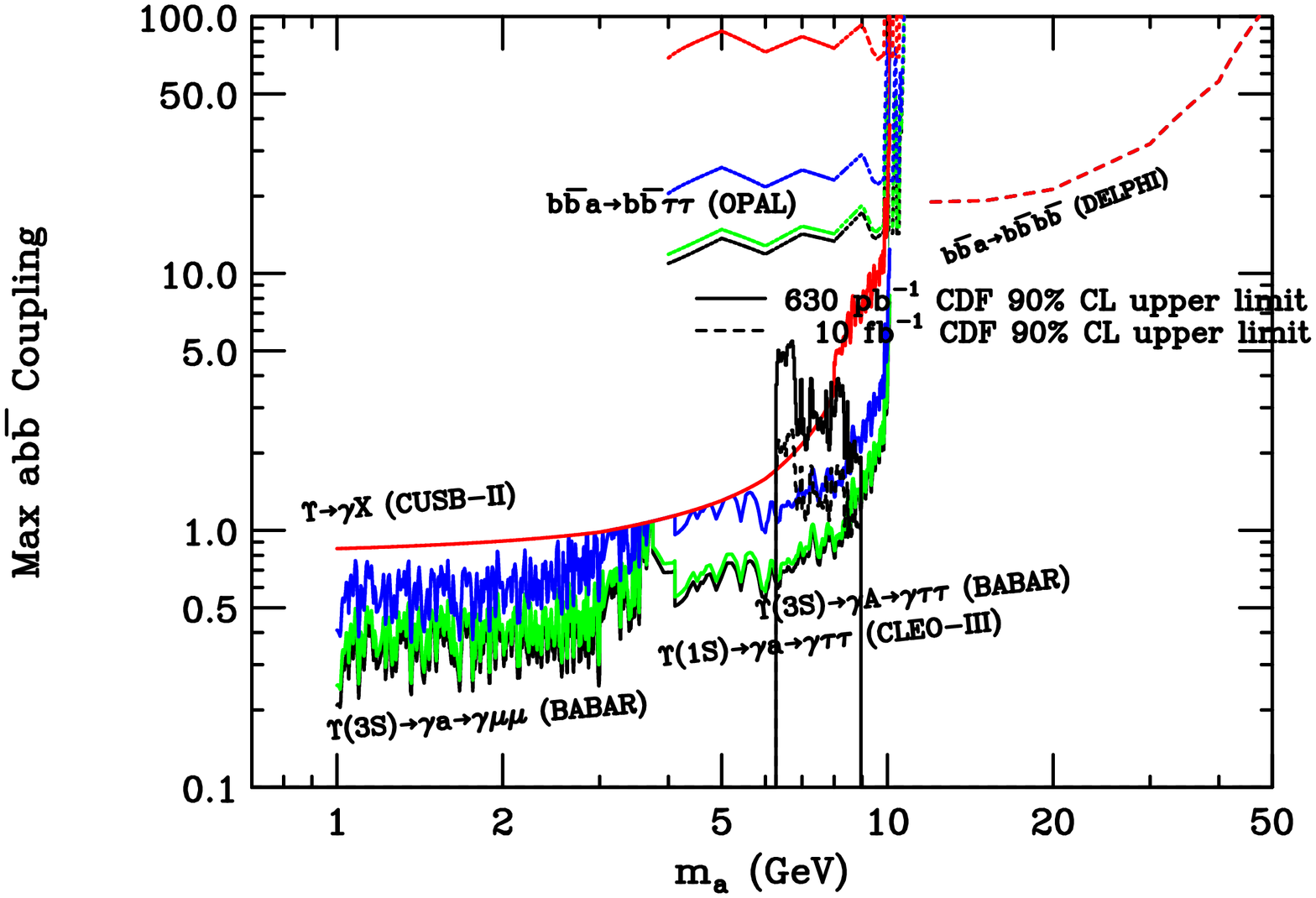}
\end{center}
\caption{Upper limit, $\cmax$, on $|\cabb|$ as a function of $\ma$ for
  a variety of $\tanb$ values coming directly from experimental
  data. The highest (red) curve is for $\tanb=0.5$, the other curves,
  in order of decreasing $\cmax$ are for $\tanb=1$, $\tanb=2$ and
  $\tanb\geq 3$.}
\label{coupmaxall}
\end{figure}

Our ultimate goal is to use the $\upsiii$ limits in combination with
other available limits to extract limits on $|\cabb|$. The older
experiments that provide the most useful constraints are as follows.
Prior to the recent BaBar data, for $2\mtau<\ma<9.2\gev$ the recent
CLEO-III~\cite{cleoiii} limits on $\Upsilon(1S)\to \gam a\to \gam
\tauptaum$ were the strongest.  For $9.2\gev<\ma<\mups$, mixing of the
$a$ with various $\eta_b$ and $\chi_0$ bound states becomes
crucial~\cite{Drees:1989du}.  Ref.~\cite{cleoiii} gives results for
$\cmax$ in this $\ma$ range without taking this mixing into account
but notes that their limits cannot be relied upon for $\ma>9.2\gev$.
Whether additional limits can be extracted from lepton
non-universality studies in the $9.2<\ma<\mups$ region is being
studied~\cite{SanchisLozano:2002pm}.  OPAL
limits~\cite{Abbiendi:2001kp} (which assume $\br(a\to \tauptaum)=1$)
on $\epem\to b\anti b\tauptaum$ become numerically relevant for
roughly $9\gev<\ma<2m_B$.  Ref.~\cite{Abbiendi:2001kp} converts these
limits to limits on the $a b\anti b $ coupling using the modeling of
\cite{Drees:1989du}.  These are the only LEP limits in the
$\mupsiii<\ma<2m_B$ range and continue to be relevant up to $12\gev$.
Above $\ma=2m_B$ these $ab\anti b$ coupling limits become quite weak
due to the $\eta_b-a$ mixing uncertainties and the decrease of
$\br(a\to \tauptaum)$. For $\ma\geq 12\gev$, limits on the $ab\anti b$
coupling can be extracted from $\epem\to b\anti b a\to b\anti b b\anti
b$~\cite{delphi}. One should also keep in mind that values of
$|\cabb|$ above 50 raise issues of non-perturbativity of the $a b\anti
b$ coupling and are likely to be in conflict with Tevatron limits on
$b\anti b a$ production~\cite{Aaltonen:2009vf}. The limits, $\cmax$,
on $\cabb$ coming from all data, including the recent BaBar results,
are plotted in Fig.~\ref{coupmaxall} for various $\rbt\equiv
\sqrt{\cabb/\catt}$ values.  (In a 2HDM model type-II context,
$\rbt=\tanb$.) Note the rapid deterioration as $\ma \to \mupsiii$. The
variation with $\rbt$ arises because $\br(a\to\tauptaum)$ varies with
$\rbt$ as shown in Fig.~\ref{bratautau}. Basically, for $\tanb>1$ the
BaBar results provide the most stringent limits. For $\tanb=0.5$ the
$a$ decays to a complicated mix of channels and the old CUSB-II limits
(which were independent of the exact $a$ final state) are strongest
for $\ma\lsim 8\gev$.

In Fig.~\ref{coupmaxall}, we have also plotted limits
extracted~\cite{Dermisek:2009fd} from Tevatron data using a
reinterpretation of a CDF analysis performed over the range
$6.3\gev\leq \mmumu\leq
9\gev$~\cite{Apollinari:2005fy,Aaltonen:2009rq}. This analysis placed
limits on the ratio $R={\sigma(\eps)\br(\eps\to \mupmum)\over
  \sigma(\upsi)\br(\upsi\to \mupmum)}$, where $\eps$ was a narrow
resonance produced in the same manner as the $\upsi$.  Fluctuations of
$R$ above a smooth fit to the overall spectrum were searched for and
90\% CL limits were placed on $R$.  It is relatively straightforward
to apply this analysis to place limits on $R={\sigma(a)\br(a\to
  \mupmum)\over \sigma(\upsi)\br(\upsi\to \mupmum)}$. The 90\% CL
limits on $R$ corresponding to the available $L=630\pbi$ data set are
then easily converted to limits on $|\cabb|$.  These limits as a
function of $\ma$ are those plotted as the solid histogram.  A simple
statistical extrapolation of these limits to $L=10\fbi$ (an integrated
luminosity that will soon be available) is shown as the dashed
histogram. These limits hold for $\tanb>2$. We see that these limits
improve rapidly as $\ma$ increases.  While the $L=630\pbi$ limits are
not quite competitive with the limits from BaBar data at $\ma\sim
9\gev$, we observe that the $L=10\fbi$ limits will actually be
slightly better if the extrapolation holds.

While $\Upsilon(nS)$-based limits are kinematically limited and become
weak for $\ma\gsim 9.6\gev$, there is no such kinematic limitation for
limits based on hadronic collider data. In fact,
CDF measured the $\mmumu$ spectrum above $9\gev$, but did not
perform  the easily reinterpreted $R$ analysis in the region $\mmumu>9\gev$.
In~\cite{Dermisek:2009fd}, we estimated the 90\% CL limits from the
$L=630\pbi$ measurements in the $\mmumu>9\gev$ region (out to
$\mmumu=12\gev$) and found that, in the range $9.6\gev\lsim \ma\lsim 2m_B$,
implied limits on $|\cabb|$ were of order $|\cabb|<1.6-1.8$ for $\ma$ outside
the $\upsii$ and $\upsiii$ peaks.  At both peaks we found
$|\cabb|\lsim 2$. For $L=10\fbi$, these limits should come down to
$|\cabb|\lsim 1$, and begin to constrain the most preferred NMSSM
parameter regions, especially for large $\tanb$.

\begin{figure}
\begin{center}
\includegraphics[width=0.65\textwidth,angle=90]{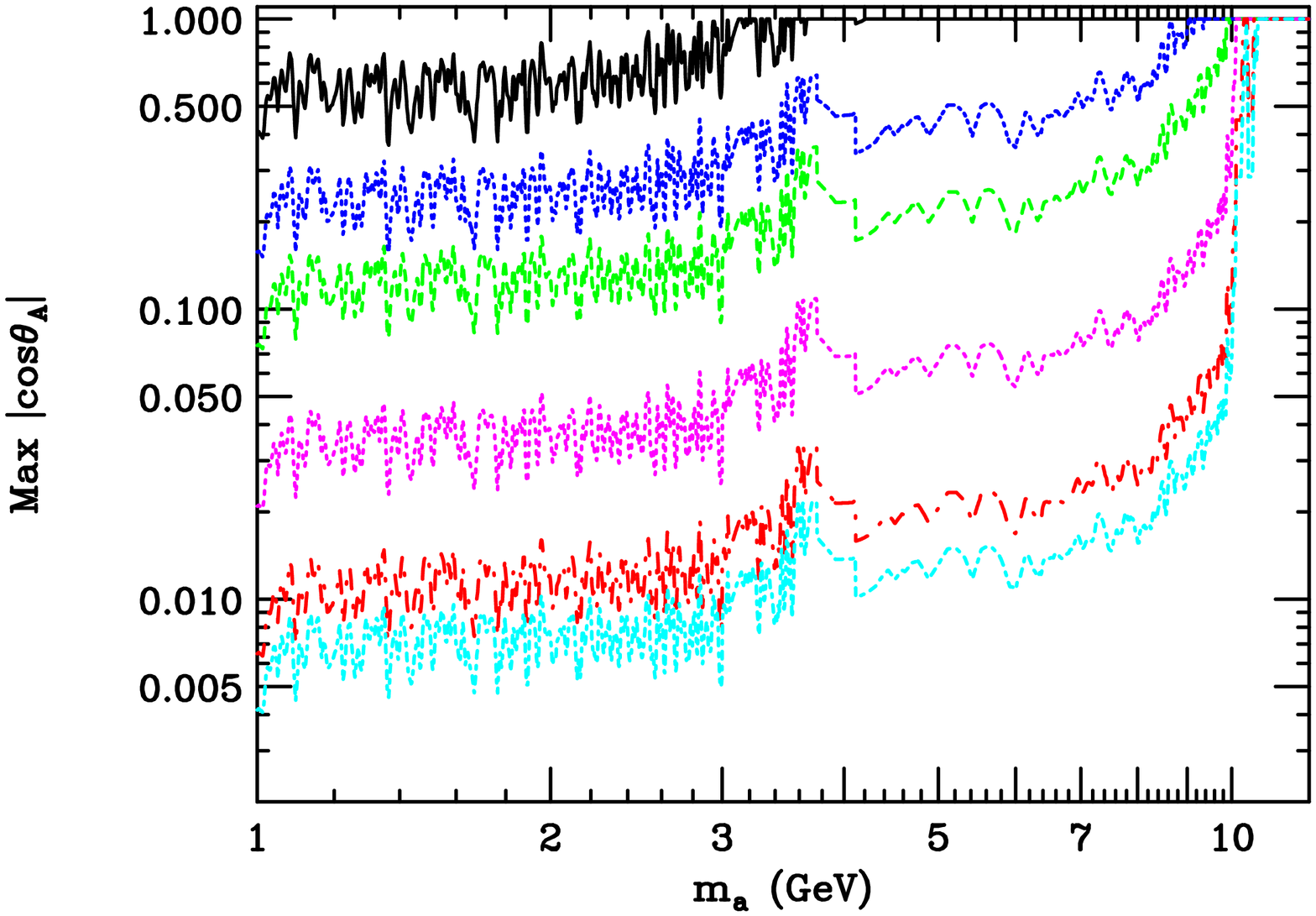}
\end{center}
\caption{$\ctamax$ in the NMSSM (where $\cabb=\cta\tanb$) as a
  function of $\ma$.  The different curves correspond to $\tanb=1$
  (upper curve), $1.7$, $3$, $10$, $32$ and $50$ (lowest curve). CDF/Tevatron
  constraints do not affect this plot.  }
\label{ctamaxvsma}
\end{figure}

\section{Implications of general $ab\anti b$ limits for NMSSM
  scenarios}

In the NMSSM, we note that it is always possible to choose $\cta$ so
that the limits on $\caibb$ as a function of $\tanb$ are satisfied. The
maximum allowed value of $|\cta|$, $\ctamax$, as a function of $\ma=\mai$
for various $\tanb$ values is plotted in
Fig.~\ref{ctamaxvsma}. Constraints are strongest for $\ma\lsim 9\gev$
for which Upsilon limits are relevant, and deteriorate rapidly above
that.  As seen in Fig.~\ref{coupmaxall}, currently the limits from the
Tevatron/CDF data are not as strong as those from the BaBar $\upsiii$
data and do not affect this plot.

As an aside regarding the general 2HDM(II) model, we note that any
point for which $\ctamax$ is smaller than $1$ corresponds to an $\ma$
and $\tanb$ choice that is not consistent with the experimental
limits.  Disallowed regions emerge in the range $\ma\lsim 2\mtau$ for
$\tanb=1$, rising quickly to $\ma\lsim 9\gev$ for $\tanb=1.7$ and
$\ma\lsim 10\gev$ for $\tanb\geq 3$.  These excluded regions apply to
any light doublet CP-odd Higgs boson, including the beyond the MSSM
scenarios of~\cite{Dermisek:2008id, Dermisek:2008sd, Bae:2010cd} which
are consistent with other experimental constraints for $\tan \beta
\lsim 2.5$.

\begin{figure}
\begin{center}
\includegraphics[width=0.4\textwidth,angle=90]{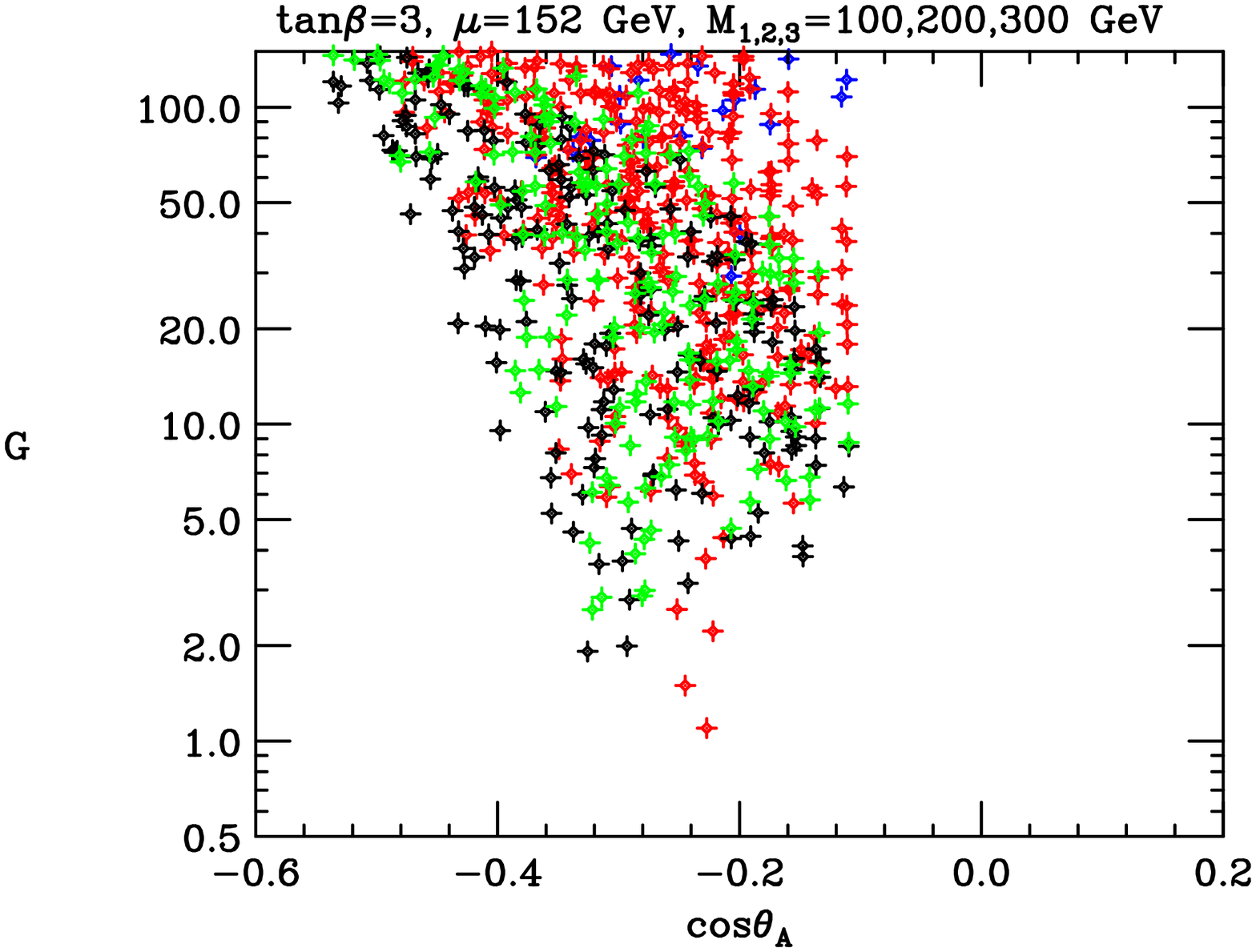}\includegraphics[width=0.4\textwidth,angle=90]{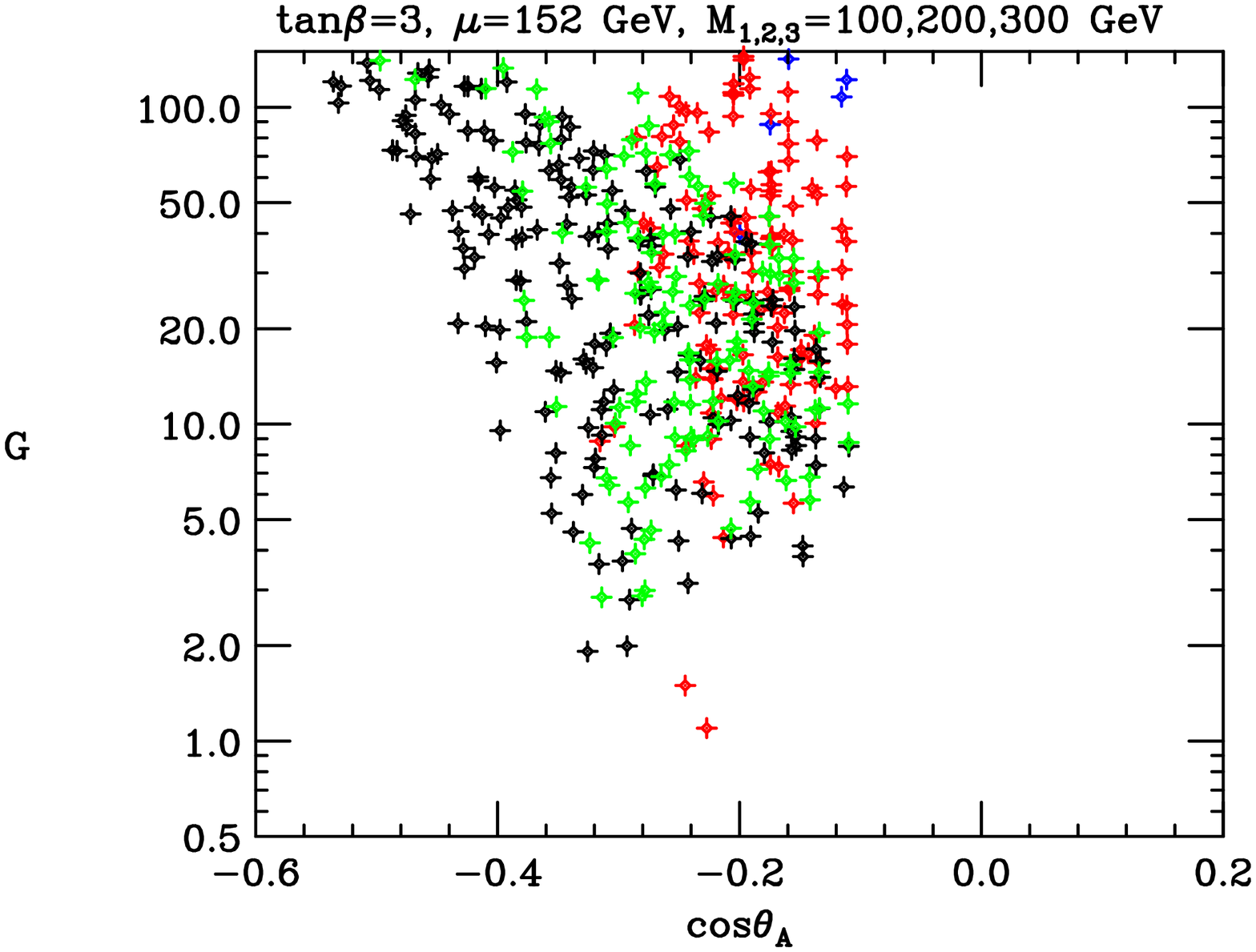}
\end{center}
\caption{Light-$\ai$ finetuning measure $G$ before and after imposing
  limits $|\cta|\leq \ctamax$. These plots are those obtained for
  ``fixed-$\mu$ scans'' with $\mu=152\gev$ and setting $\tanb=3$. Note
  that many points with low $\mai$ and large $|\cta|$ are eliminated
  by the $|\cta|<\ctamax$ requirement, including almost all the
  $\mai<2\mtau$ (blue) points and a good fraction of the
  $2\mtau<\mai<7.5\gev$ (red) points.  }
\label{tb3comp}
\end{figure}

\begin{figure}
\begin{center}
\includegraphics[width=0.4\textwidth,angle=90]{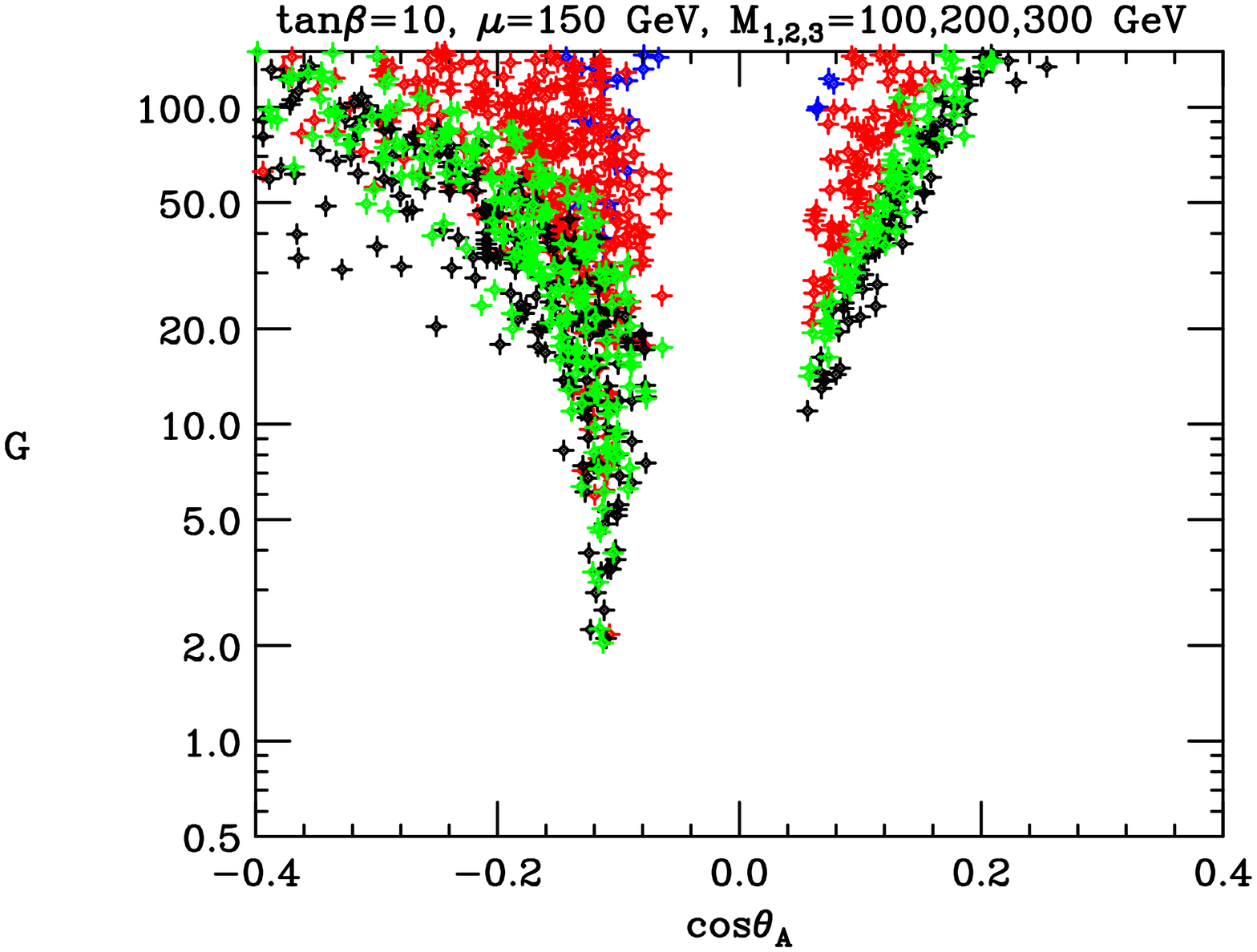}\includegraphics[width=0.4\textwidth,angle=90]{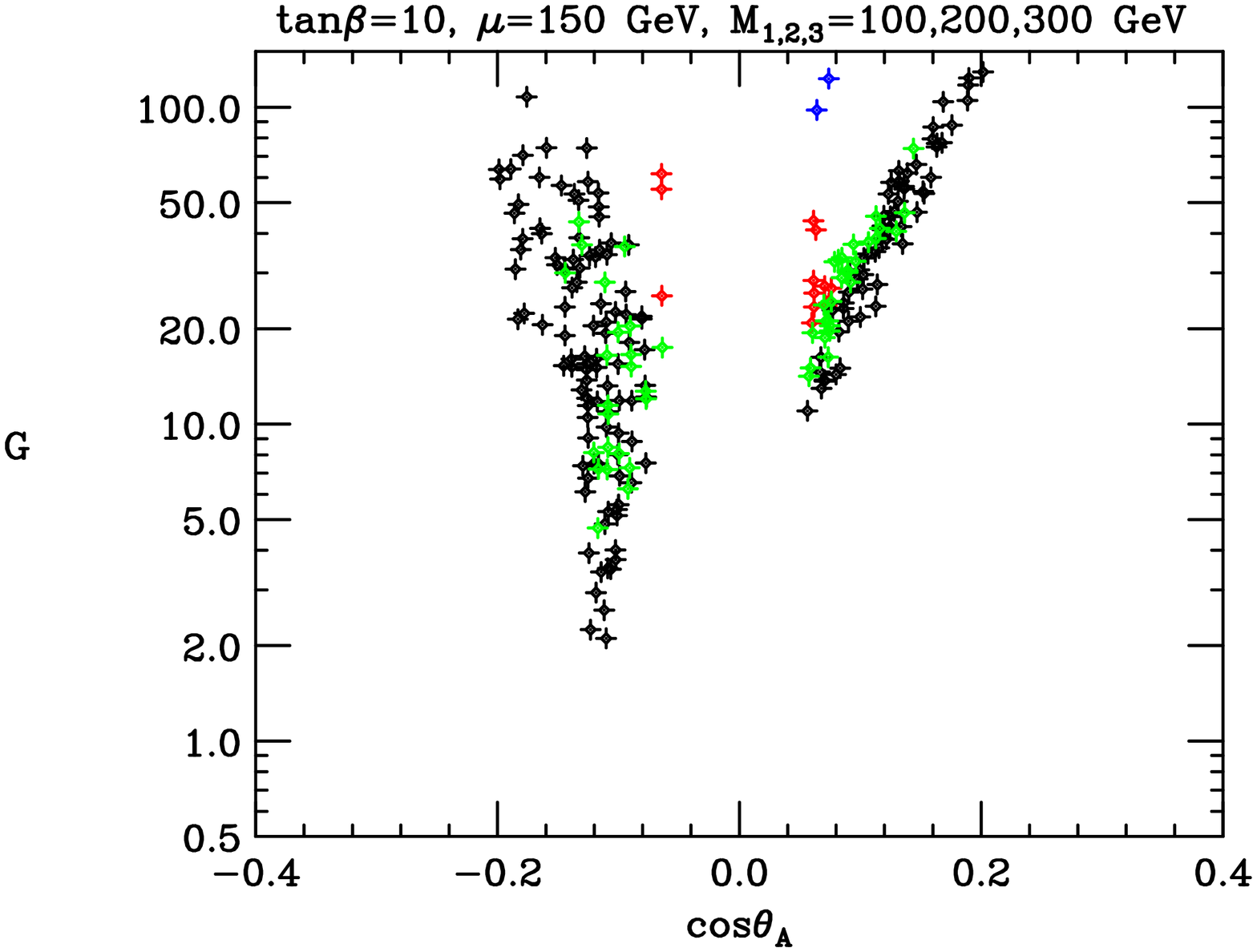}
\end{center}
\caption{As in Fig.~\protect\ref{tb3comp}, but for $\mu=150\gev$ and
  $\tanb=10$. Note that many points with low $\mai$ and large $|\cta|$
  are eliminated, including almost all the $\mai<2\mtau$ (blue) points
  and $2\mtau<\mai<7.5\gev$ (red) points.}
\label{tb10comp}
\end{figure}

\begin{figure}
\begin{center}
\includegraphics[width=0.4\textwidth,angle=90]{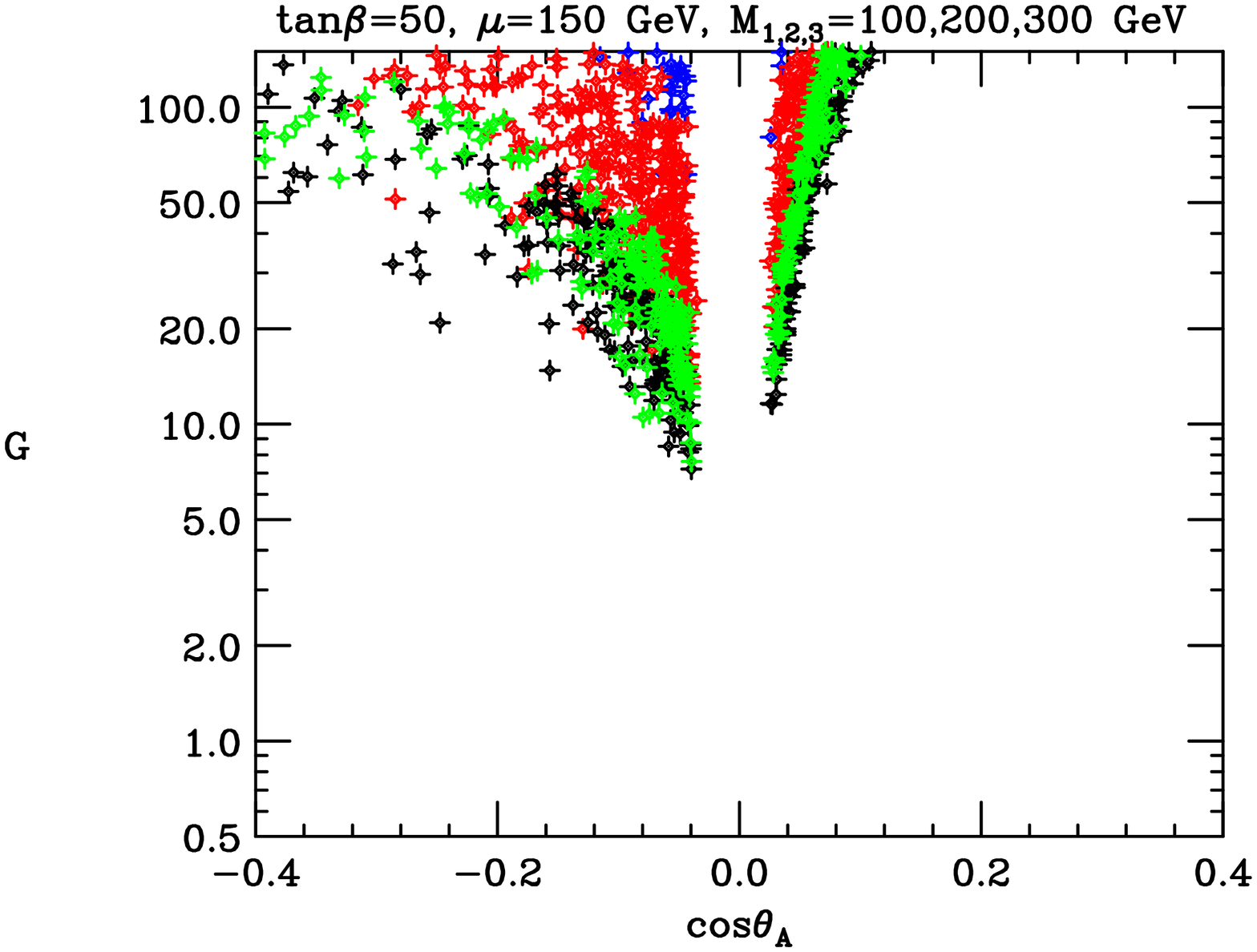}\includegraphics[width=0.4\textwidth,angle=90]{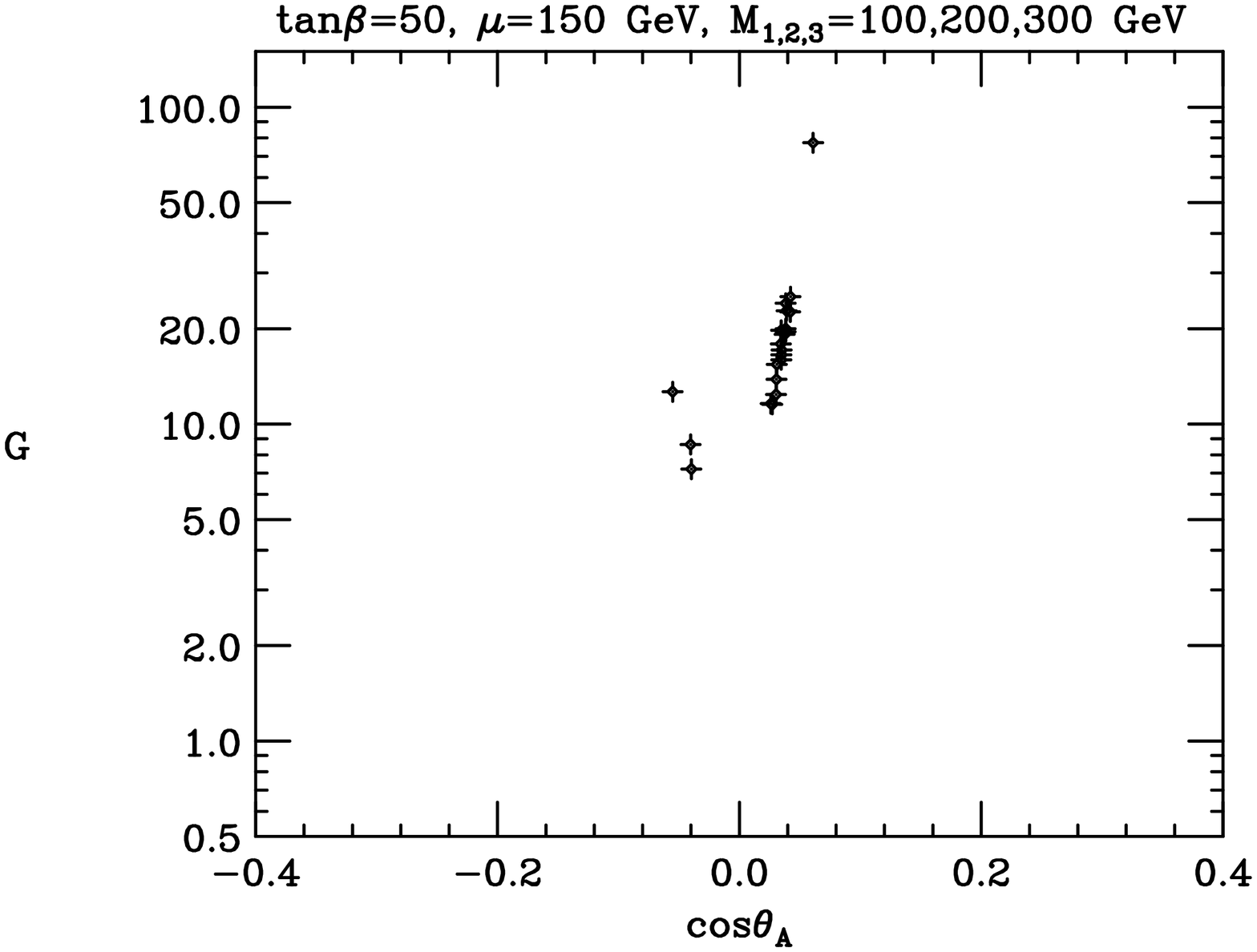}
\end{center}
\caption{As in Fig.~\protect\ref{tb3comp}, but for $\mu=150\gev$ and
  $\tanb=50$. Note that the only surviving points are those with
  $\mai>8.8\gev$ (black points) at small $|\cta|$.}
\label{tb50comp}
\end{figure}

We can illustrate the effects of the limits plotted in
Fig.~\ref{ctamaxvsma} on preferred NMSSM scenarios.  Relevant plots
appear below. The first set of plots, Figs.~\ref{tb3comp},
\ref{tb10comp} and \ref{tb50comp}, for $\tanb=3$, $10$, and $50$,
respectively, show results for ``fixed-$\mu$ scans'' (see earlier
definition).  In each figure, the left-hand plot gives the light-$\ai$
finetuning measure $G$ as a function of $\cta$ before imposing the
$\ctamax$ constraint while the right-hand plot gives $G$ as a function
of $\cta$ after imposing $\ctamax$. The point notation is according to
$\mai$: blue for $\mai<2\mtau$, red for $2\mtau<\mai<7.5\gev$, green
for $7.5\gev<\mai<8.8\gev$ and black for $8.8\gev<\mai<2m_B$. We see
that the bulk of points with $\mai<7.5\gev$ are eliminated by the
$\ctamax$ limit and that the points with $\mai>7.5\gev$ at large
$|\cta|$ are also eliminated.

\begin{figure}
\begin{center}
\includegraphics[width=0.4\textwidth,angle=90]{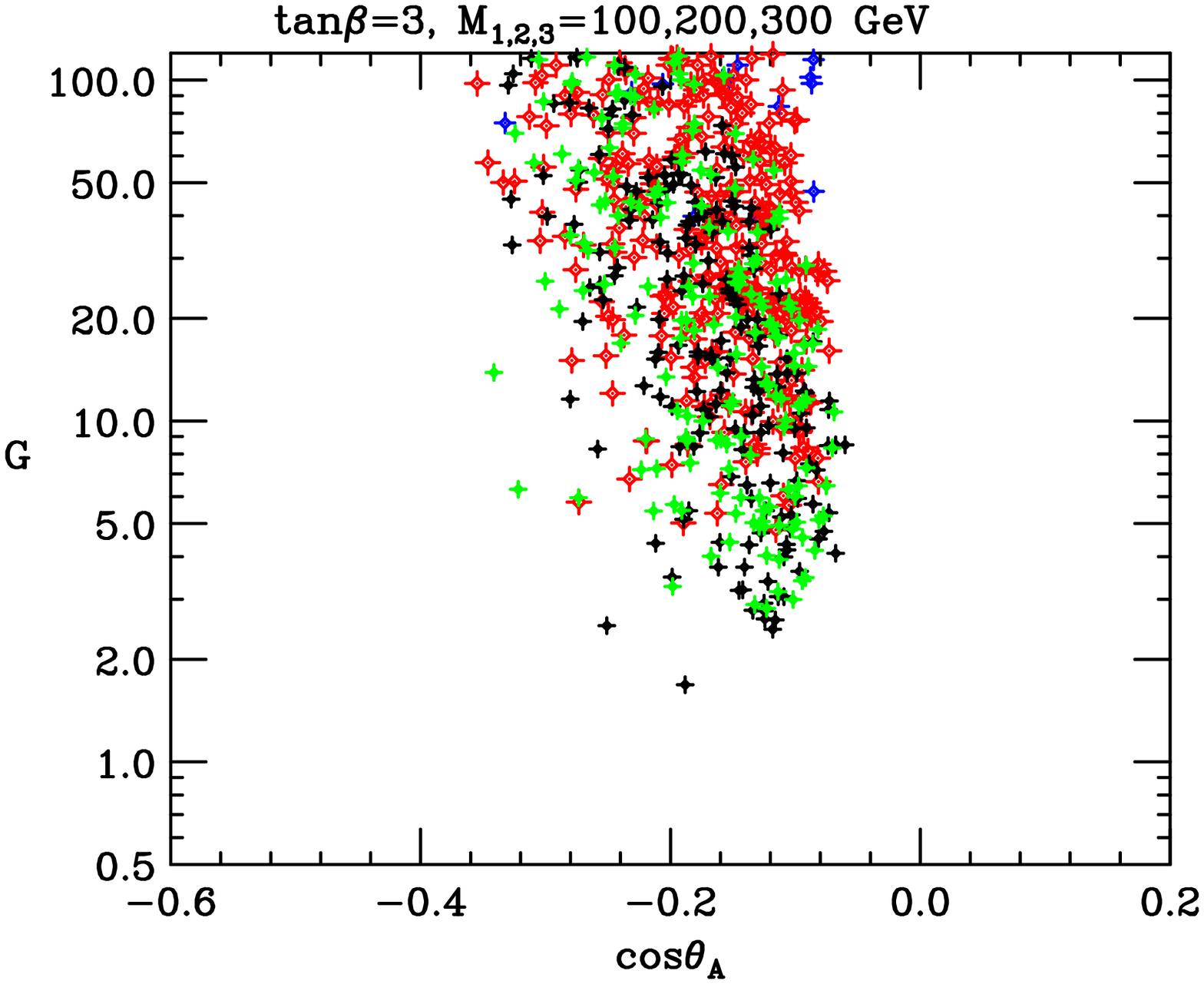}\includegraphics[width=0.4\textwidth,angle=90]{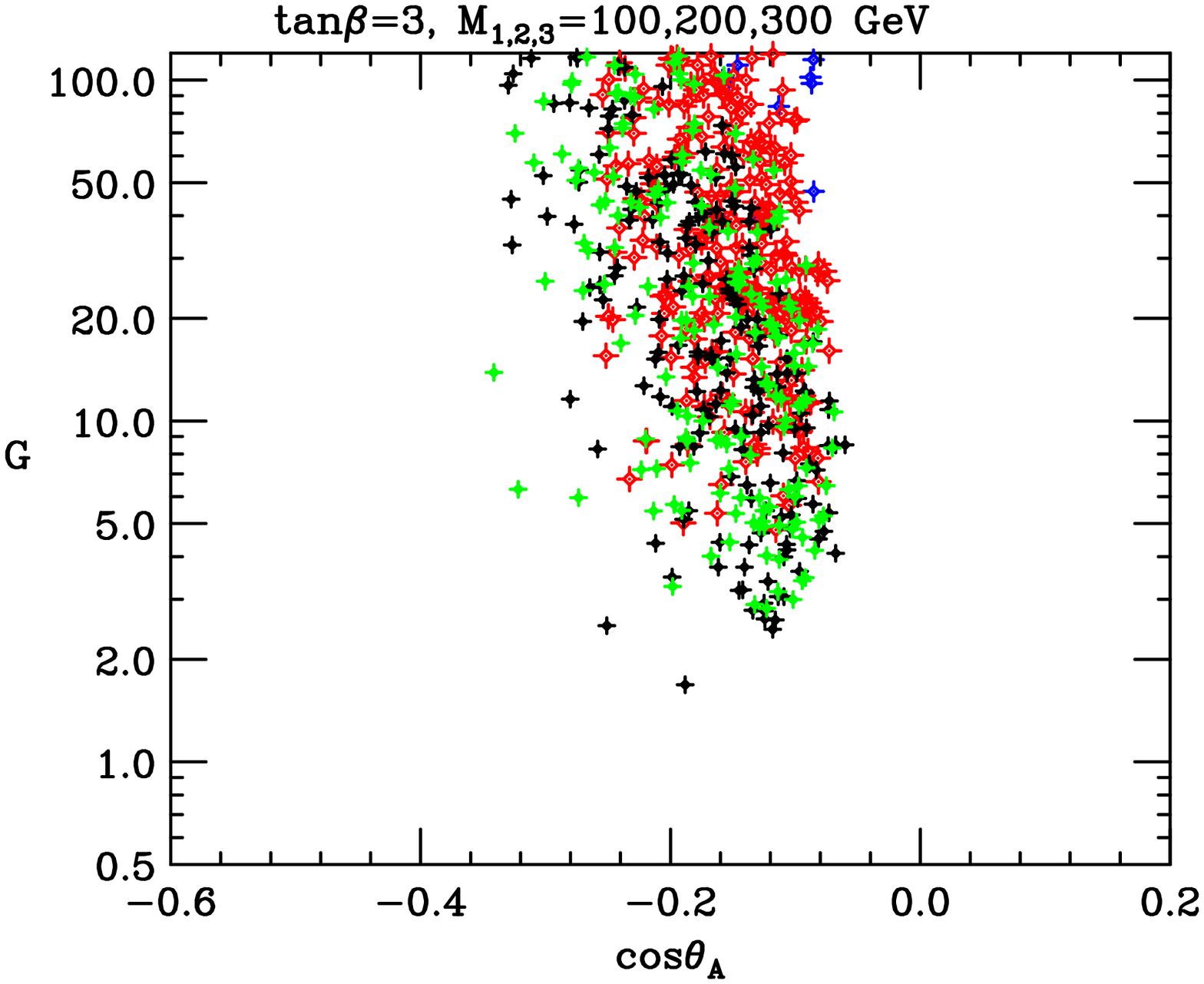}
\end{center}
\caption{Light-$\ai$ finetuning measure $G$ before and after imposing
  $|\cta|\leq \ctamax$. These are the results obtained using a ``full
  scan'' at $\tanb=3$.  Only solutions with electroweak finetuning
  measure $F<15$ are retained.  Note that a good fraction of the
  $\mai<2\mtau$ (blue) points and $2\mtau<\mai<7.5\gev$ (red) points
  are eliminated by the $\ctamax$ cut. }
\label{lowftb3comp}
\end{figure}

\begin{figure}
\begin{center}
\includegraphics[width=0.4\textwidth,angle=90]{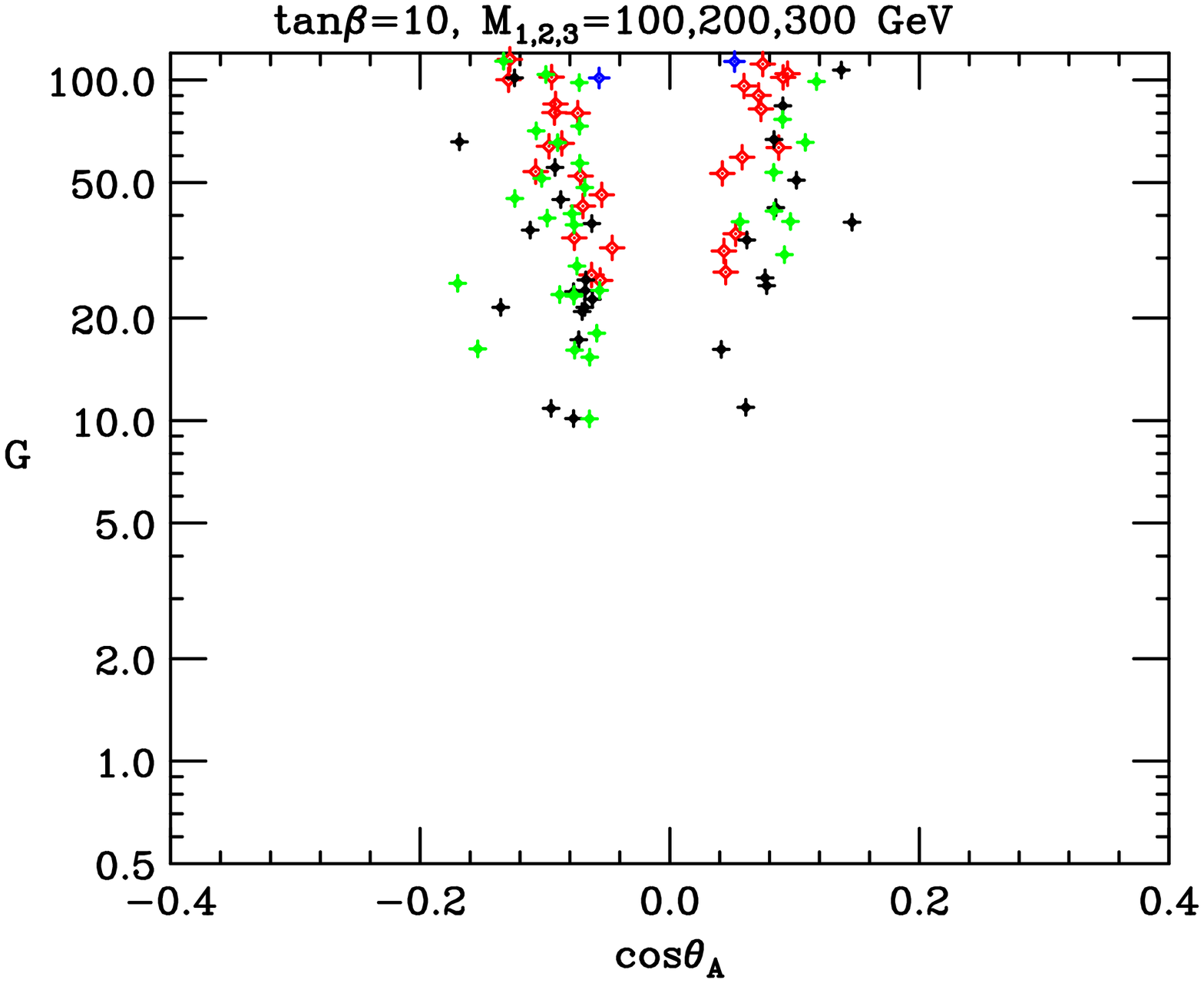}\includegraphics[width=0.4\textwidth,angle=90]{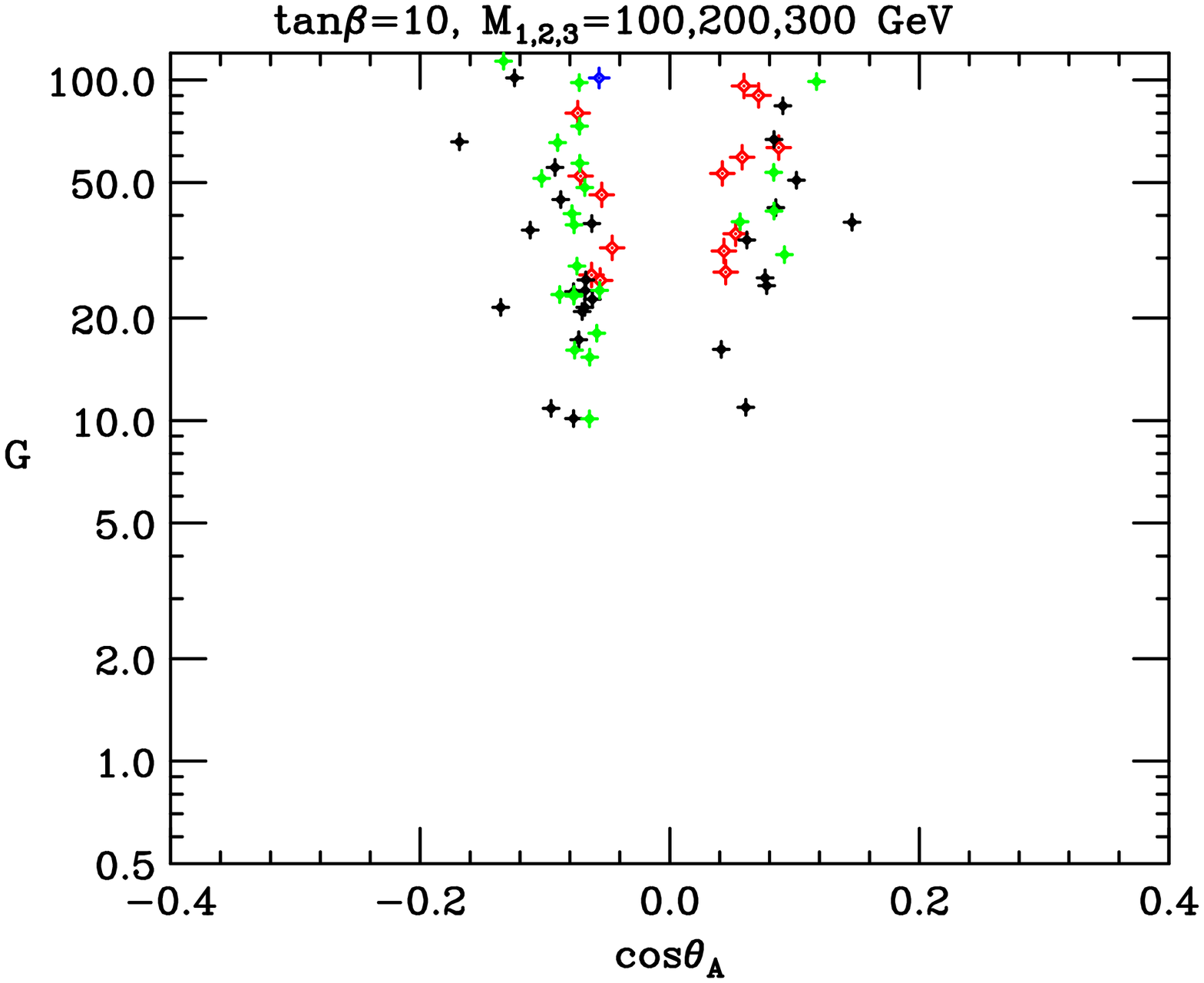}
\end{center}
\caption{As in Fig.~\protect\ref{lowftb3comp}, but for
  $\tanb=10$. Note that many points with lower $\mai$ and large
  $|\cta|$ are eliminated by the $|\cta|\leq \ctamax$ cut.}
\label{lowftb10comp}
\end{figure}

\begin{figure}
\begin{center}
\includegraphics[width=0.4\textwidth,angle=90]{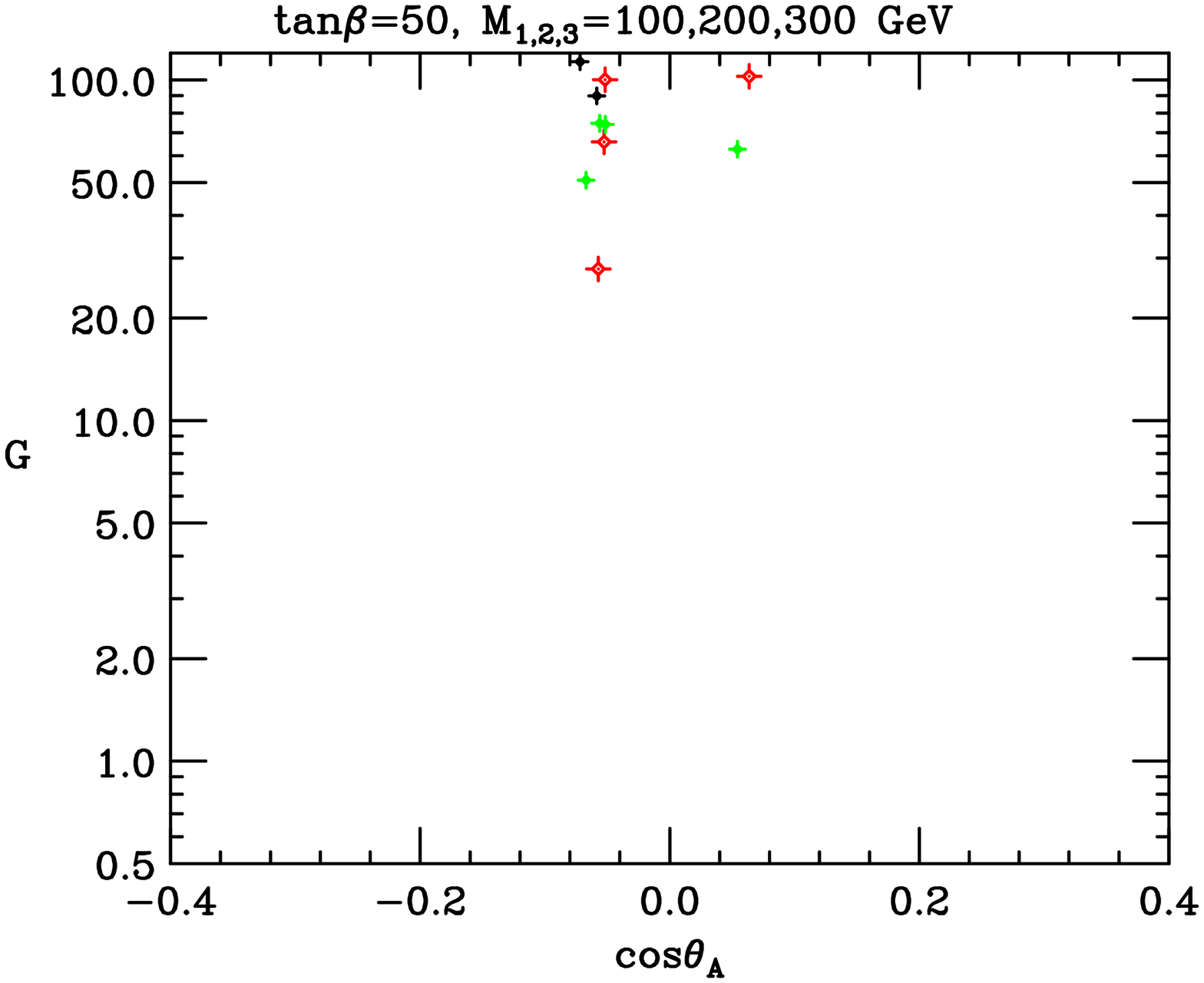}\phantom{\includegraphics[width=0.4\textwidth,angle=90]{lowfallokmaok_read_tb50_gvscta.ps}}
\end{center}
\caption{As in Fig.~\protect\ref{lowftb3comp}, but for $\tanb=50$. Note that
  no $F<15$ points found in our scans survive the $|\cta|<\ctamax$ limits.}
\label{lowftb50comp}
\end{figure}

The second set of plots below, Figs.~\ref{lowftb3comp},
\ref{lowftb10comp} and \ref{lowftb50comp}, show results for ``full
scans'', as defined previously, for $\tanb=3$, $10$, and $50$,
respectively.  Only points with electroweak finetuning measure $F$
below $15$ are plotted. As in the previous set of plots, the left-hand
plot in each figure shows the points allowed without the $\ctamax$
constraint and the right-hand plot displays the points remaining after
imposing $\ctamax$. The limited statistics for the parameter scans
that search for points with low $F$ are apparent, but the trends are
clearly the same as in the fixed $\mu$ scans presented previously.

From a theoretical perspective, an interesting pattern emerges: the
$\ctamax$ constraint eliminates those points for which the light-$\ai$
finetuning measure $G$ is never small and zeroes in on those $\cta$
values for which small $G$ is quite likely.

\section{Effective $\xi^2$ in the $h\to 4\tau$ channel for
  vector-boson fusion at the LHC and LEP $Zh$ channel constraints}

Discovery of a Higgs using vector boson fusion at the LHC or at LEP
with $2\mtau<\mai<2m_B$ (which is the only kind of point that survives with
$G<20$) is essentially determined by
\beq
\xi^2=\left({g_{VV}^h\over g_{VV}^{\hsm}}\right)^2\br(h\to
aa)[\br(a\to \tauptaum)]^2\,.
\eeq
We consider expectations for $\xi^2$ in the NMSSM ideal Higgs
scenarios with the $\ctamax$ constraint imposed in addition to the
usual constraints contained within NMHDECAY.

\begin{figure}
\begin{center}
\includegraphics[width=0.4\textwidth,angle=90]{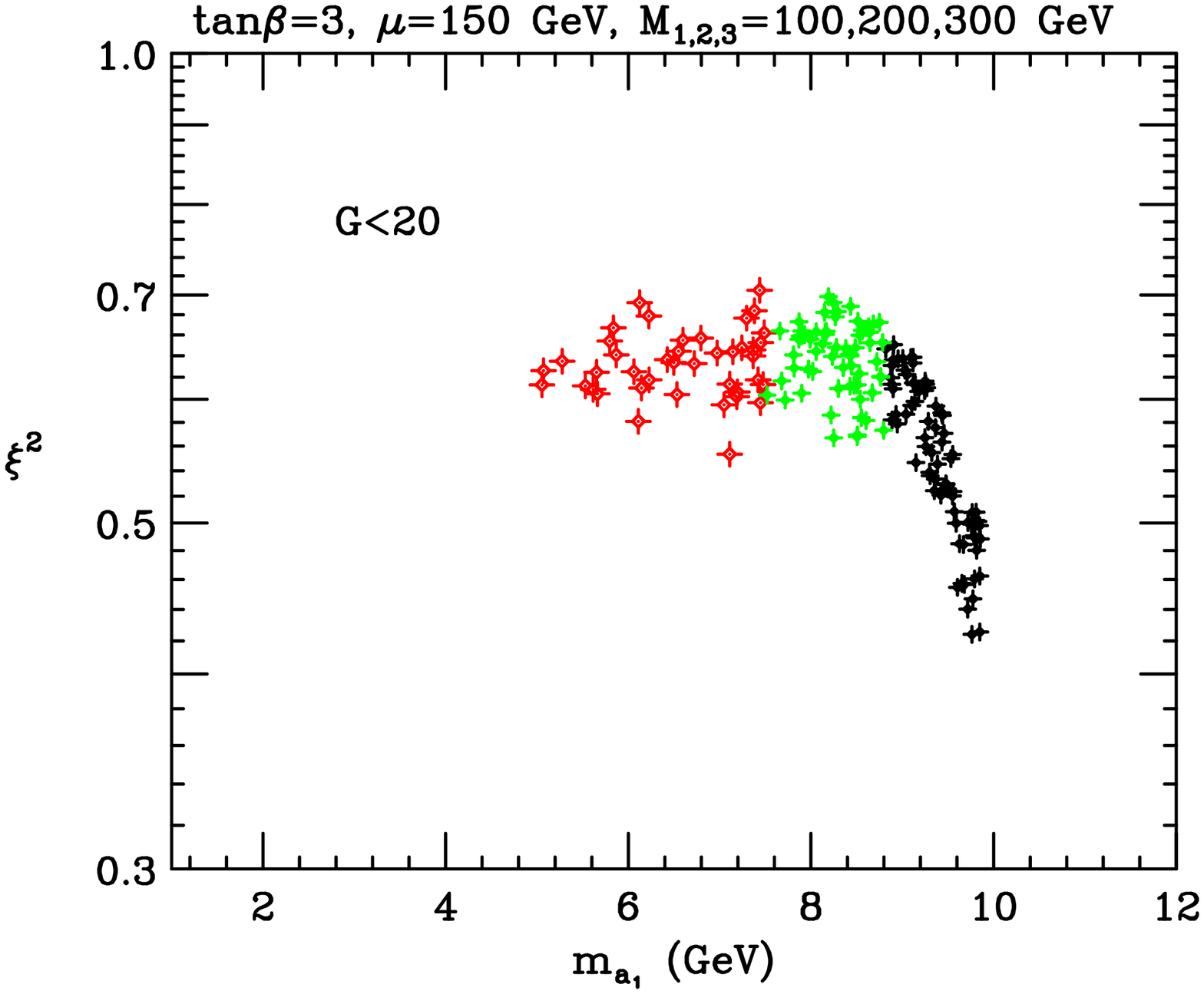}\hspace*{-.03in}\includegraphics[width=0.4\textwidth,angle=90]{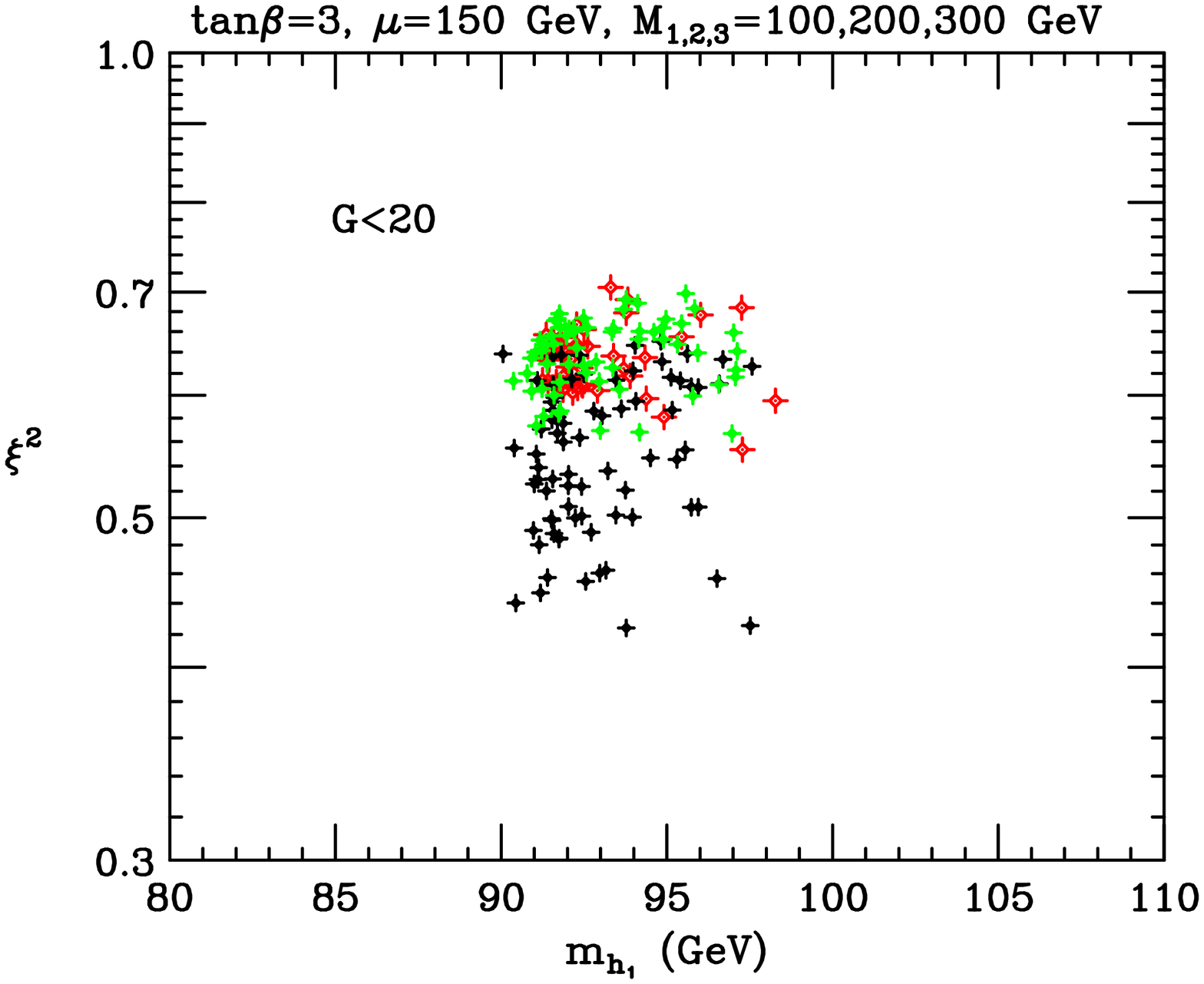}
\end{center}
\caption{$\xi^2$ for $h=\hi$ as a function of $\mai$ and $\mhi$ for
  points with $G<20$ and $|\cta|<\ctamax(\ma)$.  These plots are those
  obtained using the ``fixed-$\mu$'' scanning procedure for $\tanb=3$.}
\label{xisqtb3}
\end{figure}

\begin{figure}
\begin{center}
\includegraphics[width=0.4\textwidth,angle=90]{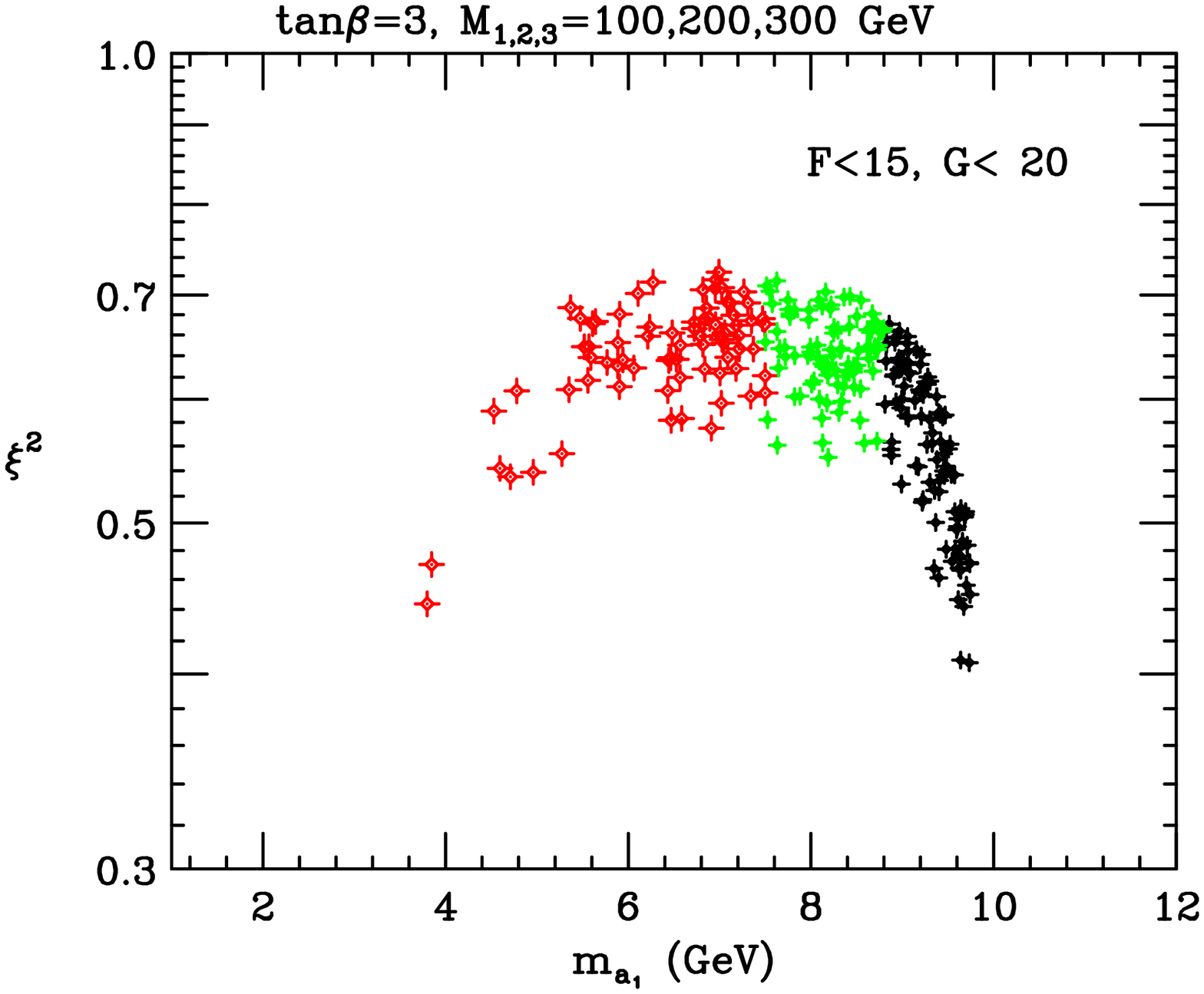}\includegraphics[width=0.4\textwidth,angle=90]{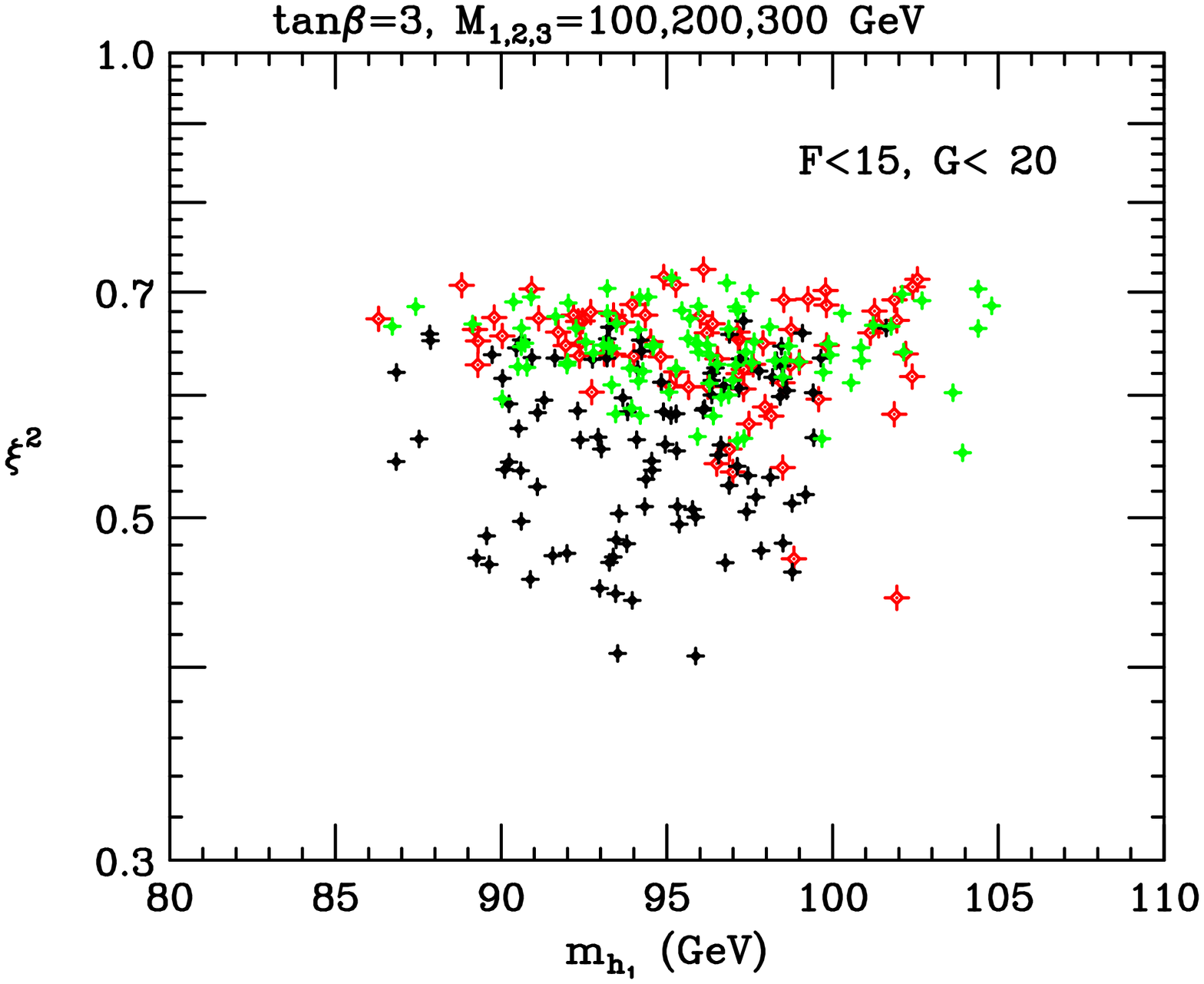}
\end{center}
\caption{$\xi^2$ for $h=\hi$ as a function of $\mai$ and $\mhi$ for
  points with $F<15$, $G<20$ and $|\cta|<\ctamax(\ma)$.  These plots
  are those obtained using the described scanning procedure for
  $\tanb=3$.}
\label{xisqtb3flt15}
\end{figure}

\begin{figure}
\begin{center}
\includegraphics[width=0.4\textwidth,angle=90]{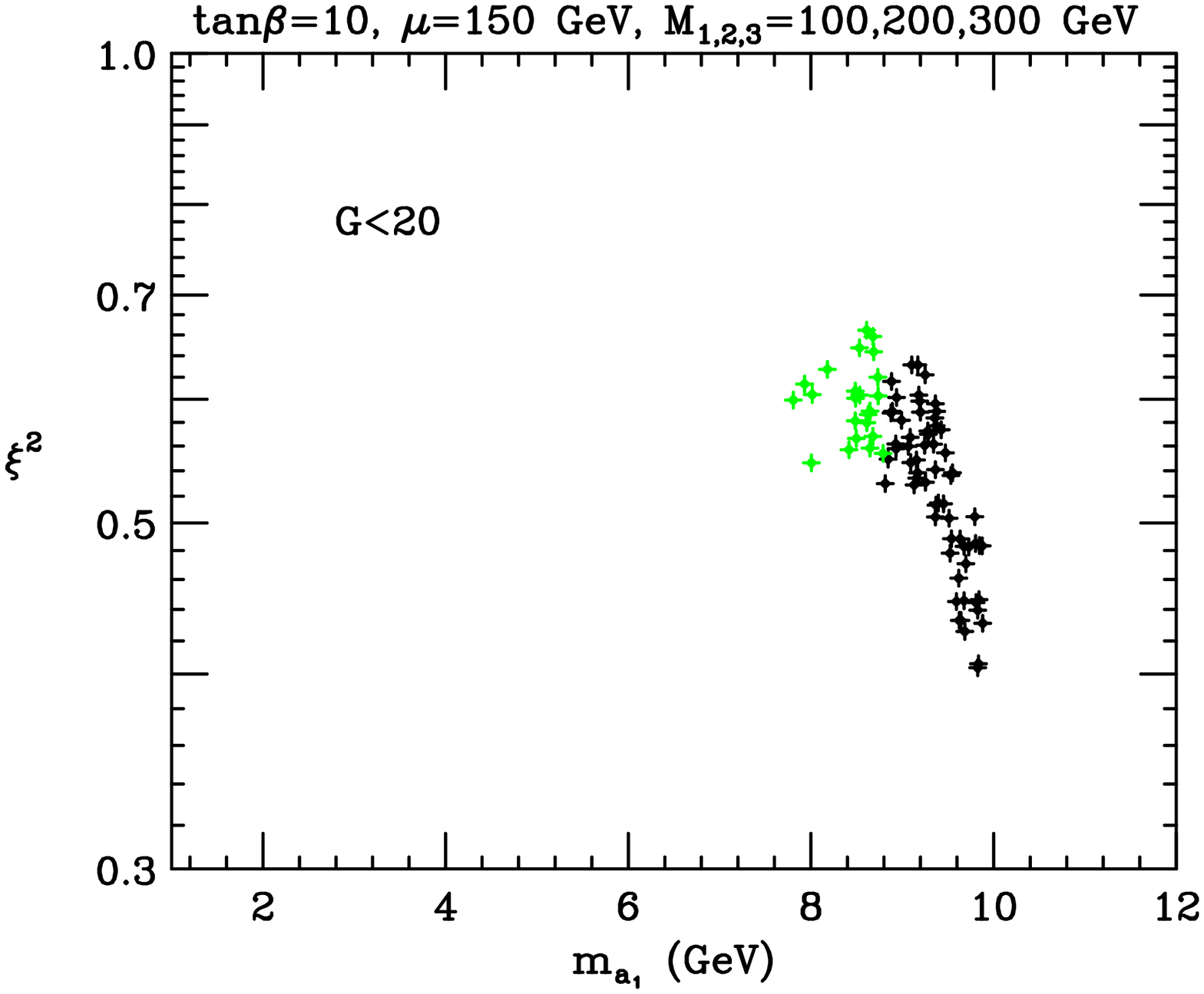}\hspace*{-.03in}\includegraphics[width=0.4\textwidth,angle=90]{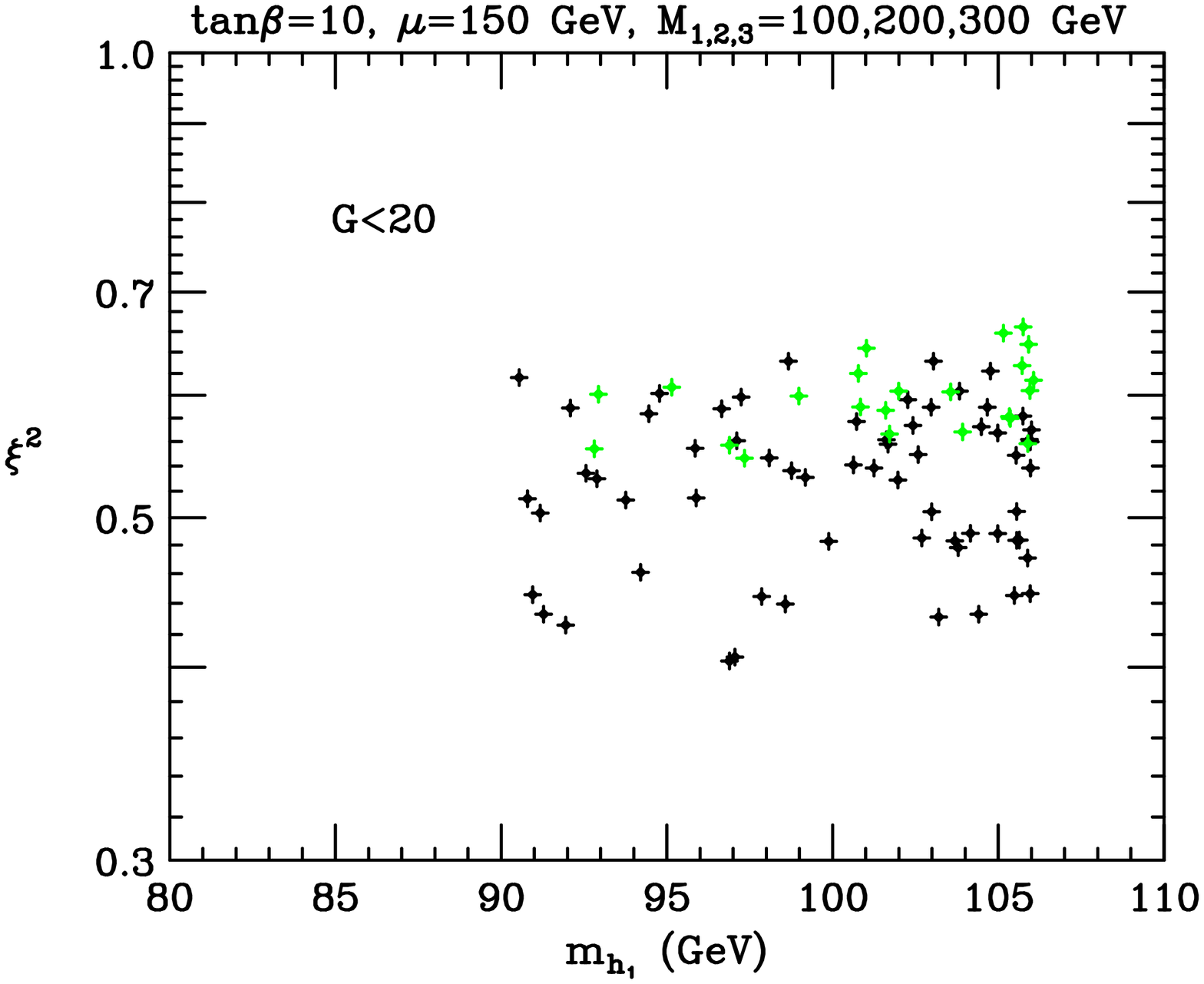}
\end{center}
\caption{$\xi^2$ for $h=\hi$ as a function of $\mai$ and $\mhi$ for
  points with $G<20$ and $|\cta|<\ctamax(\ma)$.  These plots are those
  obtained using the ``fixed-$\mu$'' scanning procedure for $\tanb=10$.}
\label{xisqtb10}
\end{figure}

\begin{figure}
\begin{center}
\includegraphics[width=0.4\textwidth,angle=90]{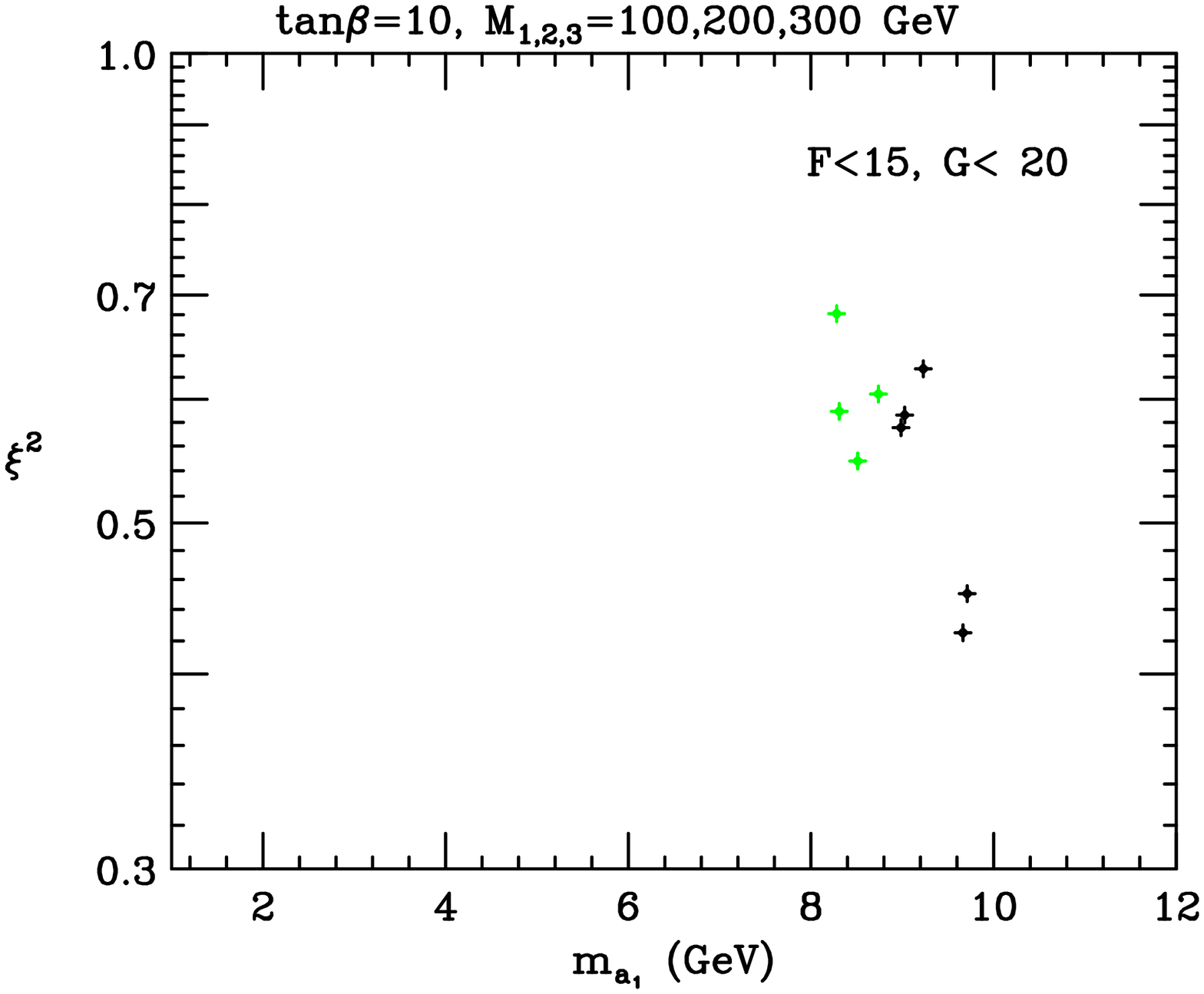}\includegraphics[width=0.4\textwidth,angle=90]{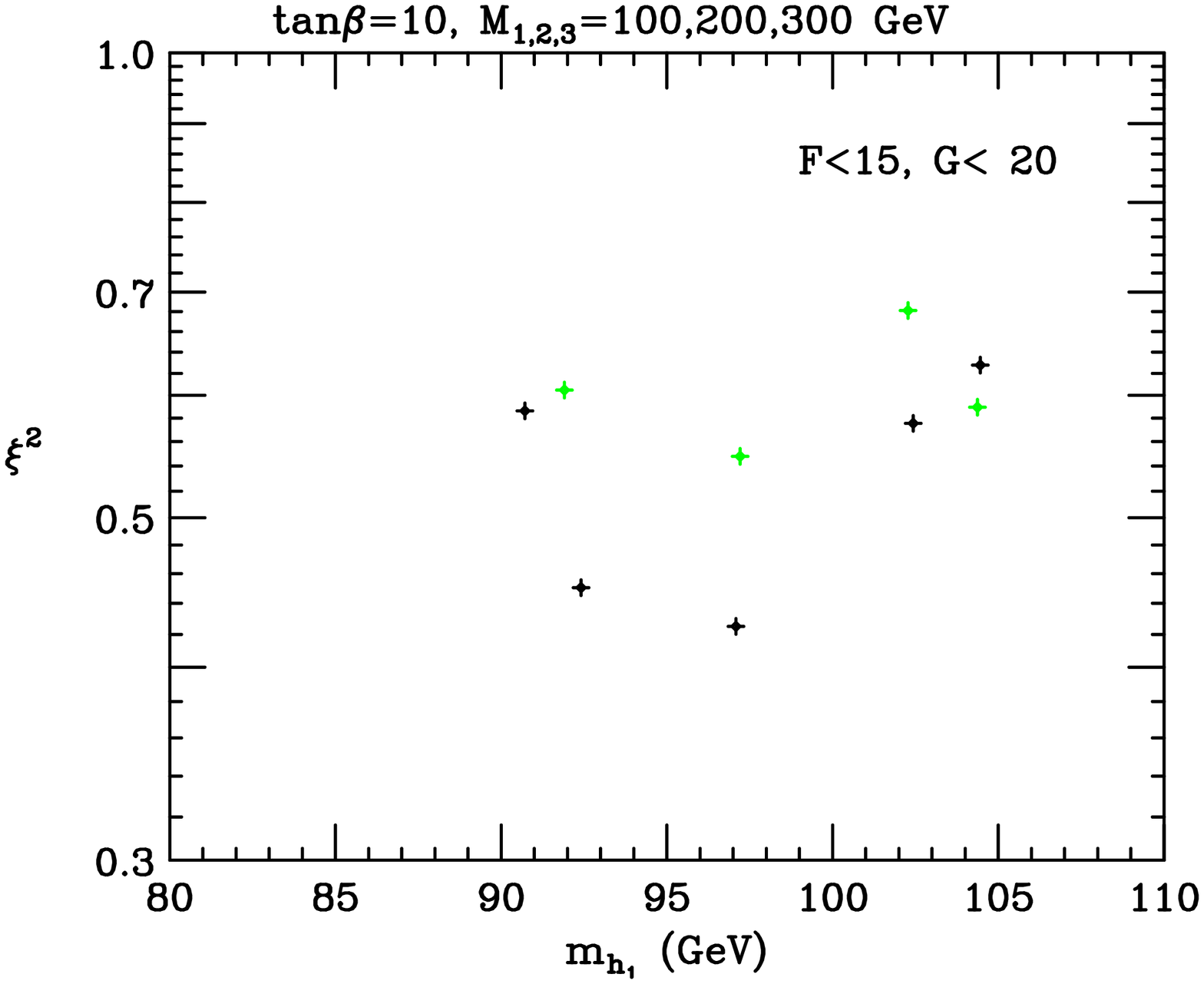}
\end{center}
\caption{$\xi^2$ for $h=\hi$ as a function of $\mai$ and $\mhi$ for
  points with $F<15$, $G<20$ and $|\cta|<\ctamax(\ma)$.  These plots
  are those obtained using a ``full scan'' for
  $\tanb=10$.}
\label{xisqtb10flt15}
\end{figure}

\begin{figure}
\begin{center}
\includegraphics[width=0.4\textwidth,angle=90]{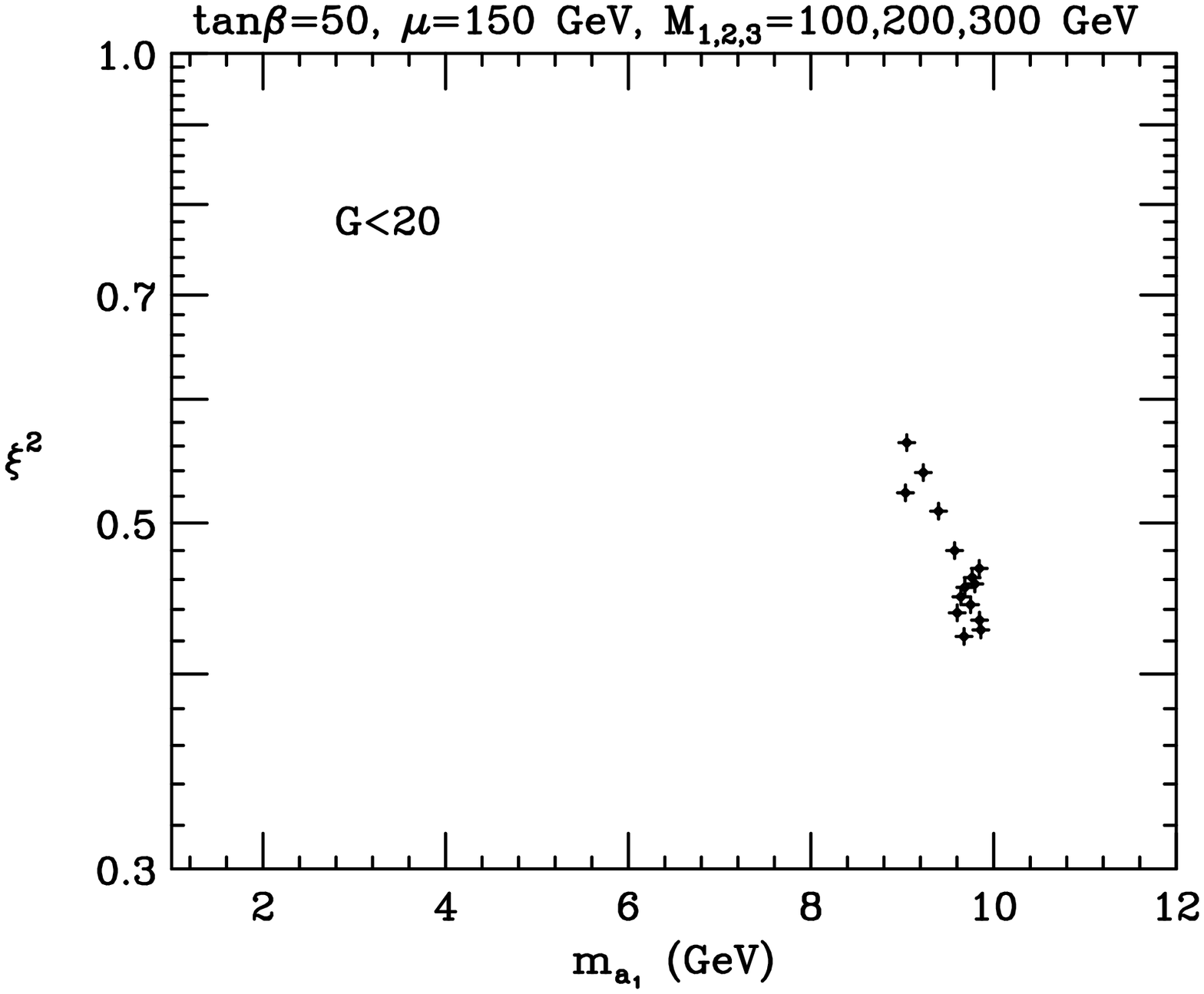}\hspace*{-.03in}\includegraphics[width=0.4\textwidth,angle=90]{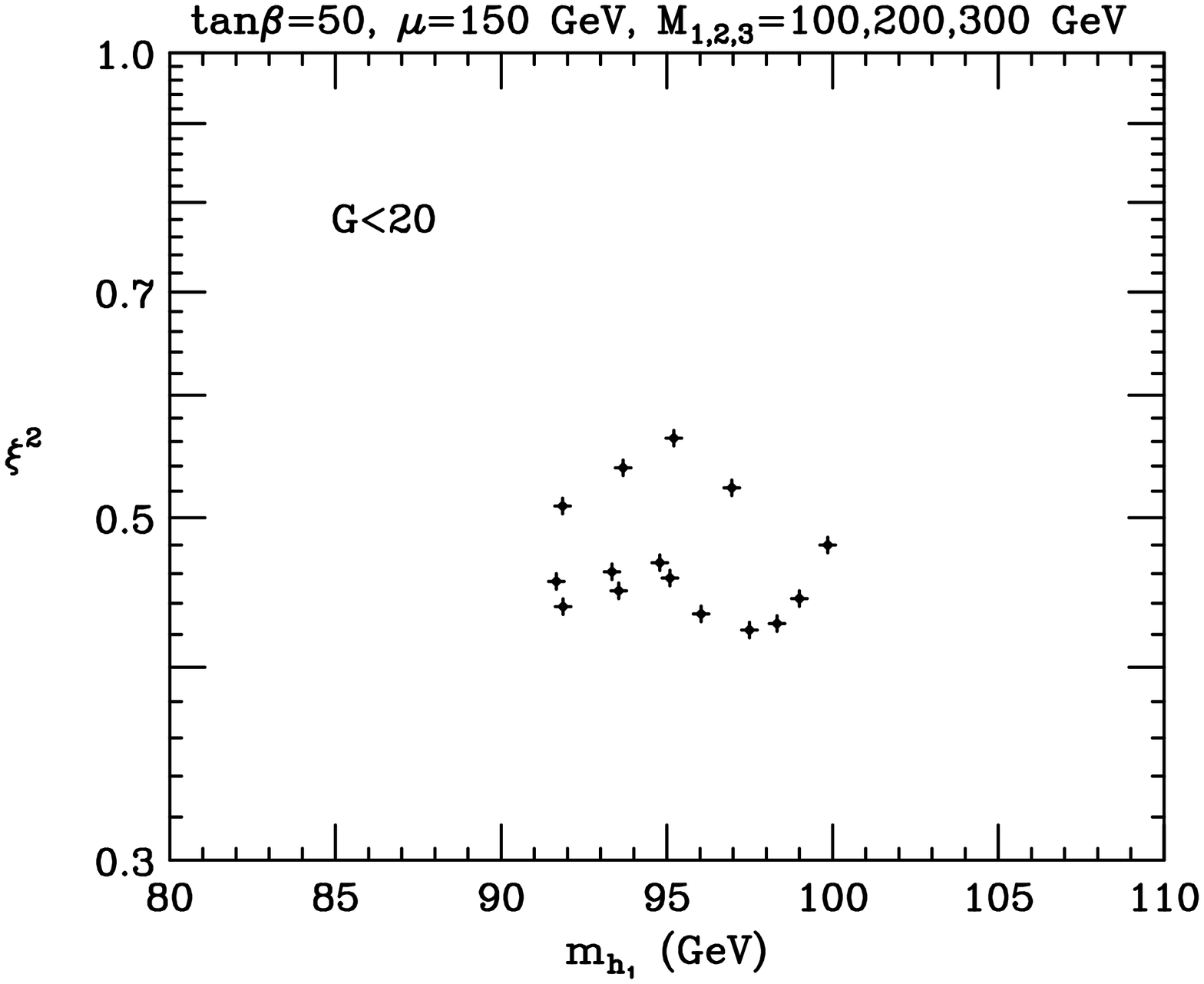}
\end{center}
\caption{$\xi^2$ for $h=\hi$ as a function of $\mai$ and $\mhi$ for
  points with $G<20$ and $|\cta|<\ctamax(\ma)$.  These plots are those
  obtained using the ``fixed-$\mu$'' scanning procedure for
  $\tanb=50$.}
\label{xisqtb50}
\end{figure}

In Fig.~\ref{xisqtb3} we take $\tanb=3$ and plot $\xi^2$ for $h=\hi$
and $a=\ai$ as a function of $\mai$ and as a function of $\mhi$ for
points coming from the fixed $\mu$ scans after imposing $G<20$ and
requiring $|\cta|<\ctamax(\ma)$. We observe that $\xi^2$ as small as
$\sim 0.42$ is possible at high $\mai$, which points tend to have
$\mhi\in[90,100]\gev$. As seen in Fig.~\ref{xisqtb3flt15}, these same
remarks apply also to the $F<15$ points obtained in our finetuning
scans when $G<20$ and $|\cta|<\ctamax(\mai)$ are imposed.  These same
remarks also apply to the $\tanb=10$ plots of Figs.~\ref{xisqtb10} and
\ref{xisqtb10flt15} as well as to the $\tanb=50$ fixed-$\mu$-scan plot of
Fig.~\ref{xisqtb50}. (Note that no $F<15$, $G<20$ points survived our
limited statistics electroweak finetuning scan in the $\tanb=50$ case
and so there is no corresponding figure.)

In addition, we have also considered $\xi^2$ expectations in scenarios
with rather low $\tanb$. These were detailed in
\cite{Dermisek:2008uu}. There, we performed fixed-$\mu$ scans as
defined earlier, with the difference that at $\tanb=1.7$ and
$\tanb=1.2$ we used different values for $\msusy$ and $A$ parameters,
which values are indicated on the figures.  At $\tanb=2$ we employed
$\msusy=-A=300\gev$ as for the fixed-$\mu$ scans for $\tanb=3,10,50$.

The main distinguishing characteristic of the
low $\tanb$ scenarios is that both $\hi$ and $\hii$ can be light with
masses not far from $100\gev$, although there are certainly choices
for the NMSSM parameters for which only $\hi$ is light while $\hii$ is
much heavier.  When $\hii$ is light, the charged Higgs $H^{\pm}$ can
also have mass close to $100\gev$.\footnote{Note that a light $H^\pm$
  can cause the NMSSM prediction for $\br(b\to s\gam)$ to
  substantially exceed the experimental value, which is only slightly
  above the SM value.  Thus, contributions from other SUSY diagrams
  must enter to cancel the $H^\pm$ diagrams. In models with low
  finetuning, SUSY is light and such cancellation is generically
  entirely possible.}  Here, our interest is in the predictions for
$\xi^2$.

\begin{figure}
\begin{center}
\includegraphics[width=0.35\textwidth,angle=90]{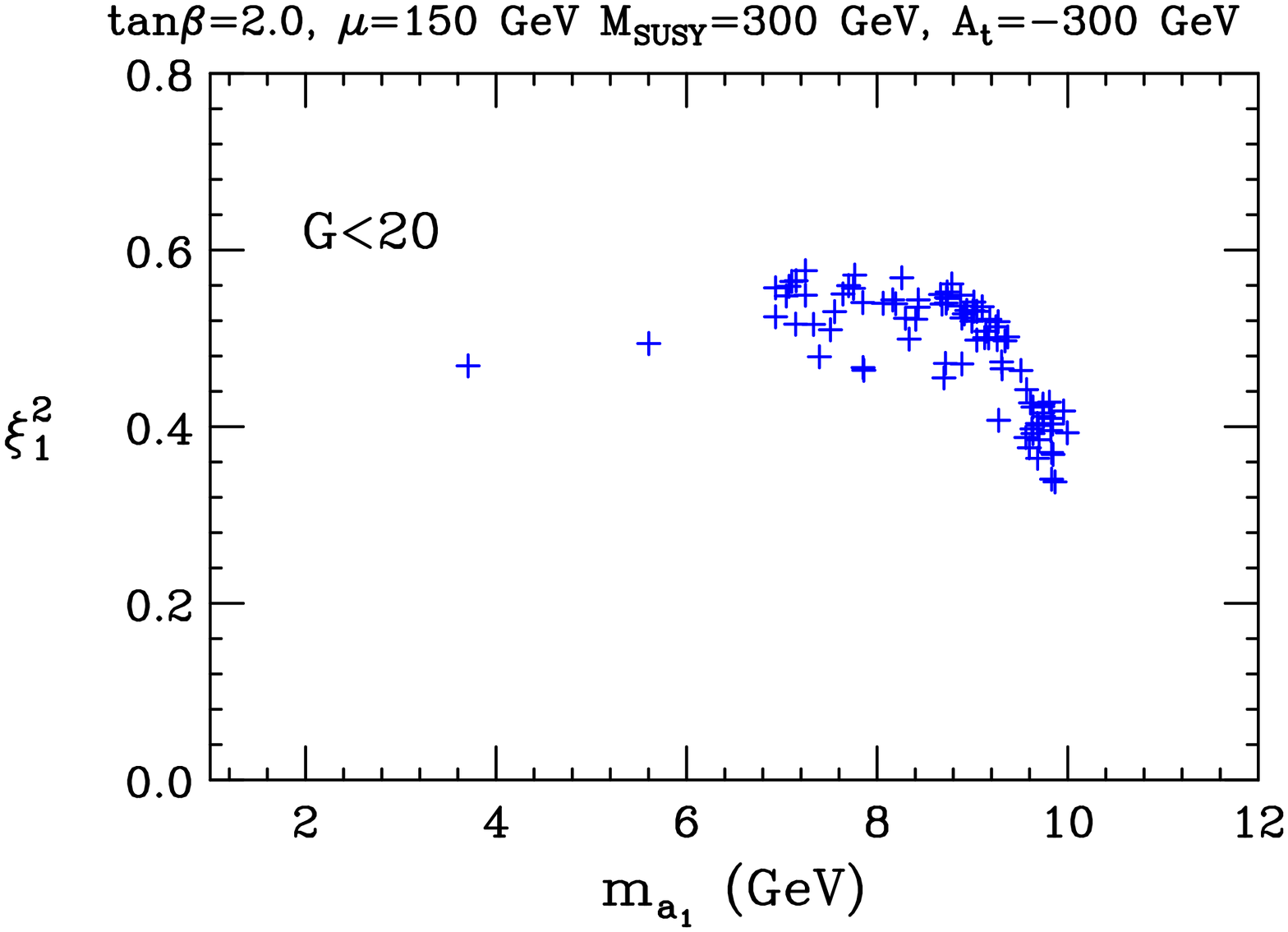}\includegraphics[width=0.35\textwidth,angle=90]{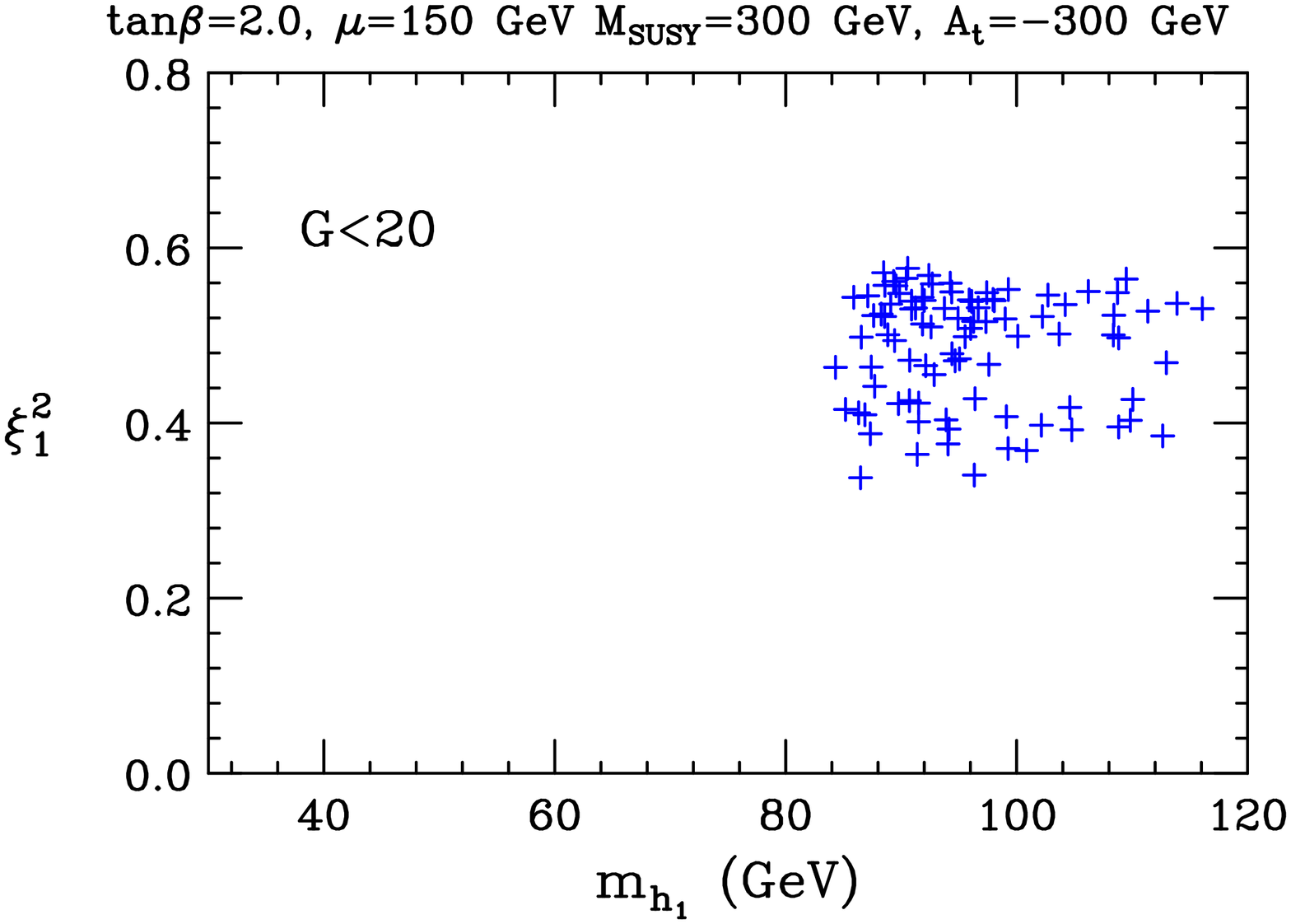}
\end{center}
\caption{$\xi^2_1$ as a function of $\mai$ and $\mhi$ for points with
  $G<20$ and $|\cta|<\ctamax(\mai)$ and $\tanb=2$.  These plots are
  those obtained using a ``fixed-$\mu$'' scanning procedure with
  the $\mu$, $\msusy$ and $A$ parameters indicated on the figure.
  We have not indicated different $\mai$ mass ranges using different
  colors in these figures.}
\label{radotb2pt0}
\end{figure}

Results for $\xi_1^2$ at $\tanb=2$ are rather similar to those found
for higher $\tanb$, as shown in Fig.~\ref{radotb2pt0}. In this figure,
the blue $+$'s are all points that satisfy the NMHDECAY constraints
--- unlike the previous figures, color coding is not employed to
distinguish different $\mai$ values.  Results for $\xi_2^2$ are not
shown; even when $\mhii$ is close to $100\gev$, $\xi_2^2$ is quite
small. This $\tanb=2$ case is similar to the $\tanb=3,10,50$ cases
also in that it is almost always the case that $VV$ couples primarily
to the $\hi$ so that when $\mhi\leq 105\gev$ we have the ``ideal''
Higgs explanation of the precision electroweak data.

\begin{figure}
\begin{center}
\includegraphics[width=0.35\textwidth,angle=90]{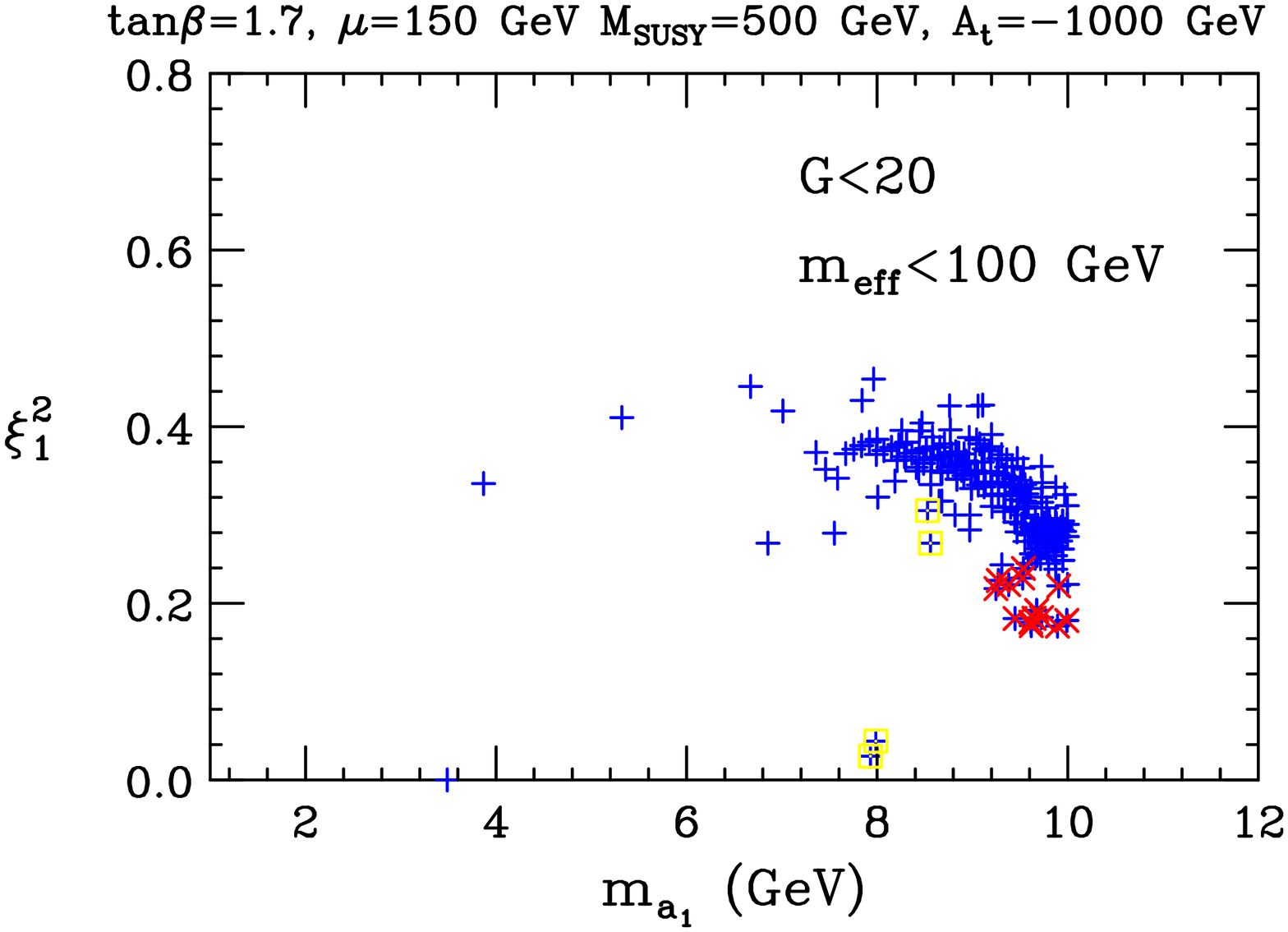}\includegraphics[width=0.35\textwidth,angle=90]{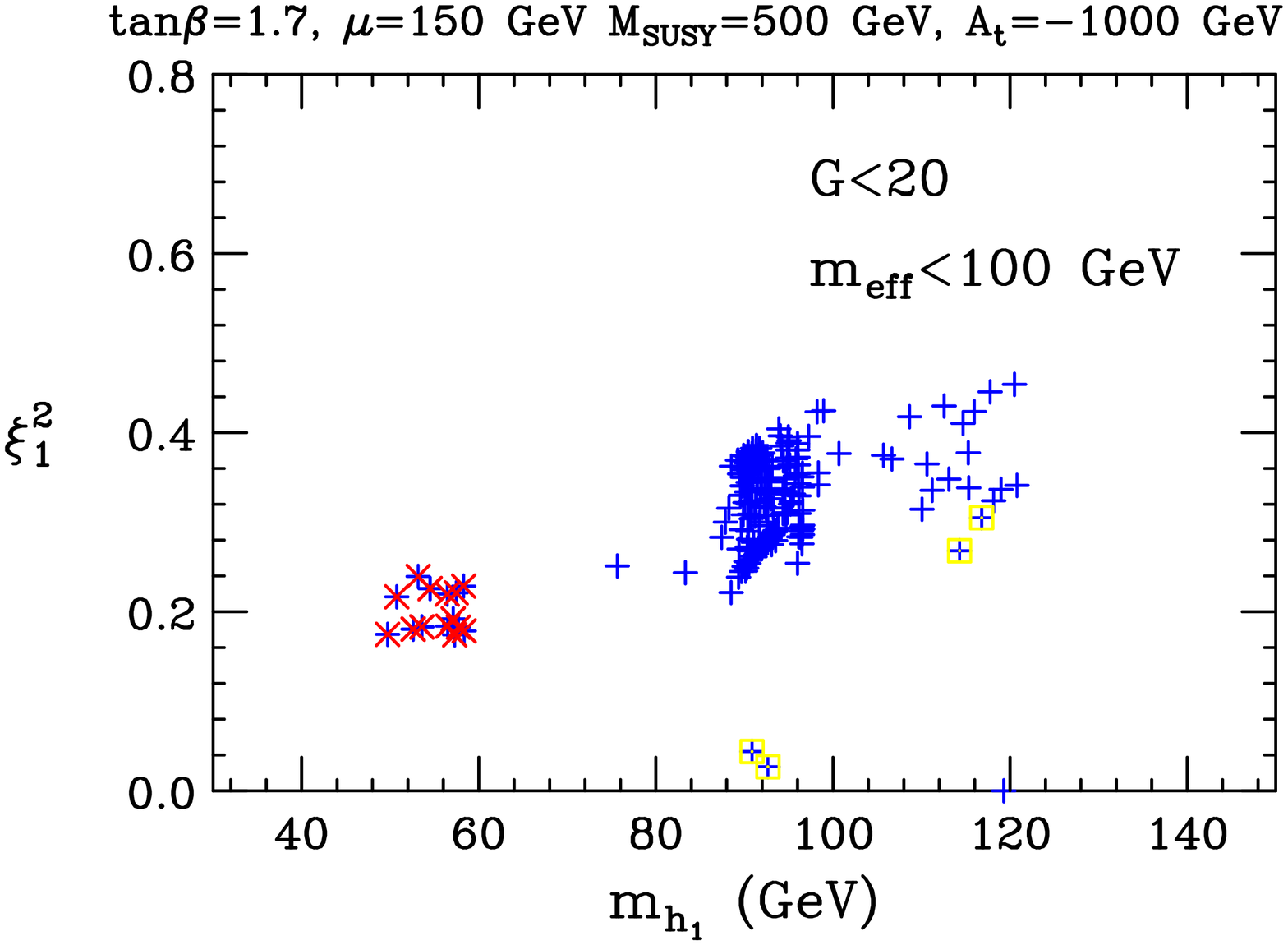}
\end{center}
\caption{$\xi^2_1$ as a function of $\mai$ and $\mhi$ for $\tanb=1.7$
  points obtained from a fixed-$\mu$ scan after requiring $G<20$,
  $\meff<100\gev$ and $|\cta|<\ctamax(\mai)$.  The point notation is
  explained in the text.}
\label{radotb1pt7}
\end{figure}

\begin{figure}
\begin{center}
d\includegraphics[width=0.35\textwidth,angle=90]{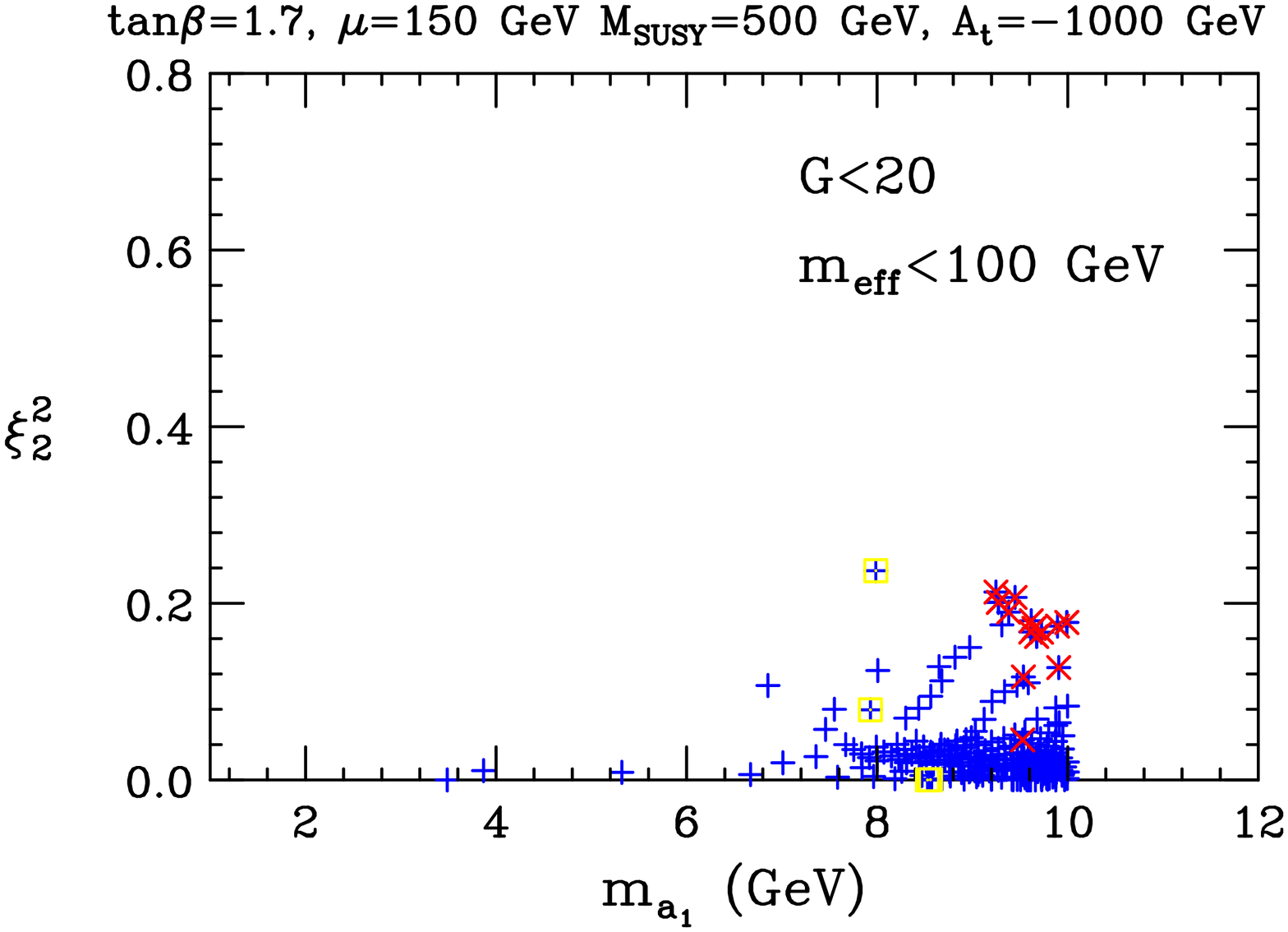}\includegraphics[width=0.35\textwidth,angle=90]{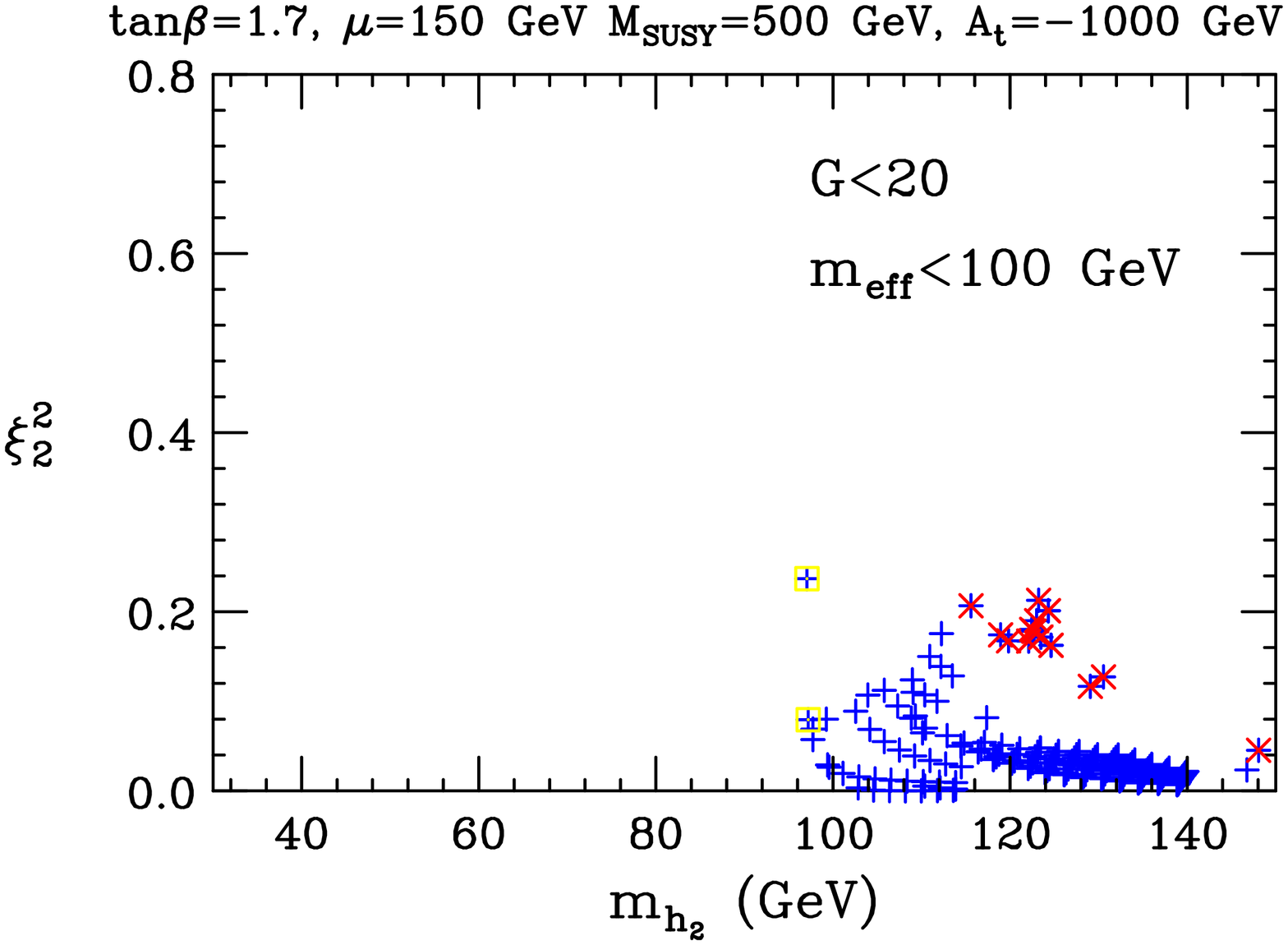}
\end{center}
\caption{$\xi^2_2$ as a function of $\mai$ and $\mhii$ for $\tanb=1.7$
  points obtained from a fixed-$\mu$ scan after requiring $G<20$,
  $\meff<100\gev$ and $|\cta|<\ctamax(\mai)$.}
\label{radotb1pt7ii}
\end{figure}

For $\tanb\lsim 1.7$, there are some interesting new subtleties
compared to $\tanb\gsim 2$. Plots of $\xi_1^2$ of the $\hi$ and
$\xi_2^2$ of $\hii$ appear in Figs.~\ref{radotb1pt7} and
\ref{radotb1pt7ii}, respectively.  In these plots, we follow the
notation established in Ref.~\cite{Dermisek:2008uu}.  In detail, the
blue $+$'s are all points that satisfy the NMHDECAY constraints. The
red crosses single out those points for which $\mhi<65\gev$.   Yellow
squares indicate points for which $\br(\hi\to \ai\ai)<0.7$.
In \cite{Dermisek:2008uu}, there were also points indicated by green diamonds
for which {\it in addition} the light CP-odd Higgs is primarily
doublet-like, $\cos^2 \theta_A > 0.5$. However, these are absent from
the present plots, not because of the improved $\ctamax$ limits from
the recent BaBar data, but rather because of the $G<20$ requirement
which very strongly disfavors large $|\cta|$ at all $\mai$, including
$\mai$ above $\mupsiii$. Of course, the BaBar data eliminates many
points with $\mai<\mupsiii$ having $\cos^2\theta_A<0.5$, certainly
more than in the analysis of \cite{Dermisek:2008uu}. 

Let us now discuss the $\tanb=1.7$ case in more detail.  We first wish
to discuss the extent to which the points that survive the NMHDECAY
scans can be  ``ideal'' in the precision electroweak sense.
Defining
\beq
\cvi=g_{VV\hi}/g_{VV\hsm}\,,\qquad \cvii=g_{VV\hii}/g_{VV\hsm}\,,
\eeq
then, noting that it is a good approximation to neglect any $\hiii$
coupling to $VV$, one has the sum rule
\beq
\cvi^2+\cvii^2\simeq 1\,.
\eeq
In this notation, the effective precision electroweak mass, $\meff$,
is given to very good approximation by
\beq
\meff=\mhi^{\cvi^2}\mhii^{\cvii^2}\,.
\eeq
In order to guarantee that all accepted points are ideal, we require
as part of our $\tanb=1.7$ scan that $\meff<100\gev$.\footnote{This
  was not imposed in the plots of \cite{Dermisek:2008uu}.}  Now, let
us describe the associated plots.  First, very low values of $\mhi$
are possible (see the red crosses).  These red cross points are such
that $\xi_1^2$ and $\xi_2^2$ are comparable and both below $0.2$.
Second, very few of the yellow square points (defined by
$\br(\hi\to\ai\ai)<0.7$) survive the ideal requirement. But, those
that do have quite small $\xi_1^2$ and $\xi_2^2$. The run-of-the-mill
blue $+$ points have somewhat larger $\xi_1^2\lsim 0.4$ and somewhat
smaller $\xi_2^2\lsim 0.2$.  Overall, the $4\tau$ final state in $\hi$
and $\hii$ decays typically has significantly smaller cross section
for $\tanb=1.7$ as compared to $\tanb\gsim 2$.

\begin{figure}
\begin{center}
\includegraphics[width=0.35\textwidth,angle=90]{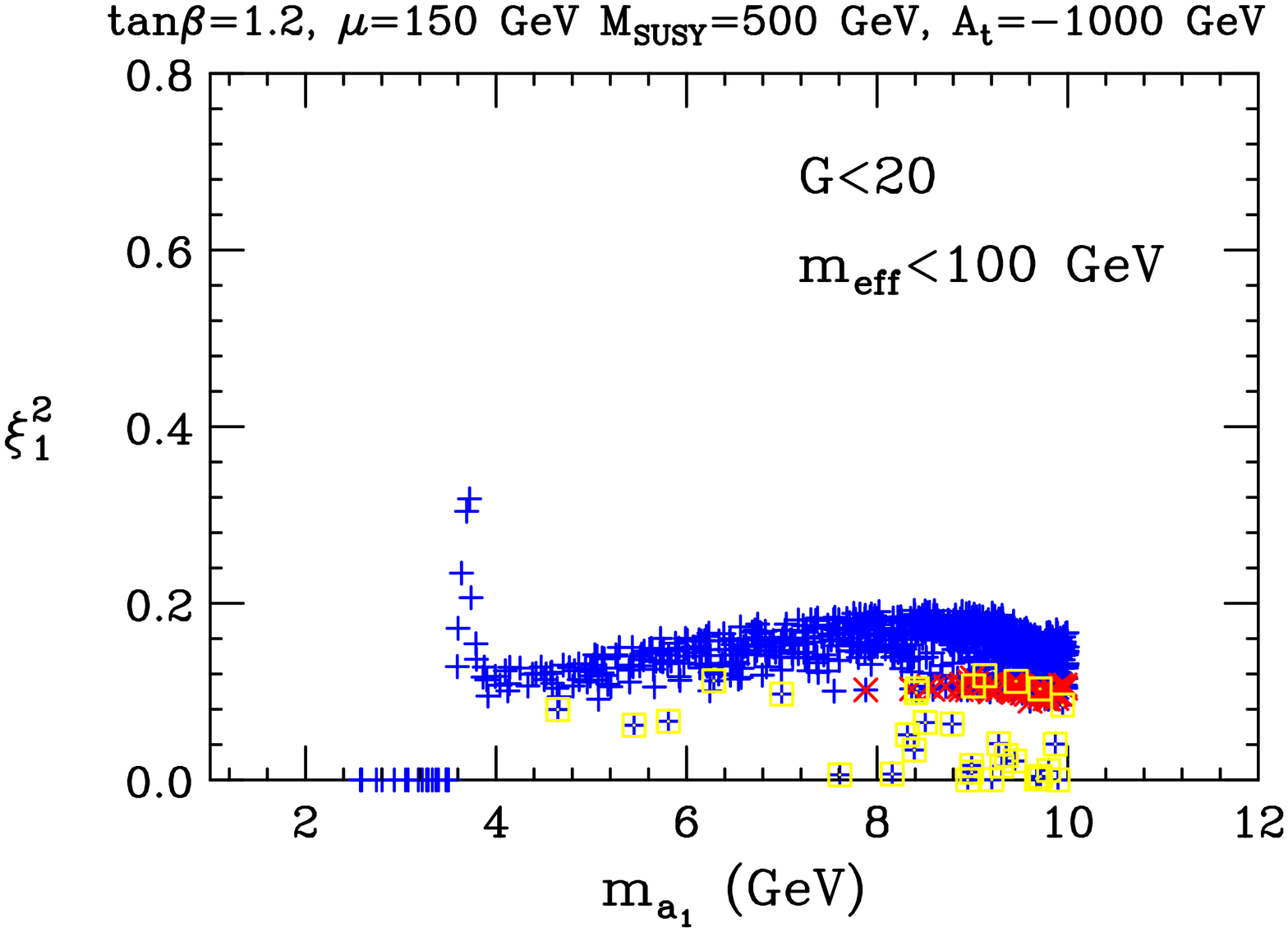}\includegraphics[width=0.35\textwidth,angle=90]{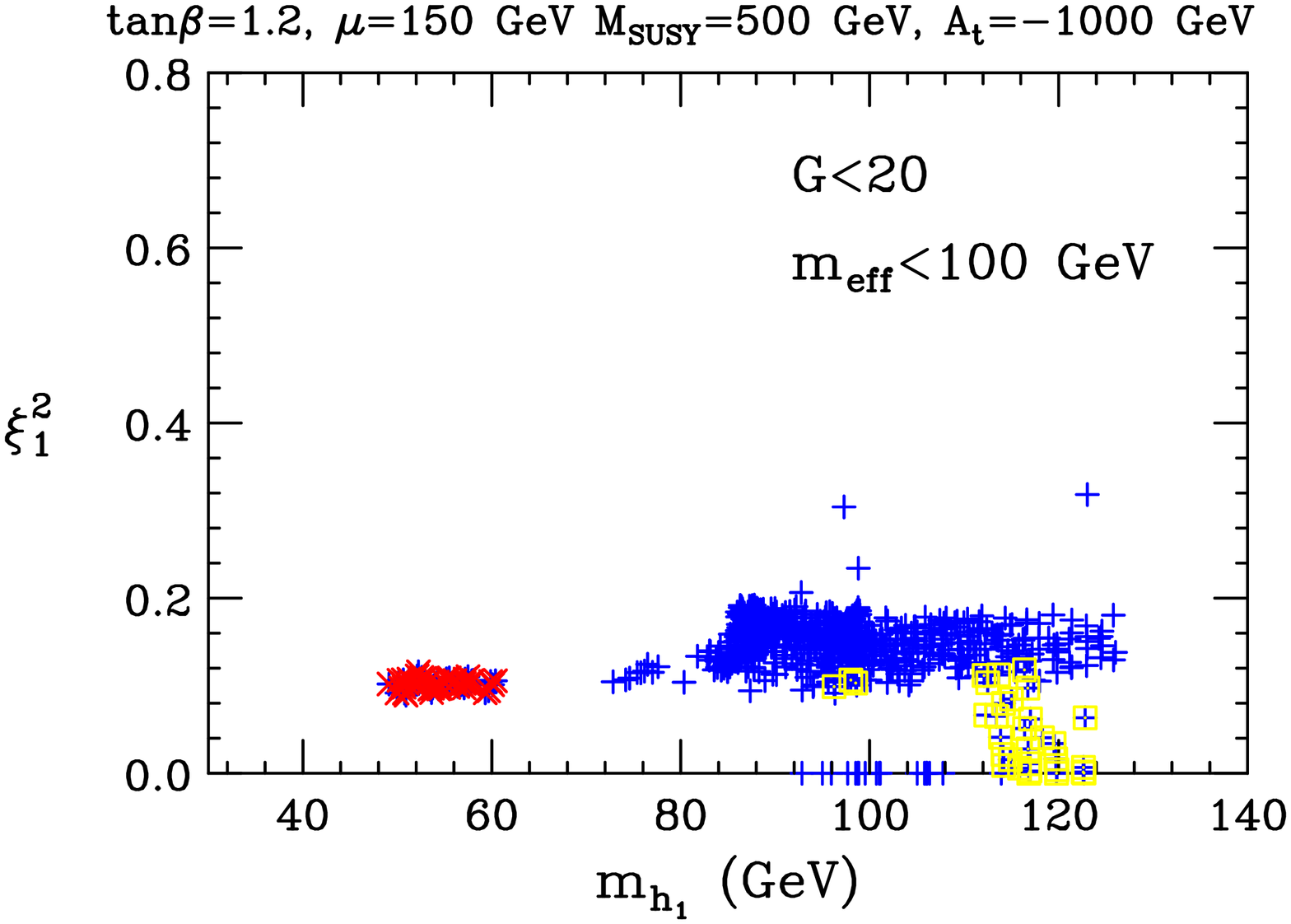}
\end{center}
\caption{$\xi^2_1$ as a function of $\mai$ and $\mhi$ for $\tanb=1.2$
  points obtained from a fixed-$\mu$ scan after requiring $G<20$,
  $\meff<100\gev$ and $|\cta|<\ctamax(\mai)$.}
\label{radotb1pt2}
\end{figure}

\begin{figure}
\begin{center}
\includegraphics[width=0.35\textwidth,angle=90]{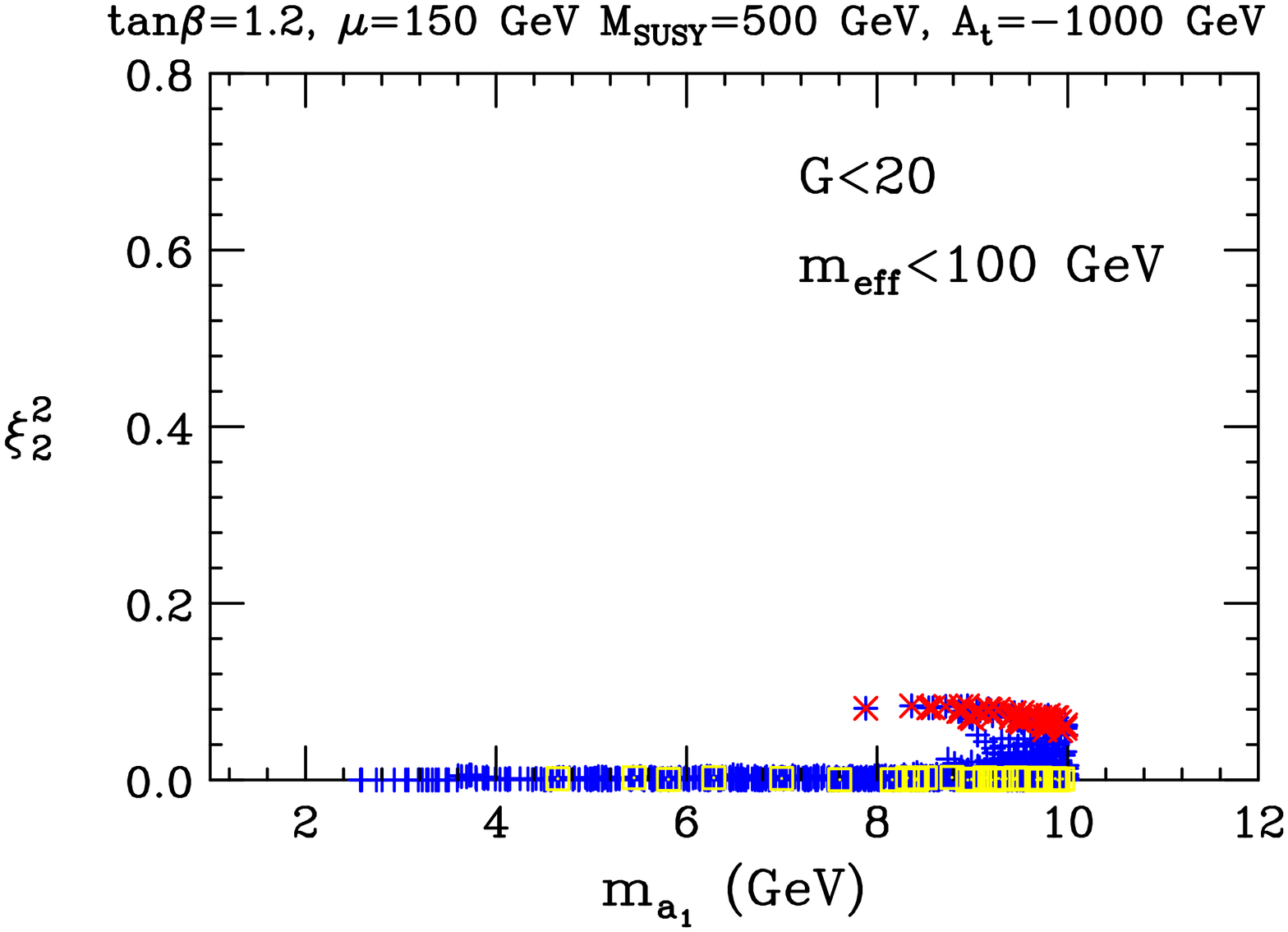}\includegraphics[width=0.35\textwidth,angle=90]{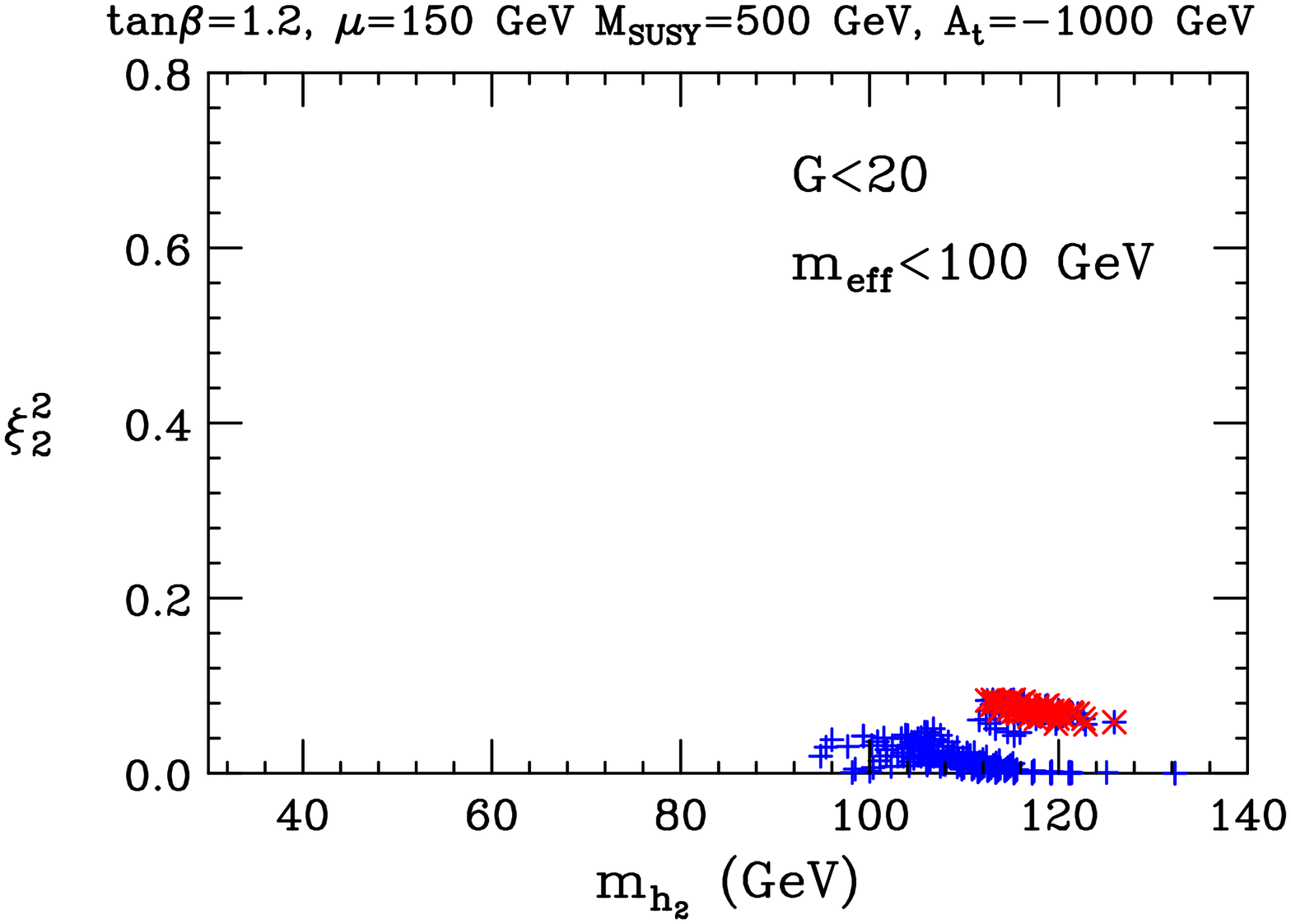}
\end{center}
\caption{$\xi^2_2$ as a function of $\mai$ and $\mhii$ for $\tanb=1.2$
  points with $G<20$, $\meff<100\gev$ and $|\cta|<\ctamax(\mai)$.
}
\label{radotb1pt2ii}
\end{figure}

The lowest value of $\tanb$ consistent with maintaining perturbativity
up to the GUT scale is $\tanb=1.2$.  $\xi_1^2$ and $\xi_2^2$ plots for
this case appear in Figs.~\ref{radotb1pt2} and \ref{radotb1pt2ii},
respectively.  In this case, the effective $\xi_1^2$ values are mostly
quite small. Relative to the $\tanb=1.7$ plots, the main thing that
has changed is that $\br(\ai\to \tauptaum)$ has declined
substantially. The majority of the $\ai$ decays are into $gg$ and
$c\anti c$, \ie\ final states that are harder to constrain.

Of course, the knowledgeable reader will recognize that all the
$\xi^2$ plots presented are aimed at comparing these NMSSM models to
the new ALEPH analysis of the $4\tau$ final
state~\cite{cranmer}.  According to the ALEPH analysis, to
have $\mhi\lsim 100\gev$, $\xi^2_1\lsim 0.52$ ($0.42$) is required if
$\mai\sim 10\gev$ ($4\gev$).  These limits rise rapidly with
increasing $\mhi$ --- for $\mhi=105\gev$ (the rough upper limit on
$\mhi$ such that electroweak finetuning remains quite small and
precision electroweak constraints are fully satisfied) the ALEPH
analysis requires $\xi^2\lsim 0.85$ ($\lsim 0.7$) at $\mai\sim
10\gev$ ($4\gev$).  These limits are such that the easily viable NMSSM
scenarios are
ones: i) with $\mai$ below but fairly close to $2m_B$,
which is, in any case, strongly preferred by minimizing the
light-$\ai$ finetuning measure $G$; and/or ii) with $\tanb$ relatively
small ($\lsim 2$).~\footnote{A similar conclusion applies to models
  beyond the MSSM with a light doublet CP-odd Higgs
  boson~\cite{Dermisek:2008id, Dermisek:2008sd, Bae:2010cd}. Since
  these scenarios are consistent with other experimental limits only
  for $\tan \beta \lsim 2.5$, the new preliminary Aleph limits only
  constrain the upper range of the allowed region of $\tan \beta$.}
These are also the scenarios for which Upsilon constraints are either
weak or absent. In particular, we note the following: a)~all
$\tanb\leq 2$ cases provide $\mhi\leq 100\gev$ scenarios that escape the
ALEPH limits; b)~there are a few $G<20$, $\tanb=3$ scenarios with
$\mhi$ as large as
$98\gev$ and $99\gev$ and with $\xi^2$  essentially equal to
the ALEPH limits of $\xi^2\leq 0.42$ and $\xi^2\leq 0.45$ applicable at
these respective $\mhi$ values; c)~$\tanb=10$ ideal scenarios easily
allow for $\mhi\sim 100-105\gev$ (because the tree-level Higgs mass is
larger at $\tanb=10 $ than at $\tanb=3$) and at $\mai\lsim 2m_B$ many
$\mhi\gsim 100\gev$ points have
$\xi^2<0.5$ in the fixed-$\mu$ scan and a few of the full-scan points
have $\xi^2<0.6$ for $\mhi\sim 105\gev$, both of which are below the
$\mai=10\gev$ ALEPH upper limits on $\xi^2$ of 0.52 at $\mhi\sim 100\gev$ and
$0.85$ at $\mhi=105\gev$; d)~at
$\tanb=50$ there are some $G<20$ points with $\mhi\sim 100\gev$ and
$\mai\lsim 2m_B$ having $\xi^2$ below the $0.52$ ALEPH limit. 
Finally, we note that for the entire range of Higgs masses
studied the ALEPH limits were actually $\sim 2\sigma$ stronger than
expected. Thus, it is not completely unreasonable to consider the
possibility that the weaker expected limits should be employed.  These
weaker limits for example allow $\xi^2$ as large as $0.52$ at $\mhi\sim
95\gev$ and $0.9$ for $\mhi\sim 100\gev$. These weaker limits allow
ample room for the majority of the $\mai\lsim 2m_B$ ideal Higgs scenarios.

\section{Conclusions}

In this paper, we have updated the constraints on the NMSSM ideal
Higgs scenarios in which $\hi$ (and for low $\tanb$, also possibly
$\hii$) has mass $\lsim 105\gev$ and decays largely (but not entirely)
via $\hi\to \ai\ai \to \tauptaum\tauptaum$.  Such low mass(es) for the
Higgs boson(s) with large $VV$ coupling are strongly preferred by
precision electroweak data and are also strongly preferred in order to
minimize electroweak finetuning. Indeed, all the NMSSM points plotted
in this paper have effective precision electroweak mass below $\sim
105\gev$.  The new data that constrains such scenarios derives from
$\upsiii\to \gam \mupmum$ and $\gam\tauptaum$ decay data from BaBar
and ALEPH studies of the $\epem \to Z 4\tau$ final state. The latter
was employed by ALEPH to place limits as a function of $\mhi$ and
$\mai$ on the quantity $\xi^2\equiv {\sigma(\hi)\over
  \sigma(\hsm)}\br(\hi\to\ai\ai)\left[\br(\ai\to \tauptaum)\right]^2$.
Although these new constraints are significant, there is still ample
room for the ideal Higgs scenarios, especially if $\tanb$ is small and
$\mai\lsim 2m_B$ (the latter region being that for which the
``light-$\ai$'' finetuning measure is minimal and also $\br(\ai\to
\tauptaum)$ is somewhat suppressed).  For $\tanb\geq 3$, it is only
the $\mai\lsim 2m_B$ points that can escape the ALEPH $\xi^2$ limits.
The case of $\tanb=3$ is the most marginal with only a few NMSSM
points with $\mhi\leq 99\gev$ (the rough upper limit on $\mhi$ at
$\tanb=3$) having $\xi^2$ essentially equal to the ALEPH limit at a
given $\mhi$.  For $\tanb=10$, one finds scenarios with $\mhi\sim
100-105\gev$ and $\xi^2\sim 0.43$ when $\mai\lsim 2m_B$, which $\xi^2$
is well below the ALEPH limit of $\sim 0.52-0.85$ for such $\mhi$ and
$\mai$. At $\tanb=50$, although our scanning statistics were limited,
we found points with $\mhi\sim 100\gev$ and $\mai\lsim 2m_B$ having
$\xi^2$ below the $0.52$ ALEPH limit. (We note that the ALEPH limits
are significantly stronger than the ALEPH collaboration was expecting.
If one were to use expected limits instead then the $\tanb\geq 3$
scenarios would be much less constrained.)  For $\tanb\lsim 2$, the
ideal-Higgs NMSSM scenarios are not particularly constrained by the
ALEPH limits.  In particular, for $\tanb=2,1.7,1.2$ one finds
$\mhi\leq 100\gev$ scenarios with $\xi^2\lsim 0.32,0.23,0.15$,
respectively.  The lower $\xi^2$ values arise because these lower
$\tanb$ values have increasingly reduced $\br(\ai\to \tauptaum)$,
which, in turn, is due to increasingly larger values of $\br(\ai\to
gg+c\anti c)$. Such $\xi^2$ values are completely consistent with the
ALEPH limits. 

The Tevatron and LHC discovery prospects for the Higgs
bosons in the low-$\tanb$ scenarios have yet to be fully analyzed.
Searches for the $\hi$ and the $\ai$ using the $\ai\to\tauptaum$ and
$\ai\to\mupmum$ decay modes will certainly become more difficult as
these branching ratios decline with decreasing $\tanb$. Such search
modes include: direct (vs. coming from $\hi\to\ai\ai$) detection of
the $\ai$ at the Tevatron and LHC in the $gg\to\ai\to\mupmum$ channel
\cite{Dermisek:2009fd}; searches for $gg\to \hi\to \ai\ai\to
\tauptaum\tauptaum$, $\tauptaum\mupmum$ and/or $\mupmum\mupmum$ at the
Tevatron~\cite{Abazov:2009yi} and LHC~\cite{Lisanti:2009uy}; and LHC
detection of $pp\to pp\hi$ with $\hi\to \ai\ai\to\tauptaum\tauptaum$
\cite{Forshaw:2007ra}.  Backgrounds in the increasingly important
channels with $\ai\to gg+c\anti c$ will undoubtedly be much larger and
will make discovery employing these latter $\ai$ decay modes quite
difficult.

As part of the NMSSM study, we first obtained updated limits on the $a
b\anti b$ coupling (assuming $\cabb=\catautau=\camumu$) that are
applicable in a wide variety of model contexts.  The main 
improvements in these general limits result from recent BaBar data.

Finally, one should not forget that the NMSSM is only the simplest
model of a general category of SUSY models having one or more singlet
scalar superfields in addition to the usual two-doublet scalar
superfields. Such models are generically very attractive in that they
allow for an NMSSM-like solution to the $\mu$ problem, while maintaining
coupling constant unification and RGE electroweak symmetry breaking
as in the MSSM. In addition, models with more than one extra singlet scalar
superfield will allow one or more light Higgs bosons with
SM-like couplings to $VV$ (a scenario having excellent agreement with
precision electroweak constraints and minimal electroweak finetuning)
that can escape Upsilon and LEP limits more easily than the NMSSM by virtue
of multiple decays channels of the Higgs$\to a_k a_j$, \ldots type.

\acknowledgments 

During the course of this work, JFG was supported by U.S. DOE grant
No. DE-FG03-91ER40674 and as a scientific associate at CERN.  We would
like to thank Y. Kolomensky, A. Mokhtar, and A. Snyder for assistance
in obtaining access to numerical tables of BaBar results and related
discussions. We also thank Kyle Cranmer for discussions regarding the
ALEPH results.

\end{document}